{}
{}

\documentclass[12pt,a4paper]{article}

\newif\ifpublic\publictrue
\newif\ifniklas\niklasfalse
\newif\ifjournal\journalfalse

\usepackage{ifpdf}
\ifniklas\ifpdf\publictrue\else\publicfalse
\PassOptionsToPackage{hypertex}{hyperref}
\PassOptionsToPackage{draft}{graphicx}
\fi\fi

\usepackage{amsmath,amssymb}
\ifpublic\else\usepackage{showkeys}\fi
\usepackage[bookmarks=true,hyperfigures=true]{hyperref}
\usepackage{graphicx}
\usepackage[nosort]{cite}
\usepackage[bulletsep]{collref}

\ifniklas

\def\showkeysrefformat#1{{\normalfont\tiny\ttfamily#1}}
\makeatletter
\def\SK@@ref#1>#2\SK@{%
 {\@inlabelfalse\leavevmode\vbox to\z@{%
 \vss\SK@refcolor\rlap{\vrule\raise .75em%
  \hbox{\showkeysrefformat{#2}}}}}}
\makeatother
\fi

\providecommand{\hypersetup}[1]{}
\providecommand{\hyperref}[2][]{#2}
\providecommand{\texorpdfstring}[2]{#1}
\providecommand{\pdfbookmark}[3][]{}

\hypersetup{plainpages=false}
\hypersetup{pdfpagemode=UseOutlines}
\hypersetup{bookmarksnumbered=true}
\hypersetup{bookmarksopen=true}
\hypersetup{pdfstartview=FitH}
\hypersetup{colorlinks=false}
\hypersetup{citebordercolor={.6 .9 .6}}
\hypersetup{urlbordercolor={.7 .8 1}}
\hypersetup{linkbordercolor={1 .7 .7}}
\hypersetup{pdfborder={0 0 .5}}

\makeatletter
\let\@keywords\@empty
\let\@subject\@empty
\providecommand{\keywords}[1]{\gdef\@keywords{#1}}
\providecommand{\subject}[1]{\gdef\@subject{#1}}
\def\thetitle{\@title}
\def\theauthor{\@author}
\def\thesubject{\@subject}
\def\thedate{\@date}
\def\thekeywords{\@keywords}
\makeatother
\AtBeginDocument{
\hypersetup{pdftitle={\thetitle}}%
\hypersetup{pdfauthor={\theauthor}}%
\hypersetup{pdfsubject={\thesubject}}%
\hypersetup{pdfkeywords={\thekeywords}}%
}

\makeatletter
\newsavebox{\apb@box}\newlength{\apb@width}
\newcommand{\autoparbox}[2][c]{\sbox{\apb@box}{#2}%
 \settowidth{\apb@width}{\usebox{\apb@box}}%
 \parbox[#1]{\apb@width}{\usebox{\apb@box}}}
\newcommand{\includegraphicsbox}[2][]{\autoparbox{\includegraphicsex[#1]{#2}}}
\makeatother

\RequirePackage{verbatim}

\makeatletter
\newwrite\bibinl@out
\newenvironment{bibtex}[1][\jobname]{%
  \immediate\openout\bibinl@out #1.bib
  \immediate\write\bibinl@out{\@percentchar generated from `\jobname' starting line \the\inputlineno^^J}%
  \def\verbatim@processline{\immediate\write\bibinl@out{\the\verbatim@line}}%
  \@bsphack\let\do\@makeother\dospecials\catcode`\^^M\active\verbatim@start
}%
{\immediate\closeout\bibinl@out\@esphack}
\makeatother

\RequirePackage{verbatim}

\makeatletter
\newwrite\mpi@out
\def\mpi@write#1{\immediate\write\mpi@out{#1}}
\immediate\openout\mpi@out\jobname.mp
\mpi@write{\@percentchar generated from `\jobname'}%
\AtEndDocument{\immediate\closeout\mpi@out}

\newcommand{\mpi@putlineno}{%
  \mpi@write{\@percentchar---------------------------------------}%
  \mpi@write{\@percentchar l.\the\inputlineno}%
}

\newcommand{\mpi@verbatim}{
  \@bsphack
  \let\do\@makeother\dospecials
  \catcode`\^^M\active
  \def\verbatim@processline{\mpi@write{\the\verbatim@line}}%
  \verbatim@start
}

\newenvironment{mpostcmd}{%
  \mpi@putlineno%
  \mpi@verbatim%
}%
{\mpi@write{}\@esphack}

\newenvironment{mpostfile}[1]{%
  \mpi@putlineno%
  \mpi@write{filenametemplate "#1";}%
  \mpi@write{beginfig(0)}%
  \mpi@verbatim%
}%
{\mpi@write{endfig;}\@esphack}

\newcommand{\includegraphicsex}[2][]{%
  \xdef\mpi@tmp{#2}%
  \IfFileExists{\mpi@tmp}%
    {\includegraphics[#1]{\mpi@tmp}}%
    {\textbf{??}\typeout{file \mpi@tmp{} missing}}%
}

\makeatother

\allowdisplaybreaks[3]

\numberwithin{equation}{section}

\usepackage[a4paper,text={450pt,650pt},centering]{geometry}

\usepackage[font=small,labelfont=bf,width=0.85\textwidth]{caption}

\expandafter\def\expandafter\bfseries\expandafter{\bfseries\ifmmode\else\boldmath\fi}
\expandafter\def\expandafter\mdseries\expandafter{\mdseries\ifmmode\else\unboldmath\fi}
\expandafter\def\expandafter\normalfont\expandafter{\normalfont\ifmmode\else\unboldmath\fi}

\usepackage{fixmath}

\newcommand{\gen}[1]{\mathfrak{#1}}

\newcommand{\geny}[1]{\mathfrak{\widehat{#1}}}
\newcommand{\alg}[1]{\mathfrak{#1}}
\newcommand{\grp}[1]{\mathrm{#1}}
\newcommand{\rep}[1]{\mathbf{#1}}

\newcommand{\superN}{\mathcal{N}}
\newcommand{\sym}{$\superN=4$ SYM}
\newcommand{\scs}{$\superN=6$ SCS}

\newcommand{\amp}{A}

\makeatletter\let\Re\@undefined\let\Im\@undefined\makeatother
\DeclareMathOperator{\Re}{Re}
\DeclareMathOperator{\Im}{Im}

\DeclareMathOperator{\sgn}{sgn}
\DeclareMathOperator{\csgn}{sgn_c}

\newcommand{\cabs}[1]{\abs{#1}\indup{c}}
\newcommand{\cdelta}{\mathop{\delta_c}}

\newcommand{\Reals}{\mathbb{R}}

\newcommand{\ff}{f\kern-5pt f}

\newcommand{\dd}{d}
\newcommand{\eps}{\varepsilon}

\newcommand{\deltad}[1]{\delta^{#1}}
\newcommand{\defeq}{\mathrel{:=}}

\ifx\genfrac\sdflkaj\else\fi
\newcommand{\sfrac}[2]{{\textstyle\frac{#1}{#2}}}
\newcommand{\half}{\sfrac{1}{2}}

\newcommand{\indup}[1]{_{\mathrm{#1}}}

\newcommand{\brk}[1]{(#1)}

\newcommand{\bigbrk}[1]{\bigl(#1\bigr)}

\newcommand{\abs}[1]{|#1|}

\newcommand{\sprods}[1]{\langle#1\rangle}

\newcommand{\nn}{\nonumber}
\newcommand{\nln}{\nonumber\\}

\def\[{\begin{equation}}
\def\]{\end{equation}}

\makeatletter
\def\mr@ignsp#1 {\ifx\:#1\@empty\else #1\expandafter\mr@ignsp\fi}%
\newcommand{\multiref}[1]{\begingroup
\xdef\mr@no@sparg{\expandafter\mr@ignsp#1 \: }%
\def\mr@comma{}%
\@for\mr@refs:=\mr@no@sparg\do{\mr@comma\def\mr@comma{,}\ref{\mr@refs}}%
\endgroup}
\makeatother

\renewcommand{\eqref}[1]{(\multiref{#1})}

\makeatletter
\newcommand{\namedref}[2]{\hyperref[#2]{#1~\ref*{#2}}}
\newcommand{\secref}{\@ifstar{\namedref{Section}}{\namedref{Section}}}
\newcommand{\fnref}{\@ifstar{\namedref{Footnote}}{\namedref{Footnote}}}
\newcommand{\appref}{\@ifstar{\namedref{Appendix}}{\namedref{Appendix}}}
\newcommand{\tabref}{\@ifstar{\namedref{Table}}{\namedref{Table}}}
\newcommand{\figref}{\@ifstar{\namedref{Figure}}{\namedref{Figure}}}
\makeatother

\begin{mpostcmd}

verbatimtex 
\documentclass{article}
\usepackage{amsfonts,amssymb}
\usepackage{latexsym}
\newcommand{\gen}[1]{\mathrm{#1}}

\begin{document}
etex

\end{mpostcmd}

\ifjournal
\begin{mpostcmd}
prologues:=3; 
picture copyrightline[],copyleftline;
copyrightline[1] := btex \sf Prepared for J. Phys. A etex;
copyrightline[2] := copyrightline[1];
copyleftline := btex $\spadesuit$ etex;
\end{mpostcmd}
\else
\begin{mpostcmd}
prologues:=2; 
picture copyrightline[],copyleftline;
copyrightline[1] := btex \copyright\ \textsf{2012 Niklas Beisert} etex;
copyrightline[2] := btex \copyright\ \textsf{2012 Tristan McLoughlin} etex;
copyleftline := btex $\circledast$ etex;
\end{mpostcmd}
\fi

\begin{mpostcmd}
def putcopyspace = 
label.bot(btex \vphantom{gA} etex scaled 0.1, lrcorner(currentpicture));
enddef;
def putcopy(expr nn) = 
label.ulft(copyrightline[nn] scaled 0.1, lrcorner(currentpicture)) withcolor 0.9white;
label.urt(copyleftline scaled 0.1, llcorner(currentpicture)) withcolor 0.9white;
currentpicture:=currentpicture shifted (10.5cm,14cm);
enddef;

ahlength:=7pt;

dotsize:=5pt;
dashdist:=6pt;
dotdist:=3pt;
xu:=1cm;
yu:=1cm;

wiggly_len := 3mm; 
wiggly_slope := 60;
curly_len := 3mm;
zigzag_len := 2mm;
zigzag_width := 1.5pt;
double_size := 2pt;
double_width := 1pt;

def colamps=withcolor 0.6white enddef;
def colampe=withcolor 0.7white enddef;
def colamp=withcolor 0.8white enddef;
def colgen=withcolor 0.7white enddef;
def colloop=withcolor 0.9white enddef;

def colblue=withcolor (0,0,0.8) enddef;
def colgreen=withcolor (0,0.6,0) enddef;
def colred=withcolor (1,0,0) enddef;

def colgrey=withcolor (0.5,0.5,0.5) enddef;
def colbgl=withcolor 0.7white enddef;

def pensize(expr s)=withpen pencircle scaled s enddef;
def thinpen=pensize(0.1pt) enddef;
def mediumpen=pensize(0.2pt) enddef;
def normalpen=pensize(0.4pt) enddef;
def thickpen=pensize(0.6pt) enddef;
def fatpen=pensize(0.8pt) enddef;

def dashdots=dashed withdots scaled (dotdist/5pt) enddef;
def dasheven=dashed evenly scaled (dashdist/6pt) enddef;
def dash(expr a,b)=dashed dashpattern(on (a*b) off ((1-a)*b)) enddef;

def onwhite(expr p)=unfill bbox p; draw p; enddef;

vardef pixlen (expr p, n) =
  for k=1 upto length(p): + segment_pixlen (subpath (k-1,k) of p, n) endfor
enddef;
vardef segment_pixlen (expr p, n) =
  for k=1 upto n: + abs (point k/n of p - point (k-1)/n of p) endfor
enddef;

vardef wiggly expr p_arg =
 save wpp;
 numeric wpp;
 wpp = ceiling (pixlen (p_arg, 10) / wiggly_len) / length p_arg;
 for k=0 upto wpp*length(p_arg) - 1:
  point k/wpp of p_arg
       {direction k/wpp of p_arg rotated wiggly_slope} ..
  point (k+.5)/wpp of p_arg
       {direction (k+.5)/wpp of p_arg rotated - wiggly_slope} ..
 endfor
 if cycle p_arg: cycle else: point infinity of p_arg fi
enddef;
vardef curly expr p =
 save cpp;
 numeric cpp;
 cpp := ceiling (pixlen (p, 10) / curly_len) / length p;
 if cycle p:
   for k=0 upto cpp*length(p) - 1:
     point (k+.33)/cpp of p
           {direction (k+.33)/cpp of p rotated 90} ..
     point (k-.33)/cpp of p
           {direction (k-.33)/cpp of p rotated -90} ..
   endfor
   cycle
 else:
   point 0 of p
         {direction 0 of p rotated -90} ..
   for k=1 upto cpp*length(p) - 1:
     point (k+.33)/cpp of p
           {direction (k+.33)/cpp of p rotated 90} ..
     point (k-.33)/cpp of p
           {direction (k-.33)/cpp of p rotated -90} ..
   endfor
   point infinity of p
         {direction infinity of p rotated 90}
 fi
enddef;
vardef zigzag expr p =
 save zpp;
 numeric zpp;
 zpp = ceiling (pixlen (p, 10) / zigzag_len) / length p;
 if not cycle p:
   point 0 of p --
 fi
 for k = 0 upto zpp*length(p) - 1:
   point (k+1/4)/zpp of p shifted
     (zigzag_width
      * dir angle (direction (k+1/4)/zpp of p rotated 90)) --
   point (k+3/4)/zpp of p shifted
     (zigzag_width
      * dir angle (direction (k+3/4)/zpp of p rotated -90)) --
 endfor
 if cycle p:
   cycle
 else:
   point infinity of p
 fi
enddef;

def midarrowperc (expr p, t) =
  fill arrowhead subpath(0,arctime(t*arclength(p)+0.5ahlength) of p) of p;
enddef;

def midarrow (expr p, t) =
  fill arrowhead subpath(0,arctime(arclength(subpath (0,t) of p)+0.5ahlength) of p) of p;
enddef;

def filldot(expr z,c)=
fill fullcircle scaled dotsize shifted z withcolor c;
draw fullcircle scaled dotsize shifted z normalpen;
enddef;

def fillbox(expr z,c)=
fill ((0.5,0.5)--(0.5,-0.5)--(-0.5,-0.5)--(-0.5,0.5)--cycle) scaled dotsize shifted z withcolor c;
draw ((0.5,0.5)--(0.5,-0.5)--(-0.5,-0.5)--(-0.5,0.5)--cycle) scaled dotsize shifted z normalpen;
enddef;

def fillsbox(expr z,c)=
fill ((0.7,0)--(0,0.7)--(-0.7,0)--(0,-0.7)--cycle) scaled dotsize shifted z withcolor c;
draw ((0.7,0)--(0,0.7)--(-0.7,0)--(0,-0.7)--cycle) scaled dotsize shifted z normalpen;
enddef;

def drawcrossg expr z=
draw ((-1,-1)--(1,1)) scaled dotsize shifted z pensize(0.5pt) colgreen;
draw ((-1,1)--(1,-1)) scaled dotsize shifted z pensize(0.5pt) colgreen;
enddef;

def drawcrossb expr z=
draw ((-1,-1)--(1,1)) scaled dotsize shifted z pensize(0.5pt) colblue;
draw ((-1,1)--(1,-1)) scaled dotsize shifted z pensize(0.5pt) colblue;
enddef;

def drawcrossr expr z=
draw ((-1,-1)--(1,1)) scaled dotsize shifted z pensize(0.5pt) colred;
draw ((-1,1)--(1,-1)) scaled dotsize shifted z pensize(0.5pt) colred;
enddef;

pair vpos[];

numeric spins[];

picture spinori[];

path paths[];
path cir;

def drawleg expr p=draw p pensize(1.0pt) enddef;
def drawdleg expr p=draw p pensize(5.0pt); undraw p pensize(2.0pt) enddef;
def drawdlegl expr p=draw p pensize(5.0pt); draw p pensize(2.0pt) colloop enddef;

def drawgluon expr p=draw (wiggly p) pensize(1.0pt) enddef;
def undrawgluon expr p=undraw (wiggly p) pensize(5pt) enddef;
def drawscalar expr p=draw p pensize(1.5pt) enddef;
def undrawscalar expr p=undraw subpath (0.2,0.8) of p pensize(5pt) enddef;
def drawfermion expr p=draw p pensize(1.5pt) dasheven enddef;
def drawany expr p=draw p pensize(1.5pt) dashdots enddef;
def drawother expr p=draw p pensize(1.5pt) dasheven enddef;
def drawcut expr p=draw p pensize(0.5pt) dashed evenly scaled (dotdist/5pt) enddef;
def drawccut expr p=draw p pensize(0.5pt) dashed withdots scaled (dotdist/5pt) enddef;

def drawdashone   expr p=draw p pensize(0.5pt) enddef;
def drawdashtwo   expr p=draw p pensize(0.5pt) dashed dashpattern(on 1bp off 1bp on 0bp off 1bp on 1bp) enddef;
def drawdashthree  expr p=draw p pensize(0.5pt) dashed dashpattern(on 2bp off 2bp on 2bp) enddef;
def drawdashzero expr p=draw p pensize(0.7pt) dashed dashpattern(off 1bp on 0bp off 1bp on 0bp off 1bp on 0bp off 1bp) enddef;

def drawvertex expr z=
fill fullcircle scaled dotsize shifted z;
enddef;

def drawfield expr z=
filldot(z, white);
enddef;

def drawcfield expr z=
filldot(z, 0.5white);
enddef;

\end{mpostcmd}


\providecommand{\href}[2]{#2}
\newcommand{\arxivlink}[1]{\href{http://arxiv.org/abs/#1}{arxiv:#1}}

\title{Conformal Anomaly for Amplitudes in
\texorpdfstring{\\}{}
\texorpdfstring{$\superN=6$}{N=6} Superconformal Chern--Simons Theory}
\author{%
Till Bargheer\texorpdfstring{$^{afg}$}{},
Niklas Beisert\texorpdfstring{$^{bdfg}$}{},
\texorpdfstring{\\}{}
Florian Loebbert\texorpdfstring{$^{cefg}$}{},
Tristan McLoughlin\texorpdfstring{$^{dgh}$}{}}

\begin{document}

\pdfbookmark[1]{Title Page}{title}


\thispagestyle{empty}

\noindent%
{\footnotesize
\begin{tabular*}{\textwidth}{@{\extracolsep{\fill}}lcr@{}}
\texttt{NSF-KITP-12-012}&\texttt{\arxivlink{1204.4406}}
&\texttt{UUITP-10/12}\\[-.1cm]%
\texttt{LPT-ENS-12-16}&\ &\texttt{AEI-2012-037}%
\end{tabular*}}

{\small
\vspace{0.5cm}
\begin{center}%
\begingroup\Large\bfseries\thetitle\par\endgroup
\vspace{.5cm}

\begingroup\scshape\theauthor\par\endgroup
\vspace{5mm}%

\begingroup\itshape
$^a$
Department of Physics and Astronomy,
Uppsala University\\
SE-751 08 Uppsala, Sweden
\vspace{3mm}

$^b$
Institut f\"ur Theoretische Physik,
Eidgen\"ossische Technische Hochschule Z\"urich\\
Wolfgang-Pauli-Strasse 27, 8093 Z\"urich, Switzerland
\vspace{3mm}

$^c$
Niels Bohr International Academy \& Discovery Center,
Niels Bohr Institute,\\
Blegdamsvej 17, 2100 Copenhagen, Denmark
\vspace{3mm}

$^d$
Max-Planck-Institut f\"ur Gravitationsphysik, 
Albert-Einstein-Institut\\
Am M\"uhlenberg 1, 14476 Potsdam, Germany
\vspace{3mm}

$^e$
Laboratoire de Physique Th\'eorique,
\'Ecole Normale Sup\'erieure\\
24 Rue Lhomond, 75005 Paris, France
\vspace{3mm}

$^f$
Kavli Institute for Theoretical Physics,
University of California\\
Santa Barbara, CA 93106, USA
\vspace{3mm}

$^g$
Nordita,\\
Roslagstullsbacken 23, 106 91 Stockholm, Sweden
\vspace{3mm}

$^h$
Centre de Physique Th{\'e}orique,
Ecole Polytechnique, CNRS\\
91128 Palaiseau cedex, France
\par\endgroup
\vspace{5mm}

\begingroup\ttfamily
\texttt{%
till.bargheer@physics.uu.se,
nbeisert@ethz.ch,\\
loebbert@nbi.dk,
tmclough@aei.mpg.de}
\par\endgroup

\vspace{.8cm}

\textbf{Abstract}\vspace{7mm}

\begin{minipage}{12.7cm}
Scattering amplitudes in three-dimensional ${\cal N}=6$ Chern--Simons
theory are shown to be non-invariant with respect to the free representation
of the $\alg{osp}(6|4)$ symmetry generators. At tree and one-loop level  these ``anomalous"
terms  occur only for non-generic,  singular configurations of the external momenta
and can be used to determine the form of the amplitudes. In particular we show 
that the symmetries predict that the one-loop six-point amplitude is non-vanishing 
and confirm this by means of an explicit calculation using generalized unitarity methods. 
We comment on the implications of this finding for any putative Wilson loop/amplitude 
duality in ${\cal N}=6$ Chern--Simons theory. 
\end{minipage}

\end{center}
}

\newpage

\setcounter{tocdepth}{2}
\hrule height 0.75pt
\pdfbookmark[1]{\contentsname}{contents}
\tableofcontents
\vspace{0.8cm}
\hrule height 0.75pt
\vspace{1cm}

\setcounter{tocdepth}{2}

\section{Introduction}

A conformal field theory has no notion of distance. Consequently, two
massless particles moving collinearly cannot be distinguished
from each other in such a theory. The standard formalism for scattering
theory, however, distinguishes the different external particles of an
amplitude---even if the particles have no mass. In the
conformal four-dimensional $\superN=4$ super Yang--Mills (SYM) theory this is reflected in the
fact that the standard, free representation of conformal symmetry on scattering
amplitudes produces an anomaly
for collinear momentum configurations. In
fact, one finds no real quantum anomaly of the symmetry but rather an anomaly of the
representation which can be cured by deformation terms 
\cite{Bargheer:2009qu,Beisert:2010gn}, see also \cite{Bargheer:2011mm}.%
\footnote{Here, and in the following, we consistently refer
to the variation of the scattering amplitude with respect to the free representation
of a symmetry generator as the anomaly. We maintain this usage even
when the anomaly may in fact be absorbed into a deformation of the representation of the
generator which 
preserves the symmetry algebra.}
In four dimensions, this superficial
violation of conformal symmetry is closely related to the holomorphic
anomaly defined by the
equation
\begin{equation}\label{eq:holanom}
\frac{\partial}{\partial \bar z}\,\frac{1}{z}=\pi \delta^2(z)~.
\end{equation}
In $(3,1)$ signature, the massless momenta of scattering amplitudes are conveniently expressed in terms
of two complex conjugate spinors $\lambda$ and $\bar \lambda$. These take the
place of $z$ and $\bar z$ in the above equation such that the naive
conformal generators (e.g.\ $\gen{\bar S}=\eta\partial/\partial\bar\lambda$) annihilate tree-level amplitudes only up to
distributional terms. Switching to $(2,2)$ signature, the solution of
the masslessness condition $p^2=0$ is given by two independent real spinors
$\lambda$ and $\tilde \lambda$ and the anomaly cannot be phrased in terms
of the above complex equation anymore. In fact, the anomaly does not
disappear in $(2,2)$ signature but it is harder to see
\cite{Beisert:2010gn}. Rewriting the amplitudes in this signature shows
that the anomaly can be captured in terms of the singularity of a signum
function with real argument $x$ (corresponding to the real $\lambda$, $\tilde\lambda$):
\begin{equation}\label{eq:signanom}
\frac{\partial}{\partial x}\sgn{x}=2\delta(x),
\end{equation}
After all, the resulting additional contributions to the invariance equations
following from symmetry
seem to be essential for fixing the complete scattering
matrix of \sym\ theory.

Certainly one can ask whether there is an analog of this anomaly of the conformal
symmetry representation in dimension number different than four.  As the naive symmetry representation
is still expected to be incompatible with the standard definition of scattering
states, the existence of a similar phenomenon is plausible. 
In this paper we study superconformal three-dimensional $\superN=6$ Chern--Simons (SCS) theory, also called ABJM theory \cite{Aharony:2008ug}, where massless momenta are described by a single real
spinor. Consequently, one cannot expect an anomaly to be of the  form of the complex equation \eqref{eq:holanom}.
We will show, however, that the anomaly in three dimensions takes the form of \eqref{eq:signanom}, cf.\ \tabref{tab:anomtab}. 
As an application we predict a non-vanishing one-loop amplitude at six points and verify this result
by an explicit unitarity construction. First, however, we give a brief motivation for the importance of the above anomalies 
followed by a short review of scattering amplitudes in ABJM theory.

\begin{table}
\begin{center}
\begin{tabular}{|c|c||c|c|}\hline
Signature&3d&Signature&4d\\\hline\rule{0pt}{2.5ex}
(2,1)&$\frac{\partial}{\partial x}\sgn{x}=2\delta(x)$&(2,2)&$\frac{\partial}{\partial x}\sgn{x}=2\delta(x)$\\[2pt]
&$\lambda$ real &&$\lambda$, $\tilde \lambda$ real\\\hline\rule{0pt}{2.5ex}
---&---&(3,1)&$\frac{\partial}{\partial \bar z}\frac{1}{z}=\pi\delta^2(z)$\\[3pt]
&&&$\lambda$, $\bar \lambda$ complex\\\hline
\end{tabular}
\caption{The conformal anomaly of scattering amplitudes depends on the signature and dimension. In four dimensions the holomorphic anomaly in $(3,1)$ signature is replaced by a sign-anomaly in $(2,2)$ signature. In three dimensions, the anomly also takes the sign form.}
\label{tab:anomtab}
\end{center}
\end{table}

The emergence of the anomalies indicated above is particularly powerful in the context of planar scattering amplitudes,
where, when combined with other recent developments, it may optimistically lead 
to an all-order understanding of these observables. Once more let us consider the case of \sym\ theory: 
Its planar S-matrix is known to possess a dual-conformal symmetry, 
both at strong \cite{Alday:2007hr} and weak \cite{Drummond:2006rz, Drummond:2008vq, Brandhuber:2008pf} coupling, which combines with the usual superconformal symmetry
into a Yangian symmetry algebra \cite{Drummond:2009fd}.
Furthermore it has been understood as originating in the self-T-duality of the
AdS/CFT-dual AdS$_5\times$S$^5$ geometry \cite{Berkovits:2008ic, Beisert:2008iq}.
Under this duality, scattering amplitudes are mapped into Wilson loops; specifically, 
it has been shown that MHV amplitudes are dual to bosonic Wilson loops  
\cite{Alday:2007hr, Drummond:2007aua,Brandhuber:2007yx,Drummond:2007cf},  
while for
superamplitudes the dual object is a generalized 
super-Wilson loop \cite{Mason:2010yk, CaronHuot:2010ek} 
(closely related objects are light-cone supercorrelation functions \cite{Eden:2011yp,Eden:2011ku}).
However, for the reasons mentioned, the discussed symmetries of scattering amplitudes 
are anomalous.%
\footnote{For the dual Wilson loops 
the origin of the anomaly is conceptually different and arises from UV effects. The functional form of the anomalies, however, is closely related.}
These anomalies, particularly those of the fermionic generators, and how they relate 
different tree-level amplitudes has been discussed 
in \cite{Bargheer:2009qu, Korchemsky:2009hm, Sever:2009aa}. 
For MHV-amplitudes/bosonic-Wilson loops, understanding the conformal anomaly provides strong
constraints at all orders in the coupling and allows the complete determination 
of the four- and five-point cases \cite{Drummond:2007au}.
The anomalies of superamplitudes/super-Wilson loops have been studied at loop level in \cite{Sever:2009aa, Beisert:2010gn, CaronHuot:2011ky} and recently, by making use of these anomalies for 
the fermionic symmetries, all-order equations relating higher-loop superamplitudes to lower-loop
ones have been found \cite{CaronHuot:2011kk, Bullimore:2011kg}.

It is natural to ask whether similar results can be obtained for scattering amplitudes in 
three dimensions and particularly for planar amplitudes in \scs\ theory, though 
here the picture is still significantly less clear.
While so far all attempts to consistently formulate a self-T-duality for the string 
background dual to the ABJM theory (cf.\ \cite{Grassi:2009yj,Adam:2009kt,Adam:2010hh,Dekel:2011qw,Bakhmatov:2010fp,Bakhmatov:2011aa,Colgain:2012ca}) have failed, there are strong indications pointing towards the existence of such a map. One of them is the discovery of Yangian symmetry of ABJM scattering amplitudes \cite{Bargheer:2010hn} and particularly its formal rewriting in terms of ``dual'' coordinates \cite{Huang:2010qy}. Furthermore, at four points the two-loop scattering amplitude was shown to match the two-loop Wilson loop \cite{Chen:2011vv,Bianchi:2011dg}. In fact, there is a remarkable congruence between the structure of the two-loop  ABJM  Wilson loop
and the one-loop \sym\ Wilson loop, which extends to arbitrary number of edges
 \cite{Wiegandt:2011uu}
and which parallels a similar relation in the spectrum of planar anomalous dimensions persisting
to all orders in the coupling. On the other hand, 
the analogy with the map in \sym\ theory is complicated due to the absence of
an analog of the four dimensional helicity classification of amplitudes: Before comparison to the Wilson loop, the MHV part of the scattering amplitude has to be stripped off. The lack of helicity and thus of an MHV scheme in three dimensions is therefore crucial for understanding a possible analogy.%
\footnote{To match the ABJM four-point Wilson loop and scattering amplitude at two loops, the tree-level part of the amplitude was stripped off. Its form is close to the one of the MHV four-point amplitude in four dimensions while the ABJM six-point tree-level amplitude resembles the four dimensional NMHV counterpart.}
It is known 
that all lightlike polygonal $n$-point Wilson loops vanish at one-loop order \cite{Henn:2010ps, Bianchi:2011rn}. Consequently it is interesting to study one-loop amplitudes at higher points
to further understand any possible map. As we will see, the structure of the anomalous symmetries
predicts that the one-loop six-point amplitude is non-zero, a fact which we
confirm by an explicit generalized unitarity calculation.%
\footnote{After publishing this work on the
\href{http://arxiv.org/abs/arXiv:1204.4406}{arXiv}, we were informed that
also Simon Caron-Huot and Yu-tin Huang independently obtained
a non-vanishing one-loop six-point ABJM amplitude.}
This poses a puzzle as to what a possible Wilson loop/amplitude 
map could look like.

After a review of scattering amplitudes in ABJM theory in \secref{sec:amp_rev}, we will 
discuss the origin and form of the anomaly for the superconformal symmetry,
\secref{sec:anomaly}. Applying the resulting anomaly vertex to the six-point amplitude
at tree and one-loop level, \secref{sec:anomaly6}, we will see how this constrains
the form of the amplitude and predicts a non-vanishing result at one loop. In 
\secref{sec:OneLoopSixPoint} we perform an explicit calculation of the one-loop 
six-point function using generalized unitarity methods and confirm the prediction
of the symmetries. We close with a summary and discussion 
of some open questions.

\paragraph{Note Added:}
As this work was being completed we were informed by the authors of
\cite{Bianchi:2012cq}
that they had obtained related results for one-loop
amplitudes in ABJM and in particular the same result for the one-loop six-point
amplitude. This work has appeared concurrently on the arXiv.

\section{Scattering Amplitudes in ABJM Theory}
\label{sec:amp_rev}
In this section, we we briefly introduce the framework that is relevant to 
 the study of scattering amplitudes in the $\superN=6$ supersymmetric
Chern--Simons theory, or ABJM theory \cite{Aharony:2008ug}, 
which was developed in \cite{Agarwal:2008pu, Bargheer:2010hn, Huang:2010qy}.

\paragraph{Fields and States.}

The matter content of ABJM theory comprises four complex scalar
fields $\phi^A$ and four complex fermion fields $\psi^a_A$, $A=1,\dots,4$,
which transform in the $(\rep{N},\rep{\bar{N}})$ representation of the
$\grp{U}(N)\times\grp{U}(N)$ gauge group. The scalars $\phi^A$ form a
fundamental multiplet of the internal $\grp{SU}(4)$ R-symmetry group, while
the fermions $\psi_A$ form an antifundamental multiplet. The Chern--Simons
gauge fields $A_\mu$, $\hat{A}_\mu$ transforming in the
$(\rep{ad},\rep{1})$, $(\rep{1},\rep{ad})$ representations of the gauge
group have no freely propagating modes and thus cannot appear as external
states in scattering amplitudes.%
\footnote{Nevertheless, the Chern--Simons zero-mode will turn out to play a
significant role, see the discussion at the end of
\secref{sec:OneLoopSixPoint} and in \secref{sec:concl}.}
As both scalar and fermion particle numbers are conserved, this in particular
implies that there are no scattering amplitudes for odd numbers of
particles.

Unlike in four dimensions, there is no helicity degree of freedom for
massless states in three dimensions. Hence, one-particle states are solely
labeled by a massless momentum $p^\mu=\gamma^\mu_{ab}\lambda^a\lambda^b$,
which is conveniently parametrized by a spacetime spinor $\lambda$ using
the real Dirac gamma matrices $\gamma^\mu$. For momenta with positive
energy $p^0>0$ in Minkowski signature, $\lambda$ has to be real. For
negative-energy momenta, $\lambda$ has to be purely imaginary.

\paragraph{Superfields and Superamplitudes.}

All free on-shell states $\phi^A(\lambda)$, $\psi_A(\lambda)$ can be
combined in a single superfield $\Phi(\lambda,\eta)$ \cite{Bargheer:2010hn}
with the help of a $\alg{u}(3)$ Gra{\ss}mann spinor $\eta^A$,
\begin{equation}
\Phi(\lambda,\eta)=
	\phi^4(\lambda)+
	\eta^A\psi_A(\lambda)+
	\half\eps_{ABC}\eta^A\eta^B\phi^C(\lambda)+
	\sfrac{1}{6}\eps_{ABC}\eta^A\eta^B\eta^C\psi_4(\lambda)\,.
\label{eq:superfield}
\end{equation}
This choice of superfield splits the internal R-symmetry into a manifest
$\alg{u}(3)$ and a non-manifest remainder, realized as multiplication and
second-order differential operators in $\eta$'s. Its virtue is that the
supersymmetry and superconformal generators take the simple form
\begin{equation}
\gen{Q}^{aA}=\lambda^a\eta^A\,,
\quad
\gen{Q}^a_A=\lambda^a\partial_A\,,
\quad
\gen{S}_a^A=\eta^A\partial_a\,,
\quad
\gen{S}_{aA}=\partial_a\partial_A\,,
\label{eq:gens}
\end{equation}
where $\partial_a$, $\partial_A$ denote derivatives with respect to
$\lambda^a$, $\eta^A$. When splitting the matter fields into mutually
conjugate components $\phi^A$, $\bar\phi_A$, $\psi_A$, $\bar\psi^A$, it is
convenient to use not the conjugate superfield
$\bar\Phi(\bar\lambda,\bar\eta)$ itself, but its Gra{\ss}mann Fourier
transform
\begin{equation}
\bar\Phi(\lambda,\eta)=
	\bar\psi^4(\lambda)+
	\eta^A\bar\phi_A(\lambda)+
	\half\eps_{ABC}\eta^A\eta^B\bar\psi^C(\lambda)+
	\sfrac{1}{6}\eps_{ABC}\eta^A\eta^B\eta^C\bar\phi^4(\lambda)\,.
\end{equation}
With the help of the superfields $\Phi$, $\bar\Phi$, scattering amplitudes
for all possible combinations of $n$ external states combine into a single
superamplitude, from which individual component amplitudes can be extracted
as coefficients of the appropriate $\eta$-monomials.

In the planar limit $N\to\infty$, scattering amplitudes can be decomposed
into color-ordered amplitudes multiplied by planar color structures. The
objects we study in this work are the color-ordered superamplitudes
\begin{equation}
A_n=A_n(\bar\Lambda_1,\Lambda_2,\bar\Lambda_3,\dots,\bar\Lambda_{n-1},\Lambda_n)\,,
\quad
\Lambda_k=(\lambda_k,\eta_k)\,,
\end{equation}
where the ordering of the arguments is significant, and the bars signify
that the respective $\Lambda_k$'s parametrize conjugate fields $\bar\Phi$.
The color decomposition requires that $\Phi$ and $\bar\Phi$ fields
alternate, and implies invariance up to a sign under cyclic double-shifts,
\begin{equation}
A_n(\bar\Lambda_3,\dots,\Lambda_n,\bar\Lambda_1,\Lambda_2)
=(-1)^{n/2-1}A_n(\bar\Lambda_1,\Lambda_2,\bar\Lambda_3,\dots,\Lambda_n)\,.
\end{equation}
By convention, conjugate superfields are put in odd arguments of the
superamplitude. The sign is due to the fact that the conjugate field
$\bar\Phi$ is fermionic. Consequently under the transformation, 
 ``$\lambda$"-parity, or its supersymmetric 
generalization \cite{Gang:2010gy},
 $\Lambda\rightarrow -\Lambda$ the superamplitude transforms
as
\[
\label{eq:l_parity}
A_n(\Lambda_1, \dots, -\Lambda_i, \dots, \Lambda_n)=(-1)^i
A_n(\Lambda_1, \dots, \Lambda_i, \dots, \Lambda_n)~.
\]
The color-ordered amplitudes have another symmetry which is due to the
reflection invariance of the fundamental vertices in the Lagrangian. This inversion
symmetry is reflected in the following transformation behavior of the $\ell$-loop amplitude:%
\footnote{We thank Marco Bianchi, Matias Leoni, Andrea Mauri, Silvia Penati 
and Alberto Santambrogio for clarification of the loop dependence.}
\begin{equation}
A_n^{(\ell)}(\bar\Lambda_1, \Lambda_2, \dots, \bar\Lambda_{n-1}, \Lambda_n)=
(-)^{n(n-2)/8+\ell} A_n^{(\ell)}(\bar\Lambda_1, \Lambda_n,\bar\Lambda_{n-1},\dots, \Lambda_2).
\label{eq:Ainversion}
\end{equation}

\paragraph{On-Shell Integration.}

Below, we will frequently need to integrate over complete sets of on-shell
states. In the superfield language, such integrals take the simple form
\begin{equation}
\int \dd\Lambda\,f(\bar\Lambda)\,g(i\Lambda)\,,
\quad
\dd\Lambda=\dd^{2|3}\Lambda=\half\dd^2\lambda\,\dd^3\eta\,,
\label{eq:Lambdaintegral}
\end{equation}
where $i\Lambda\defeq(i\lambda,i\eta)$ switches the sign of both the
momentum $\lambda^a\lambda^b$ and the supermomentum $\lambda^a\eta^A$
relative to $\Lambda$. The integration often needs to include both real and
imaginary $\lambda$, that is the domain of integration for $\lambda$ is
$\Reals^2\cup(i\Reals)^2$. The factor $1/2$ accounts for the
double-counting of states due to the $\Lambda\to-\Lambda$ symmetry.
By substituting $\Lambda\to i\Lambda$, the integration over $(i\Reals)^2$
can be converted to an integration over $\Reals^2$ and vice versa:
\begin{align}
&  \int\dd\Lambda\,f(\bar\Lambda)\,g(i\Lambda)
=-i\int\dd\Lambda\,f(i\bar\Lambda)\,g(\Lambda)\nln
&
=\int_{\Reals}\dd\Lambda\,\bigbrk{f(\bar\Lambda)\,g(i\Lambda)-if(i\bar\Lambda)\,g(\Lambda)}
=\int_{i\Reals}\dd\Lambda\,\bigbrk{f(\bar\Lambda)\,g(i\Lambda)-if(i\bar\Lambda)\,g(\Lambda)}
\,.
\label{eq:Lambdaintegralexpansion}
\end{align}
This assumes that $f(-\bar\Lambda)=-f(\bar\Lambda)$ and
$g(-\Lambda)=g(\Lambda)$, which is the case for $f$ and $g$ being
scattering amplitudes, and will always be the case below.
Note that the integration measure $\dd\Lambda$ is fermionic
and transforms according to $\dd\Lambda\to (i^2/i^3)d\Lambda=-i\dd\Lambda$
under $\Lambda\to i\Lambda$.

\section{Anomaly Vertex in Three Dimensions}
\label{sec:anomaly}

In this section we first rewrite the four-point scattering amplitude of
ABJM theory in such a way that its dependence on a specific sign function
becomes explicit. The argument of this sign function is shown to be the
spinor bracket of two neighboring external particles. The sign thus changes
when two particles become collinear. We then show that this change of sign
leads to an anomaly of the conformal symmetry. We explicitly act with the
generator $\gen{S}$ \eqref{eq:gens} on the four-point amplitude which
yields an anomaly vertex supported on collinear momentum configurations.

\paragraph{Four-Point Amplitude.}

The four-point superamplitude of ABJM theory reads \cite{Bargheer:2010hn}, see also
\cite{Agarwal:2008pu},
\begin{equation}
\amp_4(\bar 1,2,\bar 3,4)=\frac{\deltad{3}(P)\,\deltad{6}(Q)}{\sprods{12}\sprods{23}}\,.
\label{eq:4pt}
\end{equation}
For positive energies $p^0$ (incoming particles), $\lambda$ is real, while
for negative energies (outgoing particles), $\lambda$ is purely imaginary.
Assuming that particles $1$ and $2$ carry the same energy sign, and particles
$3$ and $4$ carry the opposite energy sign, we find 
\begin{align}
1&=\abs{\sprods{12}}\int\dd\alpha_3\,\dd\beta_3\,\deltad{2}(\lambda_3-i\alpha_3\lambda_1+i\beta_3\lambda_2)\,,\nn\\
1&=\abs{\sprods{12}}\int\dd\alpha_4\,\dd\beta_4\,\deltad{2}(\lambda_4-i\alpha_4\lambda_1+i\beta_4\lambda_2)\,.
\label{eq:34as12}
\end{align}
Inserting these identities into the amplitude, the momentum delta function
becomes
\begin{align}
\deltad{3}(P)
&=
\deltad{3}\bigbrk{
	 \lambda_1\lambda_1(1-\alpha_3^2-\alpha_4^2)
	+\lambda_2\lambda_2(1-\beta_3^2-\beta_4^2)
	+(\lambda_1\lambda_2+\lambda_2\lambda_1)(\alpha_3\beta_3+\alpha_4\beta_4)
}\nn\\
&=
\frac{1}{\abs{\sprods{12}}^3}\,\delta(1-\alpha_3^2-\alpha_4^2)\,\delta(1-\beta_3^2-\beta_4^2)\,\delta(\alpha_3\beta_3+\alpha_4\beta_4)\,,
\end{align}
where the last equality holds as long as $\lambda_1$ and $\lambda_2$ are
linearly independent.
The four-point amplitude thus can be written as%
\footnote{The superscript $1,2\leftrightarrow3,4$ signifies how the four
particles are split into incoming (positive energy) and outgoing (negative
energy) particles. Expressions for different energy distributions are given below.}
\begin{multline}
\amp_4^{1,2\leftrightarrow3,4}(\bar 1,2,\bar 3,4)=-\frac{\deltad{6}(Q)}{\abs{\sprods{12}}\sprods{12}^2}\int\dd\alpha_3\,\dd\alpha_4\,\dd\beta_3\,\dd\beta_4\,\frac{1}{i\alpha_3}
	\cdot\\\cdot
	\delta(1-\alpha_3^2-\alpha_4^2)\,\delta(1-\beta_3^2-\beta_4^2)\,\delta(\alpha_3\beta_3+\alpha_4\beta_4)
	\cdot\\\cdot
	\deltad{2}(\lambda_3-i\alpha_3\lambda_1-i\beta_3\lambda_2)\,\deltad{2}(\lambda_4-i\alpha_4\lambda_1-i\beta_4\lambda_2)\,.
\end{multline}
Introducing polar coordinates
\begin{equation}
\alpha_3=r_\alpha\sin\alpha\,,\quad
\alpha_4=r_\alpha\cos\alpha\,,\quad
\beta_3=r_\beta\sin\beta\,,\quad
\beta_4=r_\beta\cos\beta\,,
\end{equation}
the radial integrations can be evaluated, leaving behind a Jacobi factor of
$1/4$:
\begin{multline}
\amp_4^{1,2\leftrightarrow3,4}(\bar 1,2,\bar 3,4)=-\frac{\deltad{6}(Q)}{\abs{\sprods{12}}\sprods{12}^2}\int\dd\alpha\,\dd\beta\,\frac{1}{4i\sin\alpha}\,\delta(\sin\alpha\sin\beta+\cos\alpha\cos\beta)
	\cdot\\\cdot
	\deltad{2}(\lambda_3-i\sin\alpha\,\lambda_1-i\sin\beta\,\lambda_2)\,\deltad{2}(\lambda_4-i\cos\alpha\,\lambda_1-i\cos\beta\,\lambda_2)\,.
\end{multline}
The first delta function localizes at
$
\beta=\alpha+s_1(2-s_2)\pi/2$,
with
$s_1,s_2=\pm 1$,
where
\begin{equation}
\sin\beta=s_1 s_2 \cos\alpha\,,\qquad
\cos\beta=-s_1 s_2\sin\alpha\,.
\label{eq:betasol}
\end{equation}
Here we can collect the two signs into one which yields an overall factor of 2:
\begin{multline}
\amp_4^{1,2\leftrightarrow3,4}(\bar 1,2,\bar 3,4)=-\frac{\deltad{6}(Q)}{\abs{\sprods{12}}\sprods{12}^2}\sum_{s=\pm1}\int_0^{2\pi}\dd\alpha\,\frac{1}{2i\sin\alpha}
	\cdot\\\cdot
	\deltad{2}(\lambda_3-i\sin\alpha\,\lambda_1-is\cos\alpha\,\lambda_2)\,
	\deltad{2}(\lambda_4-i\cos\alpha\,\lambda_1+is\sin\alpha\,\lambda_2)\,.
\end{multline}
Moving $\deltad{6}(Q)$ under the integral sign, contracting $Q$ once
with $\lambda_3$ and once with $\lambda_4$, and expanding
$\lambda_{3,4}$ in terms of $\lambda_{1,2}$ shows that
\begin{align}
\deltad{6}(Q)
&=\sprods{34}^{-3}\,\deltad{3}(\sprods{31}\eta_1+\sprods{32}\eta_2+\sprods{34}\eta_4)
                  \,\deltad{3}(\sprods{41}\eta_1+\sprods{42}\eta_2+\sprods{43}\eta_3)\nn\\
&=-s\sprods{12}^3\,\deltad{3}(\eta_3-i\sin\alpha\,\eta_1-is\cos\alpha\,\eta_2)
                 \,\deltad{3}(\eta_4-i\cos\alpha\,\eta_1+is\sin\alpha\,\eta_2)\,.
\end{align}
The four-point amplitude hence reads
\begin{multline}
\amp_4^{1,2\leftrightarrow3,4}(\bar 1,2,\bar 3,4)=\sgn(\sprods{12})\sum_{s=\pm1}\int_0^{2\pi}\dd\alpha\,\frac{s}{2i\sin\alpha}
	\cdot\\\cdot
	\deltad{2|3}(\Lambda_3-i\sin\alpha\,\Lambda_1-is\cos\alpha\,\Lambda_2)\,\deltad{2|3}(\Lambda_4-i\cos\alpha\,\Lambda_1+is\sin\alpha\,\Lambda_2)\,,
\end{multline}
where $\Lambda_k=(\lambda_k,\eta_k)$.
Reverting the direction of integration in one of the two terms under the
sum shows that $s$ accounts for a possible reflection in the rotation of
$\Lambda_{1,2}$ into $\Lambda_{3,4}$:
\begin{multline}
\amp_4^{1,2\leftrightarrow3,4}(\bar 1,2,\bar 3,4)=\sgn(\sprods{12})\sum_{s=\pm1}\int_0^{2\pi}\dd\alpha\,\frac{1}{2i\sin\alpha}
	\cdot\\\cdot
	\deltad{2|3}(\Lambda_3-is\brk{\sin\alpha\,\Lambda_1+\cos\alpha\,\Lambda_2})\,
	\deltad{2|3}(\Lambda_4-i\brk{\cos\alpha\,\Lambda_1-\sin\alpha\,\Lambda_2})\,.
\label{eq:A4ppmm}
\end{multline}
In the above derivation, it was assumed that particles $1$ and $2$ carry the
same energy sign, and particles $3$ and $4$ carry the opposite energy sign.
A similar analysis (carried out in \appref{app:mixedenergy}) shows what the
amplitude becomes when incoming/outgoing particles are distributed
differently:%
\footnote{In both cases, $\sprods{12}$ is purely imaginary.}
\begin{multline}
\amp_4^{1,3\leftrightarrow2,4}(\bar 1,2,\bar 3,4)=-i\csgn(\sprods{12})\sum_{s_\alpha,s_\beta=\pm1}\int_{-\infty}^\infty\dd\alpha\,\frac{1}{4\sinh\alpha}
	\cdot\\\cdot
	\deltad{2|3}(\Lambda_3-s_\beta(s_\alpha\sinh\alpha\,\Lambda_1+i\cosh\alpha\,\Lambda_2))\,
	\deltad{2|3}(\Lambda_4-is_\alpha\cosh\alpha\,\Lambda_1+\sinh\alpha\,\Lambda_2)\,,
\label{eq:A4pmpm}
\end{multline}
\begin{multline}
\amp_4^{1,4\leftrightarrow2,3}(\bar 1,2,\bar 3,4)=i\csgn(\sprods{12})\sum_{s_\alpha,s_\beta=\pm1}\int_{-\infty}^\infty\dd\alpha\,\frac{s_\beta}{4i\cosh\alpha}
	\cdot\\\cdot
	\deltad{2|3}(\Lambda_3-is_\alpha\cosh\alpha\,\Lambda_1+\sinh\alpha\,\Lambda_2)\,
	\deltad{2|3}(\Lambda_4-s_\beta(s_\alpha\sinh\alpha\,\Lambda_1+i\cosh\alpha\,\Lambda_2))\,,
\label{eq:A4pmmp}
\end{multline}
Here and for the following it is helpful to introduce 
generalizations of the sign
and absolute value functions to the real and imaginary axes:
\[
\csgn(x):=\begin{cases}
+1 & \text{for } x \in \Reals^+,i\Reals^+,\\
-1 & \text{for } x \in \Reals^-,i\Reals^-.\\
\end{cases}
\qquad
\cabs{x}:=\begin{cases}
|\Re x| & \text{for } x \in \Reals,\\
i|\Im x| & \text{for } x \in i\Reals.
\end{cases}
\label{eq:csgncabs}
\]
In conclusion, we can write the four-point amplitude as
\begin{equation}
\amp_4^{k_1,k_2\leftrightarrow k_3,k_4}(\bar 1,2,\bar 3,4)=\csgn(\sprods{12}) F^{k_1,k_2\leftrightarrow k_3,k_4}(\bar 1,2,\bar 3,4),
\label{eq:genformA4}
\end{equation}
where $F$ denotes a function whose explicit parametrization depends on the
distribution of energies.

\paragraph{Anomaly.}

Looking at the explicit form of \eqref{eq:A4ppmm,eq:A4pmpm,eq:A4pmmp}, it
is immediate that the action of
$\gen{S}^A_a=\sum_k\eta_k^A\partial/\partial\lambda_k^a$ on the super delta
functions produces terms of the form
$x\,\delta(x)$, thus the
function $F$ in \eqref{eq:genformA4} is annihilated.
The signum factor, however, produces a non-vanishing contribution
whenever the momenta $1$ and $2$ are collinear:
\begin{equation}
\gen{S}^A_a\csgn(\sprods{12})=2\eps_{ab}\brk{\eta_1^A\lambda_2^b-\eta_2^A\lambda_1^b}\,\cdelta(\sprods{12})\,.
\end{equation}
Here we consistently define
\[
\cdelta(x):=\begin{cases}
\delta(\Re x) & \text{for } x \in \Reals,\\
-i\delta(\Im x) & \text{for } x \in i\Reals,\\
\end{cases}
\label{eq:cdelta}
\]
The anomaly vertex thus takes the form
\begin{equation}
\gen{S}_a^A\amp_4^{k_1,k_2\leftrightarrow k_3,k_4}(\bar 1,2,\bar 3,4)=2\eps_{ab}\brk{\eta_1^A\lambda_2^b-\eta_2^A\lambda_1^b}\,\cdelta(\sprods{12})\,F^{k_1,k_2\leftrightarrow k_3,k_4}(\bar 1,2,\bar 3,4),,
\end{equation}
where $F$ is anomaly free
and on the support of $\cdelta(\sprods{12})$, all four momenta are
collinear.  We believe this expression furnishes the building block for all
anomaly contributions of higher-point and higher-loop amplitudes.

Let us evaluate the vertex for the configuration $(1,2\leftrightarrow3,4)$ 
in order to obtain a more symmetric expression. It is convenient to parametrize the collinear particles in terms of the massless momentum
$\lambda_{12}\lambda_{12}=\lambda_1\lambda_1+\lambda_2\lambda_2$. To this
end, we use the identity
\begin{equation}
\delta(\sprods{12})=\int\dd\Lambda_{12}\dd\Lambda'\dd\beta\deltad{2}(\lambda')\deltad{2|3}_1\deltad{2|3}_2\,,
\end{equation}
with
\begin{align}
\deltad{2|3}_1&=\deltad{2|3}(\Lambda_1-\brk{\sin\beta\,\Lambda_{12}+\cos\beta\,\Lambda'})\,,
\nn\\
\deltad{2|3}_2&=\deltad{2|3}(\Lambda_2-\brk{\cos\beta\,\Lambda_{12}-\sin\beta\,\Lambda'})\,,
\end{align}
which, after shifting $\alpha\to\alpha-\beta$, turns the delta functions in
\eqref{eq:A4ppmm} into
\begin{align}
\deltad{2|3}_3&=\deltad{2|3}(\Lambda_3-is\brk{\sin\alpha\,\Lambda_{12}+\cos\alpha\,\Lambda'})\,,
\nn\\
\deltad{2|3}_4&=\deltad{2|3}(\Lambda_4-i \brk{\cos\alpha\,\Lambda_{12}-\sin\alpha\,\Lambda'})\,.
\end{align}
Using $\eta_1^A\lambda_2^b-\eta_2^A\lambda_1^b=\lambda_{12}^b\eta'^A$ under
$\deltad{2}(\lambda')$, the anomaly vertex then reads
\begin{equation}
\gen{S}^A_a\amp_4^{1,2\leftrightarrow3,4}(\bar 1,2,\bar 3,4)=\sum_{s=\pm1}\int\dd\Lambda_{12}\dd\Lambda'\dd\alpha\,\dd\beta\,
	\deltad{2}(\lambda')\frac{1}{i\sin(\alpha-\beta)}\eps_{ab}\lambda_{12}^b\eta'^A\deltad{2|3}_1\deltad{2|3}_2\deltad{2|3}_3\deltad{2|3}_4\,.
\label{eq:anomaly}
\end{equation}
%

\section{Conformal Symmetry of the Six-Point Amplitude}
\label{sec:anomaly6}

Let us now apply the conformal anomaly vertex to six-point amplitudes
at tree level and one loop. The tree-level six-point amplitudes were first calculated
by using the superconformal symmetries and explicit Feynman diagram calculations
in \cite{Bargheer:2010hn}. Subsequently they were rederived from the orthogonal Gra{\ss}mannian
of \cite{Lee:2010du}, and given a perhaps more congenial form in \cite{Gang:2010gy} 
by means of the three-dimensional analog of the BCFW recursion relations.

In order to generalize the above anomaly four-vertex
$\gen{S}_4=\gen{S}A_4$  to higher-point and to loop amplitudes, we employ
the following strategy: We first obtain the
imaginary part of the respective amplitude using unitarity (the optical
theorem), then determine
the anomaly of the imaginary part, and finally find the anomaly of the
amplitude by requiring that acting with the anomalous generator
$\gen{S}$ and taking the imaginary part commute, $\gen{S}\Im
A=\Im\gen{S}A$.

\begin{mpostfile}{FigCut60.mps}
paths[1]:=fullcircle scaled 0.8xu shifted (-1.5xu,0);
paths[2]:=fullcircle scaled 3xu shifted (-1.5xu,0);
paths[3]:=fullcircle scaled 0.8xu shifted (+1.5xu,0);
paths[4]:=fullcircle scaled 3xu shifted (+1.5xu,0);

fill paths[1] colamp;
draw paths[1];

fill paths[3] colamp;
draw paths[3];

label(btex $\bar 4$ etex scaled 1.5, center(paths[1]));
label(btex $4$ etex scaled 1.5, center(paths[3]));

drawleg (point 0 of paths[1])--(point 4 of paths[3]);
midarrow ((point 0 of paths[1])--(point 4 of paths[3]), 0.7);

for i=-1 upto 1:
  drawleg (point (4+8*i/6) of paths[1])--(point (4+8*i/6) of paths[2]);
  drawleg (point (0+8*i/6) of paths[3])--(point (0+8*i/6) of paths[4]);
endfor;

drawcfield point (4-8/6) of paths[1];
drawfield point (4) of paths[1];
drawcfield point (4+8/6) of paths[1];
drawfield point (0) of paths[1];

drawfield point (-8/6) of paths[3];
drawcfield point (0) of paths[3];
drawfield point (+8/6) of paths[3];
drawcfield point (4) of paths[3];

label.lrt(btex $4$ etex, point (-8/6) of paths[4]);
label.rt(btex $\bar 5$ etex, point (0) of paths[4]);
label.urt(btex $6$ etex, point (+8/6) of paths[4]);
label.ulft(btex $\bar 1$ etex, point (4-8/6) of paths[2]);
label.lft(btex $2$ etex, point (4) of paths[2]);
label.llft(btex $\bar 3$ etex, point (4+8/6) of paths[2]);
label.top(btex a\vphantom{g} etex, point 0.7 of ((point 0 of paths[1])--(point 4 of paths[3])));

drawcut (0,-1.5xu)--(0,+1.5xu);

putcopy(1);
\end{mpostfile}
\begin{mpostfile}{FigCutAnom60a.mps}
paths[1]:=fullcircle scaled 0.8xu shifted (-1.5xu,0);
paths[2]:=fullcircle scaled 3xu shifted (-1.5xu,0);
paths[3]:=fullcircle scaled 0.8xu shifted (+1.5xu,0);
paths[4]:=fullcircle scaled 3xu shifted (+1.5xu,0);

fill paths[1] colgen;
draw paths[1];

fill paths[3] colamp;
draw paths[3];

label(btex $\ast$ etex scaled 1.5, center(paths[1]));
label(btex $4$ etex scaled 1.5, center(paths[3]));

drawleg (point 0 of paths[1])--(point 4 of paths[3]);
midarrow ((point 0 of paths[1])--(point 4 of paths[3]), 0.7);

for i=-1 upto 1:
  drawleg (point (4+8*i/6) of paths[1])--(point (4+8*i/6) of paths[2]);
  drawleg (point (0+8*i/6) of paths[3])--(point (0+8*i/6) of paths[4]);
endfor;

drawcfield point (4-8/6) of paths[1];
drawfield point (4) of paths[1];
drawcfield point (4+8/6) of paths[1];
drawfield point (0) of paths[1];

drawfield point (-8/6) of paths[3];
drawcfield point (0) of paths[3];
drawfield point (+8/6) of paths[3];
drawcfield point (4) of paths[3];

label.lrt(btex $4$ etex, point (-8/6) of paths[4]);
label.rt(btex $\bar 5$ etex, point (0) of paths[4]);
label.urt(btex $6$ etex, point (+8/6) of paths[4]);
label.ulft(btex $\bar 1$ etex, point (4-8/6) of paths[2]);
label.lft(btex $2$ etex, point (4) of paths[2]);
label.llft(btex $\bar 3$ etex, point (4+8/6) of paths[2]);
label.top(btex a\vphantom{g} etex, point 0.7 of ((point 0 of paths[1])--(point 4 of paths[3])));

drawcut (0,-1.5xu)--(0,+1.5xu);

putcopy(1);
\end{mpostfile}
\begin{mpostfile}{FigCutAnom60b.mps}
paths[1]:=fullcircle scaled 0.8xu shifted (-1.5xu,0);
paths[2]:=fullcircle scaled 3xu shifted (-1.5xu,0);
paths[3]:=fullcircle scaled 0.8xu shifted (+1.5xu,0);
paths[4]:=fullcircle scaled 3xu shifted (+1.5xu,0);

fill paths[1] colamp;
draw paths[1];

fill paths[3] colgen;
draw paths[3];

label(btex $4$ etex scaled 1.5, center(paths[1]));
label(btex $\ast$ etex scaled 1.5, center(paths[3]));

drawleg (point 0 of paths[1])--(point 4 of paths[3]);
midarrow ((point 0 of paths[1])--(point 4 of paths[3]), 0.7);

for i=-1 upto 1:
  drawleg (point (4+8*i/6) of paths[1])--(point (4+8*i/6) of paths[2]);
  drawleg (point (0+8*i/6) of paths[3])--(point (0+8*i/6) of paths[4]);
endfor;

drawcfield point (4-8/6) of paths[1];
drawfield point (4) of paths[1];
drawcfield point (4+8/6) of paths[1];
drawfield point (0) of paths[1];

drawfield point (-8/6) of paths[3];
drawcfield point (0) of paths[3];
drawfield point (+8/6) of paths[3];
drawcfield point (4) of paths[3];

label.lrt(btex $4$ etex, point (-8/6) of paths[4]);
label.rt(btex $\bar 5$ etex, point (0) of paths[4]);
label.urt(btex $6$ etex, point (+8/6) of paths[4]);
label.ulft(btex $\bar 1$ etex, point (4-8/6) of paths[2]);
label.lft(btex $2$ etex, point (4) of paths[2]);
label.llft(btex $\bar 3$ etex, point (4+8/6) of paths[2]);
label.top(btex a\vphantom{g} etex, point 0.7 of ((point 0 of paths[1])--(point 4 of paths[3])));

drawcut (0,-1.5xu)--(0,+1.5xu);

putcopy(1);
\end{mpostfile}
\begin{mpostfile}{FigAnom60.mps}
paths[1]:=fullcircle scaled 0.8xu shifted (-1.5xu,0);
paths[2]:=fullcircle scaled 3xu shifted (-1.5xu,0);
paths[3]:=fullcircle scaled 0.8xu shifted (+1.5xu,0);
paths[4]:=fullcircle scaled 3xu shifted (+1.5xu,0);

fill paths[1] colgen;
draw paths[1];

fill paths[3] colamp;
draw paths[3];

label(btex $\ast$ etex scaled 1.5, center(paths[1]));
label(btex $4$ etex scaled 1.5, center(paths[3]));

drawleg (point 0 of paths[1])--(point 4 of paths[3]);

for i=-1 upto 1:
  drawleg (point (4+8*i/6) of paths[1])--(point (4+8*i/6) of paths[2]);
  drawleg (point (0+8*i/6) of paths[3])--(point (0+8*i/6) of paths[4]);
endfor;

drawcfield point (4-8/6) of paths[1];
drawfield point (4) of paths[1];
drawcfield point (4+8/6) of paths[1];
drawfield point (0) of paths[1];

drawfield point (-8/6) of paths[3];
drawcfield point (0) of paths[3];
drawfield point (+8/6) of paths[3];
drawcfield point (4) of paths[3];

label.lrt(btex $4$ etex, point (-8/6) of paths[4]);
label.rt(btex $\bar 5$ etex, point (0) of paths[4]);
label.urt(btex $6$ etex, point (+8/6) of paths[4]);
label.ulft(btex $\bar 1$ etex, point (4-8/6) of paths[2]);
label.lft(btex $2$ etex, point (4) of paths[2]);
label.llft(btex $\bar 3$ etex, point (4+8/6) of paths[2]);
label.top(btex a\vphantom{g} etex, point 0.5 of ((point 0 of paths[1])--(point 4 of paths[3])));

putcopy(1);
\end{mpostfile}

\paragraph{Tree Level.}

\begin{figure}
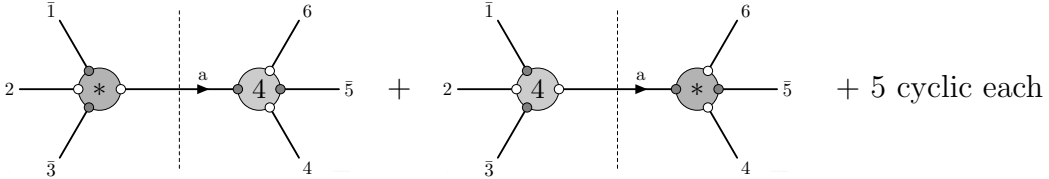
\centering
\includegraphicsbox[scale=0.7]{FigCutAnom60a.mps}\quad $+$\quad \includegraphicsbox[scale=0.7]{FigCutAnom60b.mps} 
\quad $+$ 5 cyclic each
\caption{Anomaly of the imaginary part of the tree-level six-point amplitude}
\label{fig:CutAnom6Tree}
\end{figure}

By the optical theorem, the imaginary part of the
tree-level six-point amplitude in ABJM reads, see
\figref{fig:Cut6Tree},
%
\begin{figure}\centering
\includegraphicsbox[scale=0.7]{FigCut60.mps} \quad $+$ 5 cyclic
\caption{Imaginary part of the tree-level six-point amplitude}
\label{fig:Cut6Tree}
\end{figure}
%
\begin{align}
\label{eq:cutsixtree}
2\Im \amp^{(0)}_6(\bar 1,2,\bar 3,4,\bar 5,6)=\mathord{}&
\int_{\Reals} d\Lambda\indup{a}\,
\bar\amp_4(\bar 1,2,\bar 3,i\mathrm{a})\,
\amp_4(\bar{\mathrm{a}},4,\bar 5,6)
\nln&
-i\int_{\Reals} d\Lambda\indup{a}\,
\bar\amp_4(i\bar{\mathrm{a}},2,\bar 3,4)\,
\amp_4(\bar 5,6,\bar 1,\mathrm{a})
+\text{4 cyclic}\,,
\nln\mathord{}=\mathord{}&
\int d\Lambda\indup{a}\,
\amp_4(\bar 1,2,\bar 3,i\mathrm{a})\,
\amp_4(\bar{\mathrm{a}},4,\bar 5,6)
+\text{2 cyclic}\,,
\end{align}
where the first two integrations run only over real $\lambda\indup{a}$, 
that is the internal momenta have a definite energy sign; see
\eqref{eq:Lambdaintegral} for the definition of the measure $\dd\Lambda$.%
\footnote{\label{fn:signsigh}%
Unfortunately, the prefactors of the various contributions 
are hard to come by. For example, 
reality requires a factor of $\pm i$ 
in the second term in the above unitarity relation.
Only for this (relative) prefactor two contributions will combine 
using \eqref{eq:Lambdaintegralexpansion}
into an integral with undirected energy flow
as desirable for tree level unitarity.
The correct overall signs follow, ultimately, from evaluating statistics and color 
of the S-matrix operator carefully.
In the figures here and below, we conveniently absorb such factors of $\pm1$, $\pm i$ 
originating from the on-shell integration measure $\dd\Lambda$
into a more intuitive definition of the diagram.}
A bar over an amplitude denotes complex conjugation, but note
that in fact the four-point amplitude is real, $\bar\amp_4=\amp_4$.
Furthermore note that the second term is real despite its imaginary prefactor:
The amplitude $\bar\amp_4$ is an odd function of the (imaginary) first argument
by construction.
Both four-point amplitudes on the r.h.s.\ are anomalous,
and they will contribute to the anomaly 
of the six-point tree-level amplitude. 
That is their anomalies in the integral will give 
the imaginary part of the six-point anomaly, see \figref{fig:CutAnom6Tree}.
%
%
In fact, we do not expect further contributions:
An off-shell propagator joining two four-point vertices 
yields a rational function. 
In four dimensions, the holomorphic anomaly at poles was responsible for
anomalous transformations. 
Conversely, in three dimensions,
the anomaly arises from derivatives of step functions. 

A simple ansatz to reproduce the anomaly of 
the above imaginary part reads
%
\begin{figure}
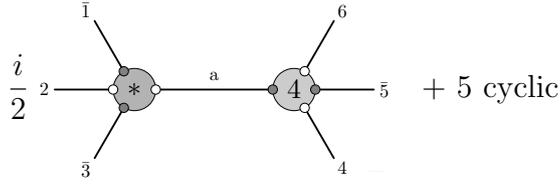
\centering
$\displaystyle\frac{i}{2}$ \includegraphicsbox[scale=0.7]{FigAnom60.mps}\quad $+$ 5 cyclic
\caption{Anomaly of the tree-level six-point amplitude}
\label{fig:Anom6Tree}
\end{figure}
%
\begin{align}
\gen{S}\amp_6^{(0)}(\bar 1,2,\bar 3,4,\bar 5,6)
=-\mathord{}&
\frac{i}{2}\int d\Lambda\indup{a}\,
\gen{S}_4(\bar 1,2,\bar 3,i\mathrm{a})
\,\amp_4(\bar{\mathrm{a}},4,\bar 5,6)
+\text{5 cyclic}, 
\label{eq:A60anom}
\end{align}
which is graphically represented in \figref{fig:Anom6Tree}, and where now
the integration includes both real and imaginary $\lambda\indup{a}$. 
The minus sign on the right hand side results from passing the fermionic 
generator $\gen{S}$ through the fermionic measure.
The correctness of this expression can be confirmed by computing its
imaginary part (by complex conjugation) and comparing to the anomaly of
\eqref{eq:cutsixtree}.
Note that the four-point anomaly is real, $\bar{\gen{S}}_4=\gen{S}_4$
(for a real choice of generator $\gen{S}$).
The anomaly is supported on configurations where three adjacent momenta are collinear.

\begin{mpostfile}{FigCutAnom61a.mps}
paths[1]:=fullcircle scaled 0.8xu shifted (-1.5xu,0);
paths[2]:=fullcircle scaled 3xu shifted (-1.5xu,0);
paths[3]:=fullcircle scaled 1.0xu shifted (+1.5xu,0);
paths[4]:=fullcircle scaled 3xu shifted (+1.5xu,0);

fill paths[1] colgen;
draw paths[1];

fill paths[3] colamp;
draw paths[3];

label(btex $\ast$ etex scaled 1.5, center(paths[1]));
label(btex $6$ etex scaled 1.5, center(paths[3]));

for i=0 upto 1:
drawleg (point (4+8*(i-0.5)/6) of paths[1])--(point (4+8*(i-0.5)/6) of paths[2]);
endfor
for i=0 upto 3:
drawleg (point (8*(i-1.5)/8) of paths[3])--(point (8*(i-1.5)/8) of paths[4]);
endfor

drawleg (point -1 of paths[1]){dir -45}..{dir +45}(point 5 of paths[3]);
drawleg (point +1 of paths[1]){dir +45}..{dir -45}(point 3 of paths[3]);
midarrow((point -1 of paths[1]){dir -45}..{dir +45}(point 5 of paths[3]),0.7);
midarrow((point +1 of paths[1]){dir +45}..{dir -45}(point 3 of paths[3]),0.7);

drawcfield point (4-8*(+0.5)/6) of paths[1];
drawfield point (4-8*(-0.5)/6) of paths[1];
drawcfield point (-1) of paths[1];
drawfield point (+1) of paths[1];
drawcfield point (3) of paths[3];
drawfield point (5) of paths[3];
drawcfield point (-8*(+1.5)/8) of paths[3];
drawfield point (-8*(+0.5)/8) of paths[3];
drawcfield point (-8*(-0.5)/8) of paths[3];
drawfield point (-8*(-1.5)/8) of paths[3];

label.lft(btex $\bar 1$ etex, point (4-8*(+0.5)/6) of paths[2]);
label.lft(btex $2$ etex, point (4-8*(-0.5)/6) of paths[2]);
label.bot(btex $\bar 3$ etex, point (-8*(+1.5)/8) of paths[4]);
label.rt(btex $4$ etex, point (-8*(+0.5)/8) of paths[4]);
label.rt(btex $\bar 5$ etex, point (-8*(-0.5)/8) of paths[4]);
label.top(btex $6$ etex, point (-8*(-1.5)/8) of paths[4]);
label.bot(btex a\vphantom{$\hat g$} etex, point 0.7 of ((point -1 of paths[1]){dir -45}..{dir +45}(point 5 of paths[3])));
label.top(btex b\vphantom{$\hat g$} etex, point 0.7 of ((point +1 of paths[1]){dir +45}..{dir -45}(point 3 of paths[3])));

drawcut (0,-1.5xu)--(0,+1.5xu);

putcopy(1);
\end{mpostfile}
\begin{mpostfile}{FigCutAnom61d.mps}
paths[1]:=fullcircle scaled 0.8xu shifted (-1.5xu,0);
paths[2]:=fullcircle scaled 3xu shifted (-1.5xu,0);
paths[3]:=fullcircle scaled 0.8xu shifted (+0.75xu,1.3xu);
paths[4]:=fullcircle scaled 3xu shifted (+0.75xu,1.3xu);
paths[5]:=fullcircle scaled 0.8xu shifted (+0.75xu,-1.3xu);
paths[6]:=fullcircle scaled 3xu shifted (+0.75xu,-1.3xu);

fill paths[1] colgen;
draw paths[1];

fill paths[3] colamp;
draw paths[3];

fill paths[5] colamp;
draw paths[5];

label(btex $\ast$ etex scaled 1.5, center(paths[1]));
label(btex $4$ etex scaled 1.5, center(paths[3]));
label(btex $4$ etex scaled 1.5, center(paths[5]));

for i=0 upto 1:
drawleg (point (4+8*(i-0.5)/6) of paths[1])--(point (4+8*(i-0.5)/6) of paths[2]);
endfor
for i=0 upto 1:
drawleg (point (4/3+8*(i-0.5)/6) of paths[3])--(point (4/3+8*(i-0.5)/6) of paths[4]);
endfor
for i=0 upto 1:
drawleg (point (-4/3+8*(i-0.5)/6) of paths[5])--(point (-4/3+8*(i-0.5)/6) of paths[6]);
endfor

drawleg (point (4/6) of paths[1])--(point (4+4/6) of paths[3]);
drawleg (point (-4/6) of paths[1])--(point (4-4/6) of paths[5]);
drawleg (point (-2) of paths[3])--(point (+2) of paths[5]);
midarrow((point (4/6) of paths[1])--(point (4+4/6) of paths[3]),0.7);
midarrow((point (-4/6) of paths[1])--(point (4-4/6) of paths[5]),0.7);

drawcfield point (4-8*(+0.5)/6) of paths[1];
drawfield point (4-8*(-0.5)/6) of paths[1];
drawcfield point (-4/6) of paths[1];
drawfield point (+4/6) of paths[1];
drawcfield point (4+4/6) of paths[3];
drawfield point (-2) of paths[3];
drawcfield point (4/6) of paths[3];
drawfield point (2) of paths[3];
drawcfield point (2) of paths[5];
drawfield point (4-4/6) of paths[5];
drawcfield point (-2) of paths[5];
drawfield point (-4/6) of paths[5];

label.lft(btex $\bar 1$ etex, point (4-8*(+0.5)/6) of paths[2]);
label.lft(btex $2$ etex, point (4-8*(-0.5)/6) of paths[2]);
label.bot(btex $\bar 3$ etex, point (-4/3-8*(+0.5)/6) of paths[6]);
label.rt(btex $4$ etex, point (-4/3-8*(-0.5)/6) of paths[6]);
label.rt(btex $\bar 5$ etex, point (4/3-8*(+0.5)/6) of paths[4]);
label.top(btex $6$ etex, point (4/3-8*(-0.5)/6) of paths[4]);
label.bot(btex a\vphantom{$\hat g$} etex, point 0.7 of ((point (-4/6) of paths[1])--(point (4-4/6) of paths[5])));
label.top(btex b\vphantom{$\hat g$} etex, point 0.7 of ((point (4/6) of paths[1])--(point (4+4/6) of paths[3])));
label.rt(btex \,c\vphantom{$\hat g$} etex, point 0.5 of ((point (+2) of paths[5])--(point (-2) of paths[3])));

drawcut (-0.3xu,-2xu)--(-0.3xu,+2xu);

putcopy(1);
\end{mpostfile}
\begin{mpostfile}{FigCutAnom61e.mps}
paths[1]:=fullcircle scaled 0.8xu shifted (-1.0xu,0);
paths[2]:=fullcircle scaled 3xu shifted (-1.0xu,0);
paths[3]:=fullcircle scaled 0.8xu shifted (+1.0xu,0);
paths[4]:=fullcircle scaled 3xu shifted (+1.0xu,0);
paths[5]:=fullcircle scaled 0.8xu shifted ((+2.0xu,0) rotated 30) shifted (+1.0xu,0);
paths[6]:=fullcircle scaled 3xu shifted ((+2.0xu,0) rotated 30) shifted (+1.0xu,0);

fill paths[1] colgen;
draw paths[1];

fill paths[3] colamp;
draw paths[3];

fill paths[5] colamp;
draw paths[5];

label(btex $\ast$ etex scaled 1.5, center(paths[1]));
label(btex $4$ etex scaled 1.5, center(paths[3]));
label(btex $4$ etex scaled 1.5, center(paths[5]));

for i=0 upto 1:
drawleg (point (4+8*(i-0.5)/6) of paths[1])--(point (4+8*(i-0.5)/6) of paths[2]);
endfor
drawleg (point (8*(-0.5)/6) of paths[3])--(point (8*(-0.5)/6) of paths[4]);
drawleg (point (8*(+0.5)/6) of paths[3])--(point (8*(-2.5)/6) of paths[5]);

for i=-1 upto 1:
drawleg (point (8*(i)/12) of paths[5])--(point (8*(i)/12) of paths[6]);
endfor

drawleg (point -1 of paths[1]){dir -45}..{dir +45}(point 5 of paths[3]);
drawleg (point +1 of paths[1]){dir +45}..{dir -45}(point 3 of paths[3]);
midarrow((point -1 of paths[1]){dir -45}..{dir +45}(point 5 of paths[3]),0.7);
midarrow((point +1 of paths[1]){dir +45}..{dir -45}(point 3 of paths[3]),0.7);

drawcfield point (4-8*(+0.5)/6) of paths[1];
drawfield point (4-8*(-0.5)/6) of paths[1];
drawcfield point (-1) of paths[1];
drawfield point (+1) of paths[1];
drawcfield point (3) of paths[3];
drawfield point (5) of paths[3];
drawcfield point (-8*(+0.5)/6) of paths[3];
drawfield point (-8*(-0.5)/6) of paths[3];
drawfield point (-8*(-1)/12) of paths[5];
drawcfield point (-8*(0)/12) of paths[5];
drawfield point (-8*(+1)/12) of paths[5];
drawcfield point (-8*(+2.5)/6) of paths[5];

label.lft(btex $\bar 1$ etex, point (4-8*(+0.5)/6) of paths[2]);
label.lft(btex $2$ etex, point (4-8*(-0.5)/6) of paths[2]);
label.rt(btex $\bar 3$ etex, point (-8*(+0.5)/6) of paths[4]);
label.rt(btex $4$ etex, point (8*(-1)/12) of paths[6]);
label.rt(btex $\bar 5$ etex, point (8*(0)/12) of paths[6]);
label.rt(btex $6$ etex, point (8*(+1)/12) of paths[6]);
label.bot(btex a\vphantom{$\hat g$} etex, point 0.7 of ((point -1 of paths[1]){dir -45}..{dir +45}(point 5 of paths[3])));
label.top(btex b\vphantom{$\hat g$} etex, point 0.7 of ((point +1 of paths[1]){dir +45}..{dir -45}(point 3 of paths[3])));
label.ulft(btex c\vphantom{$\hat g$} etex, point 0.5 of ((point (8*(+0.5)/6) of paths[3])--(point (8*(-2.5)/6) of paths[5])));

drawcut (0,-1.5xu)--(0,+1.5xu);

putcopy(1);
\end{mpostfile}
\begin{mpostfile}{FigCutAnom61f.mps}
paths[1]:=fullcircle scaled 0.8xu shifted (-1.0xu,0);
paths[2]:=fullcircle scaled 3xu shifted (-1.0xu,0);
paths[3]:=fullcircle scaled 0.8xu shifted (+1.0xu,0);
paths[4]:=fullcircle scaled 3xu shifted (+1.0xu,0);
paths[5]:=fullcircle scaled 0.8xu shifted ((+2.0xu,0) rotated -30) shifted (+1.0xu,0);
paths[6]:=fullcircle scaled 3xu shifted ((+2.0xu,0) rotated -30) shifted (+1.0xu,0);

fill paths[1] colgen;
draw paths[1];

fill paths[3] colamp;
draw paths[3];

fill paths[5] colamp;
draw paths[5];

label(btex $\ast$ etex scaled 1.5, center(paths[1]));
label(btex $4$ etex scaled 1.5, center(paths[3]));
label(btex $4$ etex scaled 1.5, center(paths[5]));

for i=0 upto 1:
drawleg (point (4+8*(i-0.5)/6) of paths[1])--(point (4+8*(i-0.5)/6) of paths[2]);
endfor
drawleg (point (8*(+0.5)/6) of paths[3])--(point (8*(+0.5)/6) of paths[4]);
drawleg (point (8*(-0.5)/6) of paths[3])--(point (8*(-3.5)/6) of paths[5]);

for i=-1 upto 1:
drawleg (point (8*(i)/12) of paths[5])--(point (8*(i)/12) of paths[6]);
endfor

drawleg (point -1 of paths[1]){dir -45}..{dir +45}(point 5 of paths[3]);
drawleg (point +1 of paths[1]){dir +45}..{dir -45}(point 3 of paths[3]);
midarrow((point -1 of paths[1]){dir -45}..{dir +45}(point 5 of paths[3]),0.7);
midarrow((point +1 of paths[1]){dir +45}..{dir -45}(point 3 of paths[3]),0.7);

drawcfield point (4-8*(+0.5)/6) of paths[1];
drawfield point (4-8*(-0.5)/6) of paths[1];
drawcfield point (-1) of paths[1];
drawfield point (+1) of paths[1];
drawcfield point (3) of paths[3];
drawfield point (5) of paths[3];
drawcfield point (-8*(+0.5)/6) of paths[3];
drawfield point (-8*(-0.5)/6) of paths[3];
drawcfield point (-8*(-1)/12) of paths[5];
drawfield point (-8*(0)/12) of paths[5];
drawcfield point (-8*(+1)/12) of paths[5];
drawfield point (-8*(+3.5)/6) of paths[5];

label.lft(btex $\bar 1$ etex, point (4-8*(+0.5)/6) of paths[2]);
label.lft(btex $2$ etex, point (4-8*(-0.5)/6) of paths[2]);
label.rt(btex $\bar 3$ etex, point (8*(-1)/12) of paths[6]);
label.rt(btex $4$ etex, point (8*(0)/12) of paths[6]);
label.rt(btex $\bar 5$ etex, point (8*(+1)/12) of paths[6]);
label.rt(btex $6$ etex, point (-8*(-0.5)/6) of paths[4]);
label.bot(btex a\vphantom{$\hat g$} etex, point 0.7 of ((point -1 of paths[1]){dir -45}..{dir +45}(point 5 of paths[3])));
label.top(btex b\vphantom{$\hat g$} etex, point 0.7 of ((point +1 of paths[1]){dir +45}..{dir -45}(point 3 of paths[3])));
label.llft(btex c\vphantom{$\hat g$} etex, point 0.5 of ((point (8*(-0.5)/6) of paths[3])--(point (8*(-3.5)/6) of paths[5])));

drawcut (0,-1.5xu)--(0,+1.5xu);

putcopy(1);
\end{mpostfile}
\begin{mpostfile}{FigAnom61a.mps}
paths[1]:=fullcircle scaled 0.8xu shifted (-1.5xu,0);
paths[2]:=fullcircle scaled 3xu shifted (-1.5xu,0);
paths[3]:=fullcircle scaled 1.0xu shifted (+1.5xu,0);
paths[4]:=fullcircle scaled 3xu shifted (+1.5xu,0);

fill paths[1] colgen;
draw paths[1];

fill paths[3] colamp;
draw paths[3];

label(btex $\ast$ etex scaled 1.5, center(paths[1]));
label(btex $6$ etex scaled 1.5, center(paths[3]));

for i=0 upto 1:
drawleg (point (4+8*(i-0.5)/6) of paths[1])--(point (4+8*(i-0.5)/6) of paths[2]);
endfor
for i=0 upto 3:
drawleg (point (8*(i-1.5)/8) of paths[3])--(point (8*(i-1.5)/8) of paths[4]);
endfor

drawleg (point -1 of paths[1]){dir -45}..{dir +45}(point 5 of paths[3]);
drawleg (point +1 of paths[1]){dir +45}..{dir -45}(point 3 of paths[3]);
midarrow((point -1 of paths[1]){dir -45}..{dir +45}(point 5 of paths[3]),0.5);
midarrow((point +1 of paths[1]){dir 45}..{dir -45}(point 3 of paths[3]),0.5);

drawcfield point (4-8*(+0.5)/6) of paths[1];
drawfield point (4-8*(-0.5)/6) of paths[1];
drawcfield point (-1) of paths[1];
drawfield point (+1) of paths[1];
drawcfield point (3) of paths[3];
drawfield point (5) of paths[3];
drawcfield point (-8*(+1.5)/8) of paths[3];
drawfield point (-8*(+0.5)/8) of paths[3];
drawcfield point (-8*(-0.5)/8) of paths[3];
drawfield point (-8*(-1.5)/8) of paths[3];

label.lft(btex $\bar 1$ etex, point (4-8*(+0.5)/6) of paths[2]);
label.lft(btex $2$ etex, point (4-8*(-0.5)/6) of paths[2]);
label.bot(btex $\bar 3$ etex, point (-8*(+1.5)/8) of paths[4]);
label.rt(btex $4$ etex, point (-8*(+0.5)/8) of paths[4]);
label.rt(btex $\bar 5$ etex, point (-8*(-0.5)/8) of paths[4]);
label.top(btex $6$ etex, point (-8*(-1.5)/8) of paths[4]);
label.bot(btex a\vphantom{$\hat g$} etex, point 0.5 of ((point -1 of paths[1]){dir -45}..{dir +45}(point 5 of paths[3])));
label.top(btex b\vphantom{$\hat g$} etex, point 0.5 of ((point +1 of paths[1]){dir 45}..{dir -45}(point 3 of paths[3])));

putcopy(1);
\end{mpostfile}
\begin{mpostfile}{FigAnom61b.mps}
paths[1]:=fullcircle scaled 0.8xu shifted (-1.5xu,0);
paths[2]:=fullcircle scaled 3xu shifted (-1.5xu,0);
paths[3]:=fullcircle scaled 1.0xu shifted (+1.5xu,0);
paths[4]:=fullcircle scaled 3xu shifted (+1.5xu,0);

fill paths[1] colgen;
draw paths[1];

fill paths[3] colamp;
draw paths[3];

label(btex $\ast$ etex scaled 1.5, center(paths[1]));
label(btex $6$ etex scaled 1.5, center(paths[3]));

for i=0 upto 1:
drawleg (point (4+8*(i-0.5)/6) of paths[1])--(point (4+8*(i-0.5)/6) of paths[2]);
endfor
for i=0 upto 3:
drawleg (point (8*(i-1.5)/8) of paths[3])--(point (8*(i-1.5)/8) of paths[4]);
endfor

drawleg (point -1 of paths[1]){dir -45}..{dir +45}(point 5 of paths[3]);
drawleg (point +1 of paths[1]){dir +45}..{dir -45}(point 3 of paths[3]);
midarrow((point 5 of paths[3]){dir -135}..{dir +135}(point -1 of paths[1]),0.5);
midarrow((point 3 of paths[3]){dir 135}..{dir -135}(point +1 of paths[1]),0.5);

drawcfield point (4-8*(+0.5)/6) of paths[1];
drawfield point (4-8*(-0.5)/6) of paths[1];
drawcfield point (-1) of paths[1];
drawfield point (+1) of paths[1];
drawcfield point (3) of paths[3];
drawfield point (5) of paths[3];
drawcfield point (-8*(+1.5)/8) of paths[3];
drawfield point (-8*(+0.5)/8) of paths[3];
drawcfield point (-8*(-0.5)/8) of paths[3];
drawfield point (-8*(-1.5)/8) of paths[3];

label.lft(btex $\bar 1$ etex, point (4-8*(+0.5)/6) of paths[2]);
label.lft(btex $2$ etex, point (4-8*(-0.5)/6) of paths[2]);
label.bot(btex $\bar 3$ etex, point (-8*(+1.5)/8) of paths[4]);
label.rt(btex $4$ etex, point (-8*(+0.5)/8) of paths[4]);
label.rt(btex $\bar 5$ etex, point (-8*(-0.5)/8) of paths[4]);
label.top(btex $6$ etex, point (-8*(-1.5)/8) of paths[4]);
label.bot(btex a\vphantom{$\hat g$} etex, point 0.5 of ((point 5 of paths[3]){dir -135}..{dir +135}(point -1 of paths[1])));
label.top(btex b\vphantom{$\hat g$} etex, point 0.5 of ((point 3 of paths[3]){dir 135}..{dir -135}(point +1 of paths[1])));

putcopy(1);
\end{mpostfile}

\paragraph{One Loop.}

At one loop, the imaginary part of the six-point amplitude expands to%
\footnote{The prefactor of $-i$ is related to the integral over one conjugate leg, see \fnref{fn:signsigh}. 
We are not certain whether the overall sign is correct,
but it will be consistent with the derivations below.}
%
\[
2\Im \amp_6^{(1)}(\bar 1,2,\bar 3,4,\bar 5,6)=
-i\int_{\Reals} d\Lambda\indup{a}\, d\Lambda\indup{b}\, 
\amp_4(\bar 1,2,i\bar{\mathrm{a}},i\mathrm{b})
\,\amp_6^{(0)}(\bar{\mathrm{b}},\mathrm{a},\bar 3,4,\bar 5,6)
+\text{5 cyclic}+\text{cc.}
\label{eq:discA61}
\]
%
These are the only contributions because all
amplitudes have an even number of external particles.%
\footnote{An important exception to this rule occurs if we consider the zero-mode 
of the Chern--Simons field. We will discuss this at the end of
\secref{sec:OneLoopSixPoint} and in \secref{sec:concl}. Corresponding singularities are supported
on kinematical subspaces of greater codimensionality and are negligible for the considerations here.}
Furthermore, a single-particle cut would split off a four-point one-loop amplitude 
which is known to vanish.%
\footnote{Indeed, 
the one-loop four-point amplitude,  reduced to scalar integrals, can only
get contributions from one- and two-mass triangles, bubbles and tadpoles. 
The one- and two-mass triangle integrals vanish in three-dimensions, and there are no
bubble or tadpole diagrams in the three-dimensional, finite, superconformal theory.}

\begin{figure}
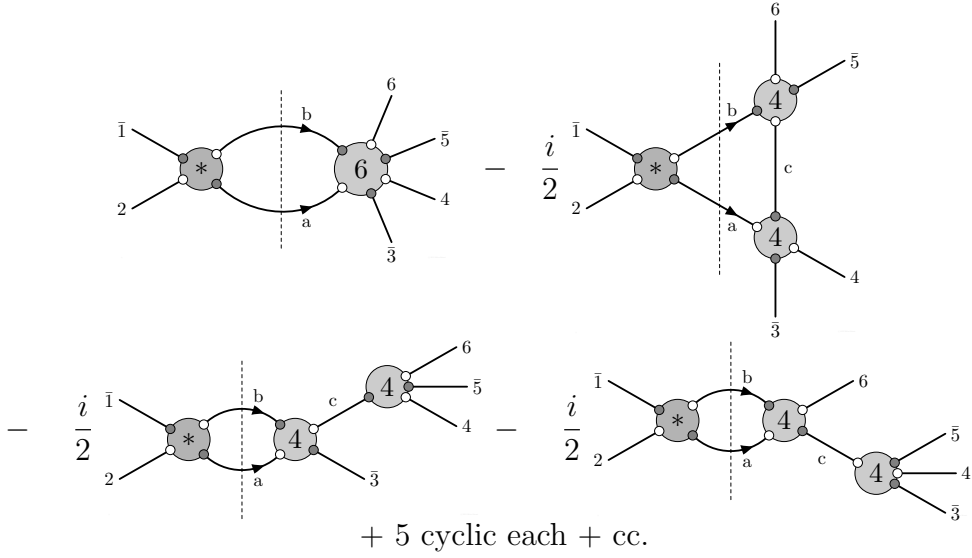
\centering
\includegraphicsbox[scale=0.7]{FigCutAnom61a.mps}\quad 
$-$\quad $\displaystyle\frac{i}{2}$ \includegraphicsbox[scale=0.7]{FigCutAnom61d.mps}\\
$-\quad~\displaystyle\frac{i}{2}$~\includegraphicsbox[scale=0.7]{FigCutAnom61e.mps}
$-$\quad~$\displaystyle\frac{i}{2}$~\includegraphicsbox[scale=0.7]{FigCutAnom61f.mps}\\
\quad
$+$ 5 cyclic each $+$ cc.
\caption{Anomaly of the imaginary part of the one-loop six-point amplitude}
\label{fig:CutAnom6Loop}
\end{figure}

We can now act with a superconformal generator on \eqref{eq:discA61} to 
find the anomaly of the imaginary part. The result is a sum of terms where
the generator $\gen{S}$ either acts on $\amp_4$, yielding
$\gen{S}\amp_4=\gen{S}_4$, or on $\amp_6^{(0)}$, yielding
\eqref{eq:A60anom}. 
The latter generates triangle-shaped diagrams as well
as contributions where two four-vertices are joined 
into a two-sided bubble with a third four-vertex attached to one of the external legs.
The triangle diagrams combine with their complex conjugates
where the energy signs of internal lines are flipped.
Collecting all the contributions, we find
(see \figref{fig:CutAnom6Loop})
%
%
\begin{align}
\gen{S}
2\Im\amp_6^{(1)}=\mathord{}&
-i\int_{\Reals} d\Lambda\indup{a}\, d\Lambda\indup{b}\, 
\gen{S}_4(\bar 1,2,i\bar{\mathrm{a}},i\mathrm{b})
\,\amp_6^{(0)}(\bar{\mathrm{b}},\mathrm{a},\bar 3,4,\bar 5,6)
\nln
&-\frac{1}{2}
\int_{\Reals} d\Lambda\indup{a}\, d\Lambda\indup{b}\, \int d\Lambda\indup{c}\, 
\gen{S}_4(\bar 1,2,i\bar{\mathrm{a}},i\mathrm{b})\,
\amp_4(\bar 5,6,\bar{\mathrm{b}},i\mathrm{c})\,
\amp_4(\bar{\mathrm{c}},\mathrm{a},\bar 3,4)
\nln
&-\frac{1}{2}
\int_{\Reals} d\Lambda\indup{a}\, d\Lambda\indup{b}\, \int d\Lambda\indup{c}\, 
\gen{S}_4(\bar 1,2,i\bar{\mathrm{a}},i\mathrm{b})\,
\amp_4(\bar{\mathrm{b}},\mathrm{a},\bar 3,i\mathrm{c})\,
\amp_4(\bar{\mathrm{c}},4,\bar 5,6)
\nln
&-\frac{1}{2}
\int_{\Reals} d\Lambda\indup{a}\, d\Lambda\indup{b}\, \int d\Lambda\indup{c}\, 
\gen{S}_4(\bar 1,2,i\bar{\mathrm{a}},i\mathrm{b})\,
\amp_4(\bar{3},4,\bar 5,i\mathrm{c})\,
\amp_4(\bar{\mathrm{c}},6,\bar{\mathrm{b}},\mathrm{a})
\nln
&+\text{5 cyclic each} + \text{cc.}
\label{eq:imA61anom}
\end{align}
Here the latter three terms arise from the action of $\gen{S}$ on $\amp_6^{(0)}$. 
They were obtained by combining with the complex conjugate contributions
and using the vanishing of the one-loop four-point amplitude.
Note that the third and the fourth term are of higher
codimension: They require not only two, but three collinear momenta.

\begin{figure}
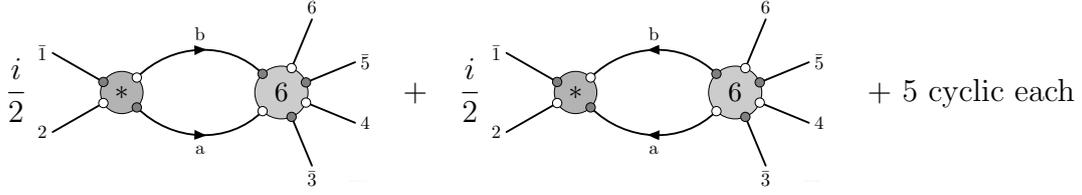
\centering
$\displaystyle\frac{i}{2}$ \includegraphicsbox[scale=0.7]{FigAnom61a.mps}\quad$+$\quad
$\displaystyle\frac{i}{2}$ \includegraphicsbox[scale=0.7]{FigAnom61b.mps}
\quad $+$ 5 cyclic each
\caption{Anomaly of the one-loop six-point amplitude}
\label{fig:Anom6Loop}
\end{figure}

Now we can make an ansatz for the six-point one-loop anomaly
$\gen{S}\amp_6^{(1)}$ consisting of on-shell bubble and triangle integrals
with six-vertices $\amp_6^{(0)}$, four-vertices $\amp_4^{(0)}$, and anomaly
vertices $\gen{S}_4$. 
We compute the imaginary part of the one-loop anomaly ansatz 
making use of the complex conjugate of the six-point tree-level amplitude
\begin{equation}
\bar\amp_6^{(0)}(\bar{\mathrm{b}},\mathrm{a},\bar 3,4,\bar 5,6)
=\amp_6^{(0)}(\bar{\mathrm{b}},\mathrm{a},\bar 3,4,\bar 5,6)
-2i\Im\amp_6^{(0)}(\bar{\mathrm{b}},\mathrm{a},\bar 3,4,\bar 5,6)\,,
\end{equation}
where $\Im\amp_6^{(0)}$ is given in \eqref{eq:cutsixtree}.
We compare it to the anomaly
\eqref{eq:imA61anom} of the one-loop imaginary part, 
and we find that the following, relatively simple expression suffices:
\begin{multline}
\label{eq:Anom6Loop}
\gen{S}\amp_6^{(1)}(\bar 1,2,\bar 3,4,\bar 5,6)=
\frac{1}{2}
\int_{\Reals} d\Lambda\indup{a}\, d\Lambda\indup{b}\, 
\gen{S}_4(\bar 1,2,i\bar{\mathrm{a}},i\mathrm{b})
\,\amp_6^{(0)}(\bar{\mathrm{b}},\mathrm{a},\bar 3,4,\bar 5,6)
\\
+\frac{1}{2}
\int_{\Reals} d\Lambda\indup{a}\, d\Lambda\indup{b}\, 
\gen{S}_4(\bar 1,2,\bar{\mathrm{a}},\mathrm{b})
\,\amp_6^{(0)}(i\bar{\mathrm{b}},i\mathrm{a},\bar 3,4,\bar 5,6)
+\text{5 cyclic each}\,.
\end{multline}
The anomaly, see \figref{fig:Anom6Loop}, is supported on configurations 
where two adjacent particles are collinear. 
In other words, for a generic configuration of particle momenta
there is no anomaly at one loop. 
%

Let us investigate the above integral for the one-loop six-point anomaly 
more explicitly:
Making use of the anomaly vertex as induced by the four-point amplitude 
in the form of \eqref{eq:A4ppmm}, 
the integrals can be straightforwardly evaluated against the delta
functions. The first term in \eqref{eq:Anom6Loop} then becomes%
\footnote{Note that
$\int\dd\Lambda'\delta(\Lambda'-\Lambda)f(\Lambda')=1/2\,f(\Lambda)$ due to
the definition of the measure \eqref{eq:Lambdaintegral}.}
\begin{multline}
\gen{S}A_6^{(1)}(\bar 1, 2, \bar 3, 4, \bar 5, 6)=
\frac{1}{8}\eps_{ab}\bigbrk{\eta_1^A\lambda_2^b-\eta_2^A\lambda_1^b}
\cdelta(\sprods{12})
\sum_{s=\pm1}\int_0^{2\pi}\dd\alpha\,
\frac{1}{i\sin(\alpha)}
\cdot\\\cdot
A_6^{(0)}(\cos\alpha\,\Lambda_1-\sin\alpha\,\Lambda_2, s\brk{\sin\alpha\,\Lambda_1+\cos\alpha\,\Lambda_2}, \bar 3, 4, \bar 5, 6)
+\text{5 cyclic}\,.
\label{eq:anomalyeval}
\end{multline}
Substituting $y=\cos(\alpha)$, and using the $\Lambda$-parity \eqref{eq:l_parity}
of $A_6^{(0)}$, this yields 
\begin{multline}
\gen{S}A_6^{(1)}(\bar 1, 2, \bar 3, 4, \bar 5, 6)=
-\frac{i}{2}\,\eps_{ab}\bigbrk{\eta_1^A\lambda_2^b-\eta_2^A\lambda_1^b}\cdelta(\sprods{12})\int_{-1}^1\dd y\,\frac{1}{1-y^2}
\cdot\\\cdot
A_6^{(0)}\bigbrk{y\Lambda_1-\sqrt{1-y^2}\Lambda_2, \sqrt{1-y^2}\Lambda_1+y\Lambda_2, \Lambda_3, \Lambda_4, \Lambda_5, \Lambda_6}
+\text{5 cyclic}\,.
\label{eq:anomalyreduced}
\end{multline}
Now we make use of the explicit expression for the six-point amplitude introduced in \cite{Bargheer:2010hn}:
\begin{equation}
\mathcal{A}_6^{(0)}=\deltad{3}(P)\deltad{6}(Q)\bigbrk{f^+\deltad{3}(\alpha^+)+f^-\deltad{3}(\alpha^-)},
\label{eq:A6treecoords}
\end{equation}
where
\begin{equation}
\alpha^{\pm,A} = \sum_{k=1}^6x_k^\pm\eta_k^A\,,
\qquad
\deltad{3}(\alpha^\pm) = \prod_{A=1,2,3}\alpha^{\pm,A}\,,
\end{equation}
and the coordinates $x_k^\pm$ can be chosen in accordance with the conditions in \cite{Bargheer:2010hn} to be
\begin{equation}
x^\pm_i=\frac{i\eps_{ijk}\sprods{jk}}{2\sqrt{2}\sqrt{p_{6,1,2}^2}},\quad i,j,k=6,1,2\,,
\qquad
x^\pm_i=\frac{\pm
\eps_{ijk}\sprods{jk}}{2\sqrt{2}\sqrt{p_{3,4,5}^2}},\quad i,j,k=3,4,5\,.
\label{eq:xpmit}
\end{equation}
With the above choice of coordinates, the expressions for $\deltad{3}(P)$, $\deltad{6}(Q)$ and $\alpha^{\pm}$ do not change under the rotation
\begin{equation}
\Lambda_1\to\Lambda_a=y\Lambda_1-\sqrt{1-y^2}\Lambda_2\,,
\qquad
\Lambda_2\to\Lambda_b=\sqrt{1-y^2}\Lambda_1+y\Lambda_2\,.
\label{eq:Lambdarot}
\end{equation}
A simple form for the functions $f^\pm$ can be found using the component amplitudes of \cite{Bianchi:2012cq}:
\begin{equation}
A_{6\psi}=A_6(\bar\psi^4,\psi_1,\bar\psi^2,\psi_4,\bar\psi^1,\psi_2)=0\,,
\quad
A_{6\phi}=A_6(\bar\phi_3,\phi^2,\bar\phi_1,\phi^3,\bar\phi_2,\phi^1)=1\,.
\end{equation}
These yield
\begin{equation}
f^\pm
=\frac{i\sqrt{2}\sqrt{p_{3,4,5}^2}}{\bigbrk{\sprods{2|p_{34}|5}\mp i\sprods{61}\sprods{34}}\bigbrk{\sprods{3|p_{45}|6}\mp i\sprods{12}\sprods{45}}}\,.
\label{eq:fpmit2}
\end{equation}
The anomaly term \eqref{eq:anomalyreduced} then becomes
\begin{multline}
\gen{S}A_6^{(1)}(\bar 1, 2, \bar 3, 4, \bar 5, 6)=\\
-\frac{i}{2}\eps_{ab}\bigbrk{\eta_1^A\lambda_2^b-\eta_2^A\lambda_1^b}\cdelta(\sprods{12})\,\deltad{3}(P)\,\deltad{6}(Q)\,
\bigbrk{i\sqrt{2}\sqrt{p_{3,4,5}^2}}\,
\bigbrk{I_{12}^+\,\deltad{3}(\alpha^+)+I_{12}^-\,\deltad{3}(\alpha^-)}\\
+\text{5 cyclic}\,.
\end{multline}
with the integrals given by
\begin{equation}
I_{12}^\pm=
\int_{-1}^1\dd y\,\frac{1}{1-y^2}\,
\frac{1}{\bigbrk{\sprods{b|p_{34}|5}\mp i\sprods{6a}\sprods{34}}\bigbrk{\sprods{3|p_{45}|6}\mp i\sprods{ab}\sprods{45}}}\,.
\end{equation}
Setting $\sprods{ab}=0$ due to the presence
of $\cdelta(\sprods{12})$, the integrals simplify to
\begin{equation}
I_{12}^\pm
=\frac{1}{\sprods{3|p_{45}|6}}
 \int_{-1}^1\dd y\,\frac{1}{1-y^2}\,
 \frac{1}{\bigbrk{\sprods{b|p_{34}|5}\mp i\sprods{6a}\sprods{34}}}\,.
\end{equation}
The integral diverges at both integration boundaries, but taking a
symmetric limit yields
\begin{equation}
I_{12}^\pm
=\frac{1}{\sprods{3|p_{45}|6}}\,
 \frac{ i s_\pm\pi}{\sprods{2|p_{34}|5}\mp i\sprods{61}\sprods{34}}\,,
\end{equation}
where $s_{\pm}$ are sign factors that may depend on the external momenta, and
whose explicit form we did not determine.
Comparing this to \eqref{eq:fpmit2} and keeping in mind that
$\sprods{12}=0$ under $\cdelta(\sprods{12})$, the final result for
\eqref{eq:anomalyreduced} is
\begin{multline}
\gen{S}A_6^{(1)}(\bar 1, 2, \bar 3, 4, \bar 5, 6)=\\
\frac{\pi}{2}\eps_{ab}\bigbrk{\eta_1^A\lambda_2^b-\eta_2^A\lambda_1^b}\cdelta(\sprods{12})
\deltad{3}(P)\deltad{6}(Q)\bigbrk{s_+f^+\deltad{3}(\alpha^+)+s_-f^-\deltad{3}(\alpha^-)}\\
+\text{5 cyclic}\,.
\end{multline}
Notably, the evaluation of the above integrals leaves
overall sign factors $s_\pm$ in this expression. 
Depending on the values of these sign factors, we
notice that the anomaly of the one-loop amplitude 
either is
proportional to the tree-level amplitude \eqref{eq:A6treecoords} itself, or
to the tree-level amplitude with a flipped sign in front of $f^-$. 
As will become more transparent below, 
flipping the relative sign between $f^+$ and $f^-$ in the tree-level
amplitude amounts to a cyclic shift of the external particles. Hence we
find
\begin{align}
\gen{S}_a^A\amp_6^{(1)}(\bar 1,2,\bar 3,4,\bar 5,6)=\mathord{}
&s \frac{\pi}{4} \bigbrk{\gen{S}_a^A\csgn{\sprods{12}}}
\begin{cases}
A_6^{(0)}(\bar 1,2,\bar 3,4,\bar 5,6) & \text{ if } s_\pm=s\\
iA_6^{(0)}(\bar 6,1,\bar 2,3,\bar 4,5) & \text{ if } s_\pm=\pm s\\
\end{cases}
\nn\\
&\qquad+\text{5  cyclic}.
\label{eq:anomalyfinal}
\end{align}

\paragraph{Discussion.}

The fact that the one-loop anomaly is singular can be used 
in connection with Yangian symmetry
to gain some easy insights into the one-loop six-point amplitude.
At six points there exist two independent  Yangian (almost) invariant functions which we call
$Y_{1,2}$.%
\footnote{They are almost invariants in the sense that they are only
invariant up to distributional terms.}
They can be constructed by an explicit Feynman diagram calculation and their Yangian invariance
checked \cite{Bargheer:2010hn}, or 
by means of a contour integral over an orthogonal Gra{\ss}mannian 
which is manifestly Yangian invariant \cite{Gang:2010gy}.
Both approaches involve the two solutions of a quadratic equation
with coefficients depending on the external momenta. 
In the Gra{\ss}mannian approach, after choosing a particular patch,  
 the Yangian invariants depend on the variables  $c_{s \bar r }$, $\bar r=1,3,5$, and 
 $s=2,4,6$, such that
\[
\lambda_{\bar r}+\sum_s \lambda_s c_{s \bar r}=0, \quad \text{and} \quad \sum_{\bar r}c_{s \bar r}c_{t\bar r}=\delta_{st},
\]
so that we label the two solutions $(c^\ast_\pm)_{s\bar r}$.
The two Yangian (almost)  invariants, in this approach, 
correspond to the Gra{\ss}mannian integrand 
evaluated on these two solutions, schematically 
$Y_{1,2}\equiv Y(\{c^\ast_\pm\}_{s\bar r})$, and 
explicit expressions can be found in \cite{Gang:2010gy}.
Significantly, in terms of the $\lambda_i$'s, these invariants are simply 
rational functions as all square roots can explicitly be performed. 

We know that Yangian symmetry is unbroken at tree level 
except for a codimension-two anomaly.%
\footnote{As Yangian level-one generator we can use 
$\geny{P}$ which is anomaly-free for finite contributions.}
Therefore the tree-level six-point amplitude must be some linear combination 
of the two invariants.
We can fix which linear combination forms the tree amplitude
by demanding the correct behavior of the amplitude 
under $\Lambda$-parity \eqref{eq:l_parity} with the odd-numbered legs
being fermionic and the even-numbered legs being bosonic, 
\[
\amp^{(0)}_6(\bar 1, 2, \bar 3, 4, \bar 5, 6)= c_6 (Y_1+Y_2)(\bar 1, 2, \bar 3, 4, \bar 5, 6)
\]
for some constant $c_6$ depending on the gauge coupling.%
\footnote{The linear combination can also be fixed by demanding that the
amplitude factorizes correctly
into four-point amplitudes with fermionic odd-numbered legs. In the Gra{\ss}mannian approach
it follows naturally from the cyclic gauge choice.}
Here, the bars over the labels on the right-hand side merely signify that
the function $(Y_1+Y_2)(\Lambda_k)$ transforms odd under sign flips of the
respective $\Lambda$'s.
The other linear combination
\begin{equation}
(Y_1-Y_2)(1, \bar 2, 3, \bar 4, 5, \bar 6)
\end{equation}
has exactly the opposite transformation property, as indicated by the
distribution of bars on the labels, and thus cannot appear in the
tree-level amplitude.%
\footnote{Phrased differently, $Y_1\to(-1)^{k+1}Y_2$ and
$Y_2\to(-1)^{k+1}Y_1$ under $\Lambda_k\to-\Lambda_k$.}
Notably, this linear combination again equals the tree-level amplitude when
shifting all labels cyclically by one,
\[
\amp^{(0)}_6(\bar 6, 1, \bar 2, 3, \bar 4, 5)= i c_6 (Y_1-Y_2)(1, \bar 2, 3, \bar 4, 5, \bar 6).
\label{eq:shifted6amp}
\]
At one-loop order, the loop momentum is still completely constrained by the
four-point anomaly vertex \eqref{eq:Anom6Loop}, and hence the anomaly does
not get smeared across all configurations of external momenta, but rather
stays distributional, see \figref{fig:Anom6Loop}. Consequently, also the
one-loop six-point amplitude has to equal a linear combination of the
tree-level Yangian (almost) invariants, with a prefactor that is constant
at least locally. However, the support of the anomalies at one loop is
different than at tree level. Most notably, the one-loop amplitude has
codimension-one anomalies, which are absent at tree level and in the
Yangian (almost) invariants. In conclusion, the one-loop amplitude has to
be a linear combination
\begin{multline}
\amp_6^{(1)}(\bar 1,2,\bar 3,4,\bar 5,6)
=c_6^+(1,2,3,4,5,6)\,(Y_1+Y_2)(\bar 1, 2, \bar 3, 4, \bar 5, 6)\\
+c_6^-(\bar 1, \bar 2, \bar 3, \bar 4, \bar 5, \bar 6)\,(Y_1-Y_2)(1, \bar 2, 3, \bar 4, 5, \bar 6)
\end{multline}
with coefficients $c_6^\pm$ that are locally constant, but discontinuous on
the support of the codimension-one anomalies---that is they have to be
piecewise constants that jump whenever two adjacent external particles
become collinear. In order to maintain the correct statistics of the
amplitude, $c_6^+$ has to transform bosonic in all labels, and $c_6^-$ has
to transform fermionic in all labels, as indicated by the bars over their
labels.
Thus the symmetries predict a non-vanishing result for the one-loop 
six-point amplitude. 

\begin{mpostfile}{FigAnom80.mps}
paths[1]:=fullcircle scaled 0.8xu shifted (-1.5xu,0);
paths[2]:=fullcircle scaled 3xu shifted (-1.5xu,0);
paths[3]:=fullcircle scaled 0.8xu shifted (+1.5xu,0);
paths[4]:=fullcircle scaled 3xu shifted (+1.5xu,0);
paths[5]:=fullcircle scaled 0.8xu shifted (0,+1.5xu);
paths[6]:=fullcircle scaled 3xu shifted (0,+1.5xu);

fill paths[1] colgen;
draw paths[1];

fill paths[3] colamp;
draw paths[3];

fill paths[5] colamp;
draw paths[5];

label(btex $4$ etex scaled 1.5, center(paths[1]));
label(btex $4 $ etex scaled 1.5, center(paths[3]));
label(btex $\ast$ etex scaled 1.5, center(paths[5]));

drawleg (point 1 of paths[1])--(point 5 of paths[5]);
midarrow ((point 5 of paths[5])--(point 1 of paths[1]), 0.5);

drawleg (point 3 of paths[3])--(point 7 of paths[5]);
midarrow ((point 7 of paths[5])--(point 3 of paths[3]), 0.5);

for i=-1 upto 1:
  drawleg (point (4+8*(i+1)/6) of paths[1])--(point (4+8*(i+1)/6) of paths[2]);
  drawleg (point (0+8*(i-1)/6) of paths[3])--(point (0+8*(i-1)/6) of paths[4]);
endfor;

for i=0 upto 1:
  drawleg (point (2+8/4*(i-1/2)) of paths[5])--(point (2+8/4*(i-1/2)) of paths[6]);
endfor;

drawcfield point (1) of paths[1];
drawfield point (4) of paths[1];
drawcfield point (4+8/6) of paths[1];
drawfield point (4+16/6) of paths[1];

drawfield point (5) of paths[5];
drawcfield point (7) of paths[5];
drawfield point (1) of paths[5];
drawcfield point (3) of paths[5];

drawfield point (3) of paths[3];
drawcfield point (0) of paths[3];
drawfield point (0-8/6) of paths[3];
drawcfield point (0-16/6) of paths[3];

label.lrt(btex $6$ etex, point (-8/6) of paths[4]);
label.rt(btex $\bar 7$ etex, point (0) of paths[4]);
label.llft(btex $5$ etex, point (-16/6) of paths[4]);
label.ulft(btex $\bar 1$ etex, point (3) of paths[6]);
label.urt(btex $8 $ etex, point (1) of paths[6]);
label.lft(btex $2$ etex, point (4) of paths[2]);
label.llft(btex $\bar 3$ etex, point (4+8/6) of paths[2]);
label.lrt(btex $ 4$ etex, point (4+16/6) of paths[2]);

putcopy(1);
\end{mpostfile}
\begin{mpostfile}{FigAnom62.mps}
paths[1]:=fullcircle scaled 0.8xu shifted (-1.5xu,0);
paths[2]:=fullcircle scaled 3xu shifted (-1.5xu,0);
paths[3]:=fullcircle scaled 1.0xu shifted (+1.5xu,0);
paths[4]:=fullcircle scaled 3xu shifted (+1.5xu,0);
paths[5]:=fullcircle scaled 0.8xu shifted (0,+1.5xu);
paths[6]:=fullcircle scaled 3xu shifted (0,+1.5xu);

fill paths[1] colgen;
draw paths[1];

fill paths[3] colamp;
draw paths[3];

label(btex $\ast$ etex scaled 1.5, center(paths[1]));
label(btex $8$ etex scaled 1.5, center(paths[3]));

drawleg (point (0) of paths[1])--(point (4) of paths[3]);
midarrow ((point (4) of paths[1])--(point (0) of paths[3]),0.5);

drawleg (point (4) of paths[1])--(point (4) of paths[2]);

for i=0 upto 4:
drawleg (point (8*(i-2)/8) of paths[3])--(point (8*(i-2)/8) of paths[4]);
endfor

drawleg (point -1 of paths[1]){dir -45}..{dir +45}(point 5 of paths[3]);
drawleg (point +1 of paths[1]){dir +45}..{dir -45}(point 3 of paths[3]);
midarrow((point -1 of paths[1]){dir -45}..{dir +45}(point 5 of paths[3]),0.5);
midarrow((point 1 of paths[1]){dir 45}..{dir -45}(point +3 of paths[3]),0.5);

drawcfield point (4) of paths[1];
drawfield point (1) of paths[1];
drawcfield point (0) of paths[1];
drawfield point (-1) of paths[1];
drawfield point (+1) of paths[1];

drawcfield point (3) of paths[3];
drawfield point (4) of paths[3];
drawcfield point (5) of paths[3];
drawfield point (-8*(+2)/8) of paths[3];
drawcfield point (-8*(+1)/8) of paths[3];
drawfield point (-8*(0)/8) of paths[3];
drawcfield point (-8*(-1)/8) of paths[3];
drawfield point (-8*(-2)/8) of paths[3];

label.lft(btex $\bar 1$ etex, point (4) of paths[2]);
label.bot(btex $2$ etex, point (-8*(+2)/8) of paths[4]);
label.lrt(btex $\bar 3$ etex, point (-8*(+1)/8) of paths[4]);
label.rt(btex $4$ etex, point (0) of paths[4]);
label.urt(btex $\bar 5$ etex, point (-8*(-1)/8) of paths[4]);
label.top(btex $6$ etex, point (2) of paths[4]);

putcopy(1);
\end{mpostfile}

\paragraph{Proposal summary.}

\begin{figure}
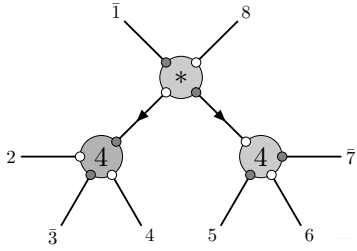
\centering
\includegraphicsbox[scale=0.7]{FigAnom80.mps}
\caption{Two internal-leg contribution to tree-level eight-point anomaly}
\label{fig:Anom8Tree}
\end{figure}

Let us summarize what we have found for the anomalies
of the scattering amplitudes and outline a proposal 
for the anomalies at higher points and more loops.
The variation of the  four-point tree-level
amplitude simply gives the distributional, 
inhomogeneous term found on the right hand side of 
equation \eqref{eq:anomaly}. That is, it is directly given by the
anomaly vertex and has support only on the codimension-two surface.%
\footnote{Naively, this anomaly cannot be considered as a deformation
of the generator acting on a lower point amplitude as there is no 
lower point amplitude. However, see the discussion in \secref{sec:concl}.}
For higher point amplitudes we attach the anomaly vertex to 
subamplitudes along a single internal leg. We saw this explicitly
in the case of the six-point tree-level amplitude, see \figref{fig:Anom6Tree}.
At eight points and beyond there are additional possible configurations, 
where two anomaly vertex legs are attached to different subamplitudes, 
see \figref{fig:Anom8Tree}.
At ten points, there are in principle configurations where 
three vertex legs can be attached along internal lines
to three different subamplitudes and at twelve points, all vertex legs can 
be attached to internal lines.

\begin{figure}
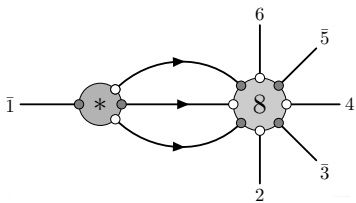
\centering
\includegraphicsbox[scale=0.7]{FigAnom62.mps}
\caption{Proposed anomaly of two-loop six-point amplitude }
\label{fig:Anom62loop}
\end{figure}

At one-loop level we also attach two anomaly vertex legs, but now both with 
the same energy and to the same subamplitude. We saw this 
for the six-point one-loop amplitude, see \figref{fig:Anom6Loop}, where
two legs were attached to a tree-level six-point amplitude. 
This gives rise to an anomaly that is also purely distributional as it
only occurs when two external legs are collinear. 
At higher loops more than two legs of the anomaly vertex 
can be attached to the same lower-loop subamplitude, see \figref{fig:Anom62loop},
with all legs having the same energy sign so 
it is necessary to extend the anomaly vertex into 
a different kinematic regime. Moreover, for these loop 
amplitudes this results in an
anomaly for generic configurations of the external momenta.  
For example at two-loops, three anomaly vertex legs 
can be attached to the same tree-level amplitude and, because
of the integrations, the resulting 
anomaly will have support 
for any value of the external momenta. This is analogous
to what happens in \sym\ at one loop
where the anomaly becomes ``smeared" by the loop integrations 
and so the amplitudes are anomalous with 
respect to the superconformal symmetries for generic states
of the external particles.

\begin{mpostfile}{FigTriple61a.mps}
paths[1]:=(fullcircle scaled 0.8xu shifted (-1.5xu,0)) rotated -90;
paths[2]:=(fullcircle scaled 3xu shifted (-1.5xu,0)) rotated -90;
paths[3]:=(fullcircle scaled 0.8xu shifted (+0.75xu,1.3xu)) rotated -90;
paths[4]:=(fullcircle scaled 3xu shifted (+0.75xu,1.3xu)) rotated -90;
paths[5]:=(fullcircle scaled 0.8xu shifted (+0.75xu,-1.3xu)) rotated -90;
paths[6]:=(fullcircle scaled 3xu shifted (+0.75xu,-1.3xu)) rotated -90;

fill paths[1] colamp;
draw paths[1];

fill paths[3] colamp;
draw paths[3];

fill paths[5] colamp;
draw paths[5];

label(btex $4$ etex scaled 1.5, center(paths[1]));
label(btex $4$ etex scaled 1.5, center(paths[3]));
label(btex $4$ etex scaled 1.5, center(paths[5]));

for i=0 upto 1:
drawleg (point (4+8*(i-0.5)/6) of paths[1])--(point (4+8*(i-0.5)/6) of paths[2]);
endfor
for i=0 upto 1:
drawleg (point (4/3+8*(i-0.5)/6) of paths[3])--(point (4/3+8*(i-0.5)/6) of paths[4]);
endfor
for i=0 upto 1:
drawleg (point (-4/3+8*(i-0.5)/6) of paths[5])--(point (-4/3+8*(i-0.5)/6) of paths[6]);
endfor

drawleg (point (4/6) of paths[1])--(point (4+4/6) of paths[3]);
drawleg (point (-4/6) of paths[1])--(point (4-4/6) of paths[5]);
drawleg (point (-2) of paths[3])--(point (+2) of paths[5]);

drawcfield point (4-8*(+0.5)/6) of paths[1];
drawfield point (4-8*(-0.5)/6) of paths[1];
drawcfield point (-4/6) of paths[1];
drawfield point (+4/6) of paths[1];
drawcfield point (4+4/6) of paths[3];
drawfield point (-2) of paths[3];
drawcfield point (4/6) of paths[3];
drawfield point (2) of paths[3];
drawcfield point (2) of paths[5];
drawfield point (4-4/6) of paths[5];
drawcfield point (-2) of paths[5];
drawfield point (-4/6) of paths[5];

label.top(btex $\bar 1$ etex, point (4-8*(+0.5)/6) of paths[2]);
label.top(btex $2$ etex, point (4-8*(-0.5)/6) of paths[2]);
label.lft(btex $\bar 3$ etex, point (-4/3-8*(+0.5)/6) of paths[6]);
label.bot(btex $4$ etex, point (-4/3-8*(-0.5)/6) of paths[6]);
label.bot(btex $\bar 5$ etex, point (4/3-8*(+0.5)/6) of paths[4]);
label.rt(btex $6$ etex, point (4/3-8*(-0.5)/6) of paths[4]);
label.lft(btex \,b\,\vphantom{$\hat g$} etex, point 0.7 of ((point (-4/6) of paths[1])--(point (4-4/6) of paths[5])));
label.rt(btex \,a\,\vphantom{$\hat g$} etex, point 0.7 of ((point (4+4/6) of paths[3])--(point (4/6) of paths[1])));
label.bot(btex c\vphantom{$\hat g$} etex, point 0.7 of ((point (+2) of paths[5])--(point (-2) of paths[3])));

drawccut (0,0)--(point 0.5 of ((point (4-8*(-0.5)/6) of paths[2])--(point (-4/3-8*(+0.5)/6) of paths[6])));
drawccut (0,0)--(point 0.5 of ((point (-4/3-8*(-0.5)/6) of paths[6])--(point (4/3-8*(+0.5)/6) of paths[4])));
drawccut (0,0)--(point 0.5 of ((point (4/3-8*(-0.5)/6) of paths[4])--(point (4-8*(+0.5)/6) of paths[2])));

putcopy(1);
\end{mpostfile}
\begin{mpostfile}{FigTriple61b.mps}
paths[1]:=(fullcircle scaled 0.8xu shifted (-1.5xu,0)) rotated 90;
paths[2]:=(fullcircle scaled 3xu shifted (-1.5xu,0)) rotated 90;
paths[3]:=(fullcircle scaled 0.8xu shifted (+0.75xu,1.3xu)) rotated 90;
paths[4]:=(fullcircle scaled 3xu shifted (+0.75xu,1.3xu)) rotated 90;
paths[5]:=(fullcircle scaled 0.8xu shifted (+0.75xu,-1.3xu)) rotated 90;
paths[6]:=(fullcircle scaled 3xu shifted (+0.75xu,-1.3xu)) rotated 90;

fill paths[1] colamp;
draw paths[1];

fill paths[3] colamp;
draw paths[3];

fill paths[5] colamp;
draw paths[5];

label(btex $4$ etex scaled 1.5, center(paths[1]));
label(btex $4$ etex scaled 1.5, center(paths[3]));
label(btex $4$ etex scaled 1.5, center(paths[5]));

for i=0 upto 1:
drawleg (point (4+8*(i-0.5)/6) of paths[1])--(point (4+8*(i-0.5)/6) of paths[2]);
endfor
for i=0 upto 1:
drawleg (point (4/3+8*(i-0.5)/6) of paths[3])--(point (4/3+8*(i-0.5)/6) of paths[4]);
endfor
for i=0 upto 1:
drawleg (point (-4/3+8*(i-0.5)/6) of paths[5])--(point (-4/3+8*(i-0.5)/6) of paths[6]);
endfor

drawleg (point (4/6) of paths[1])--(point (4+4/6) of paths[3]);
drawleg (point (-4/6) of paths[1])--(point (4-4/6) of paths[5]);
drawleg (point (-2) of paths[3])--(point (+2) of paths[5]);

drawfield point (4-8*(+0.5)/6) of paths[1];
drawcfield point (4-8*(-0.5)/6) of paths[1];
drawfield point (-4/6) of paths[1];
drawcfield point (+4/6) of paths[1];
drawfield point (4+4/6) of paths[3];
drawcfield point (-2) of paths[3];
drawfield point (4/6) of paths[3];
drawcfield point (2) of paths[3];
drawfield point (2) of paths[5];
drawcfield point (4-4/6) of paths[5];
drawfield point (-2) of paths[5];
drawcfield point (-4/6) of paths[5];

label.bot(btex $4$ etex, point (4-8*(+0.5)/6) of paths[2]);
label.bot(btex $\bar 5$ etex, point (4-8*(-0.5)/6) of paths[2]);
label.rt(btex $6$ etex, point (-4/3-8*(+0.5)/6) of paths[6]);
label.top(btex $\bar 1$ etex, point (-4/3-8*(-0.5)/6) of paths[6]);
label.top(btex $2$ etex, point (4/3-8*(+0.5)/6) of paths[4]);
label.lft(btex $\bar 3$ etex, point (4/3-8*(-0.5)/6) of paths[4]);
label.rt(btex \,a\,\vphantom{$\hat g$} etex, point 0.3 of ((point (-4/6) of paths[1])--(point (4-4/6) of paths[5])));
label.lft(btex \,c\,\vphantom{$\hat g$} etex, point 0.3 of ((point (4+4/6) of paths[3])--(point (4/6) of paths[1])));
label.top(btex b\vphantom{$\hat g$} etex, point 0.3 of ((point (+2) of paths[5])--(point (-2) of paths[3])));

drawccut (0,0)--(point 0.5 of ((point (4-8*(-0.5)/6) of paths[2])--(point (-4/3-8*(+0.5)/6) of paths[6])));
drawccut (0,0)--(point 0.5 of ((point (-4/3-8*(-0.5)/6) of paths[6])--(point (4/3-8*(+0.5)/6) of paths[4])));
drawccut (0,0)--(point 0.5 of ((point (4/3-8*(-0.5)/6) of paths[4])--(point (4-8*(+0.5)/6) of paths[2])));

putcopy(1);
\end{mpostfile}
\begin{mpostfile}{FigCut61CSa.mps}
paths[1]:=(fullcircle scaled 2xu shifted (-1.5xu,0)) rotated -90;
paths[2]:=(fullcircle scaled 4xu shifted (-1.5xu,0)) rotated -90;
paths[3]:=(fullcircle scaled 0.8xu shifted (+0.75xu,1.3xu)) rotated -90;
paths[4]:=(fullcircle scaled 3xu shifted (+0.75xu,1.3xu)) rotated -90;
paths[5]:=(fullcircle scaled 0.8xu shifted (+0.75xu,-1.3xu)) rotated -90;
paths[6]:=(fullcircle scaled 3xu shifted (+0.75xu,-1.3xu)) rotated -90;

fill paths[1] colloop;
draw paths[1];

fill paths[3] colamp;
draw paths[3];

fill paths[5] colamp;
draw paths[5];

label(btex $4$ etex scaled 1.5, center(paths[3]));
label(btex $4$ etex scaled 1.5, center(paths[5]));

for i=0 upto 1:
drawleg (point (4+8*(i-0.5)/6) of paths[1])--(point (4+8*(i-0.5)/6) of paths[2]);
endfor
for i=0 upto 1:
drawleg (point (4/3+8*(i-0.5)/6) of paths[3])--(point (4/3+8*(i-0.5)/6) of paths[4]);
endfor
for i=0 upto 1:
drawleg (point (-4/3+8*(i-0.5)/6) of paths[5])--(point (-4/3+8*(i-0.5)/6) of paths[6]);
endfor

drawleg (point (4-8*(+0.5)/6) of paths[1])--(point (+4/6) of paths[1]);
drawleg (point (4-8*(-0.5)/6) of paths[1])--(point (-4/6) of paths[1]);
filldot (point 0.5 of ((point (4-8*(+0.5)/6) of paths[1])--(point (+4/6) of paths[1])), 0white);
filldot (point 0.5 of ((point (4-8*(-0.5)/6) of paths[1])--(point (-4/6) of paths[1])), 0white);
drawgluon (point 0.5 of ((point (4-8*(+0.5)/6) of paths[1])--(point (+4/6) of paths[1])))--(point 0.5 of ((point (4-8*(-0.5)/6) of paths[1])--(point (-4/6) of paths[1])));

drawleg (point (4/6) of paths[1])--(point (4+4/6) of paths[3]);
drawleg (point (-4/6) of paths[1])--(point (4-4/6) of paths[5]);
drawleg (point (-2) of paths[3])--(point (+2) of paths[5]);

drawcfield point (4-8*(+0.5)/6) of paths[1];
drawfield point (4-8*(-0.5)/6) of paths[1];
drawcfield point (-4/6) of paths[1];
drawfield point (+4/6) of paths[1];
drawcfield point (4+4/6) of paths[3];
drawfield point (-2) of paths[3];
drawcfield point (4/6) of paths[3];
drawfield point (2) of paths[3];
drawcfield point (2) of paths[5];
drawfield point (4-4/6) of paths[5];
drawcfield point (-2) of paths[5];
drawfield point (-4/6) of paths[5];

label.top(btex $\bar 1$ etex, point (4-8*(+0.5)/6) of paths[2]);
label.top(btex $2$ etex, point (4-8*(-0.5)/6) of paths[2]);
label.lft(btex $\bar 3$ etex, point (-4/3-8*(+0.5)/6) of paths[6]);
label.bot(btex $4$ etex, point (-4/3-8*(-0.5)/6) of paths[6]);
label.bot(btex $\bar 5$ etex, point (4/3-8*(+0.5)/6) of paths[4]);
label.rt(btex $6$ etex, point (4/3-8*(-0.5)/6) of paths[4]);
label.bot(btex a\vphantom{$\hat g$} etex, point 0.7 of ((point (+2) of paths[5])--(point (-2) of paths[3])));

drawccut ((-3.5xu,0)--(+2xu,0)) rotated -90;

putcopy(1);
\end{mpostfile}
\begin{mpostfile}{FigCut61CSb.mps}
paths[1]:=fullcircle scaled 0.8xu shifted (-1.5xu,0);
paths[2]:=fullcircle scaled 3xu shifted (-1.5xu,0);
paths[3]:=fullcircle scaled 0.8xu shifted (+1.5xu,0);
paths[4]:=fullcircle scaled 3xu shifted (+1.5xu,0);

fill paths[1] colamp;
draw paths[1];

fill paths[3] colamp;
draw paths[3];

label(btex $4$ etex scaled 1.5, center(paths[1]));
label(btex $4$ etex scaled 1.5, center(paths[3]));

drawleg (point 0 of paths[1])--(point 4 of paths[3]);

for i=-1 upto 1:
  drawleg (point (4+8*i/6) of paths[1])--(point (4+8*i/6) of paths[2]);
  drawleg (point (0+8*i/6) of paths[3])--(point (0+8*i/6) of paths[4]);
endfor;

filldot (point 0.25 of ((point (4-8*1/6) of paths[1])--(point (4-8*1/6) of paths[2])), 0white);
filldot (point 0.25 of ((point (0+8*1/6) of paths[3])--(point (0+8*1/6) of paths[4])), 0white);
drawgluon (point 0.25 of ((point (4-8*1/6) of paths[1])--(point (4-8*1/6) of paths[2])))--(point 0.25 of ((point (0+8*1/6) of paths[3])--(point (0+8*1/6) of paths[4])));

drawfield point (4-8/6) of paths[1];
drawcfield point (4) of paths[1];
drawfield point (4+8/6) of paths[1];
drawcfield point (0) of paths[1];

drawcfield point (-8/6) of paths[3];
drawfield point (0) of paths[3];
drawcfield point (+8/6) of paths[3];
drawfield point (4) of paths[3];

label.lrt(btex $\bar 5$ etex, point (-8/6) of paths[4]);
label.rt(btex $6$ etex, point (0) of paths[4]);
label.urt(btex $\bar 1$ etex, point (+8/6) of paths[4]);
label.ulft(btex $2$ etex, point (4-8/6) of paths[2]);
label.lft(btex $\bar 3$ etex, point (4) of paths[2]);
label.llft(btex $4$ etex, point (4+8/6) of paths[2]);
label.bot(btex a\vphantom{g'} etex, point 0.7 of ((point 0 of paths[1])--(point 4 of paths[3])));

drawccut (0,-1.5xu)--(0,+1.5xu);

putcopy(1);
\end{mpostfile}
\begin{mpostfile}{FigCut61CSc.mps}
paths[1]:=fullcircle scaled 0.8xu shifted (-1.5xu,0);
paths[2]:=fullcircle scaled 3xu shifted (-1.5xu,0);
paths[3]:=fullcircle scaled 0.8xu shifted (+1.5xu,0);
paths[4]:=fullcircle scaled 3xu shifted (+1.5xu,0);

fill paths[1] colamp;
draw paths[1];

fill paths[3] colamp;
draw paths[3];

label(btex $4'$ etex scaled 1.5, center(paths[1]));
label(btex $4'$ etex scaled 1.5, center(paths[3]));

drawleg (point -2/3 of paths[1]){dir -30}..{dir 30}(point (4+2/3) of paths[3]);
drawgluon((point +2/3 of paths[1]){dir 30}..{dir -30}(point (4-2/3) of paths[3]));

for i=-1 upto 1:
  drawleg (point (4+8*i/6) of paths[1])--(point (4+8*i/6) of paths[2]);
  drawleg (point (0+8*i/6) of paths[3])--(point (0+8*i/6) of paths[4]);
endfor;

drawfield point (4-8/6) of paths[1];
drawcfield point (4) of paths[1];
drawfield point (4+8/6) of paths[1];
drawcfield point (-2/3) of paths[1];
filldot(point (+2/3) of paths[1], 0white);

drawcfield point (-8/6) of paths[3];
drawfield point (0) of paths[3];
drawcfield point (+8/6) of paths[3];
drawfield point (4+2/3) of paths[3];
filldot(point (4-2/3) of paths[3],0white);

label.lrt(btex $\bar 5$ etex, point (-8/6) of paths[4]);
label.rt(btex $6$ etex, point (0) of paths[4]);
label.urt(btex $\bar 1$ etex, point (+8/6) of paths[4]);
label.ulft(btex $2$ etex, point (4-8/6) of paths[2]);
label.lft(btex $\bar 3$ etex, point (4) of paths[2]);
label.llft(btex $4$ etex, point (4+8/6) of paths[2]);
label.bot(btex a\vphantom{g'} etex, point 0.7 of ((point -2/3 of paths[1]){dir -30}..{dir 30}(point (4+2/3) of paths[3])));

drawccut (0,-1.5xu)--(0,+1.5xu);

putcopy(1);
\end{mpostfile}

\section{One-Loop Six-Point Amplitude from Unitarity}
\label{sec:OneLoopSixPoint}

In this section we will apply the methods of generalized unitarity 
which have proved so useful in \sym\
\cite{Britto:2004nc,Buchbinder:2005wp} to \scs. 
In particular, we want to reconstruct the one-loop six-point 
superamplitude from tree-level amplitudes by evaluating the maximal cuts,
i.e.\ triple cuts for the case of three dimensions.
We assume%
\footnote{Using the standard arguments analogous to those in four dimensions, 
e.g.\ \cite{Bern:1993kr}, 
it is possible to show that any three-dimensional CS matter 
one-loop amplitude can be written as a linear combination of triangle, 
bubbles and tadpoles. 
For a finite, superconformal, and indeed at least at weak coupling ``dual" superconformal, 
theory such as ABJM
there will be no bubble or tadpole scalar integrals and we can use the scalar 
triangles as a basis.}
that 
an arbitrary $n$-point one-loop amplitude, $A_{n}^{(1)}$, 
can be written as a linear combination  of scalar triangle diagrams, ${\cal I}_{3,i}$, 
\[
A_{n}^{(1)}=\sum_i d_i~ {\cal I}_{3,i}
\]
and thus we can use the maximal cuts to determine the coefficients $d_i$. 

\begin{figure}
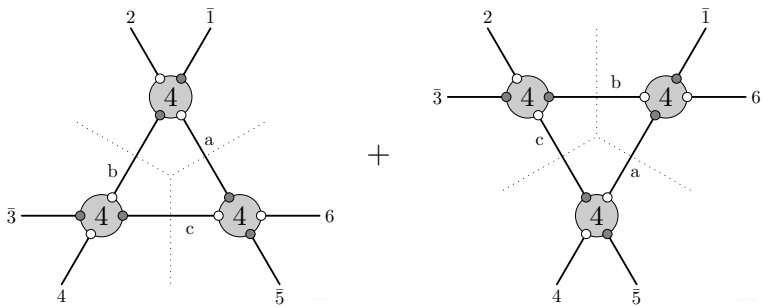
\centering
\includegraphicsbox[scale=0.7]{FigTriple61a.mps}\quad $+$ \quad \includegraphicsbox[scale=0.7]{FigTriple61b.mps}
\caption{Maximal cuts for the one-loop six-point amplitude in terms of massive triangles.}
\label{fig:max_cut_triangle}
\end{figure}

All on-shell amplitudes in \scs\ have an even number of legs,
which implies that there are at least two external legs at each corner of the triangle.
Expressions for the massive triangles in an arbitrary number of dimensions 
can be found in \cite{Boos:1987bg,Boos:1990rg}. 
For external momenta $K_1$, $K_2$, $K_3$ in $D=3-2 \epsilon$
the triangle integral evaluates to%
\footnote{For the scalar integral we choose the 
three-dimensional normalization to be $c_N=\tfrac{4\pi}{(2\pi)^3}$ corresponding
to the usual momentum measure factor and an additional factor of $4\pi$ 
(c.f.\ normalization of four-dimensional scalar box integrals in \cite{Bern:1994zx}).
This normalization is consistent with earlier sections.}
\begin{align}
\mathcal{I}_3&=c_{N}\int \frac{d^3\ell} 
{\bigbrk{-\ell^2+i\epsilon}\bigbrk{-(\ell+K_1)^2+i\epsilon}\bigbrk{-(\ell+K_1+K_2)^2+i\epsilon}}
\nln
&=
-\frac{i\pi}{2}\,\frac{1}{\sqrt{K_1^2-i\epsilon}\sqrt{K_2^2-i\epsilon}\sqrt{K_3^2-i\epsilon}}\,.
\end{align}
Obviously, the minimal number of external legs is six
implying that the four-point amplitude is uncorrected at one loop.
For six points 
there are two relevant massive triangle diagrams,
see \figref{fig:max_cut_triangle},
\[
A_6^{(1)}(\bar 1,2,\bar 3,4,\bar 5,6)=
 d_1\, {\cal I}_{3,1}
+ d_2\, {\cal I}_{3,2}.
\]
The integral ${\cal I}_{3,1}$
has external momenta 
\[
K_1=p_1+p_2,\quad
K_2=p_3+p_4,\quad
K_3=p_5+p_6
\]
and ${\cal I}_{3,2}$ correspondingly
\[
K_1=p_6+p_1,\quad
K_2=p_2+p_3,\quad
K_3=p_4+p_5.
\]
The triple cuts correspond to putting all internal propagators on-shell
and one must sum over all such momentum configurations. 
For six points the sum is thus over products of tree-level four-point amplitudes, 
where the sum over internal states can be done by performing Gra{\ss}mann integrations 
over the internal legs as in \eqref{eq:Lambdaintegral}. 

The coefficient of the first integral, ${\cal I}_{3,1}$, 
is given by the left-hand cut in \figref{fig:max_cut_triangle},
\[
d_{1}=\frac{1}{2} \sum_{\rm sol}\int 
\prod_{i=\mathrm{a},\mathrm{b},\mathrm{c}}
d^{0|3}\eta_i
~
A_{4}(\bar 1, 2, i\mathrm{\bar b}, \mathrm{a})\,
A_{4}(\bar 3, 4, i\mathrm{\bar c}, \mathrm{b})\,
A_{4}(\bar 5, 6, i\mathrm{\bar a}, \mathrm{c}).
\]
The on-shell loop momenta, $\ell_i$,  are completely fixed by the delta-functions from the cut 
propagators so there is no remaining integration but rather a sum over the
two solutions to the equations%
\footnote{One way to determine $\ell$ is 
to choose $\ell^\mu=\alpha K_1^\mu+\beta K_2^\mu +\gamma K_\times^\mu$
where $K_\times^\mu=\epsilon^{\mu\nu\rho}K_1{}_{\nu} K_2{}_{\rho}$ 
and use \eqref{eq:cutconstraints} to determine $\alpha$, $\beta$ and $\gamma$. 
One finds two solutions
\begin{equation*}
\alpha = \frac{(K_1\cdot K_2)K_2^2-K_1^2K_2^2+2 (K_1\cdot K_2)^2}{2K_\times^2}~,~~
\beta=\frac{K_1^2(K_1\cdot K_2+K_2^2)}{2K_\times^2}~,~~
\gamma=s\,\frac{\sqrt{K_1^2K_2^2K_3^2}}{2K_\times^2}\,.
\end{equation*}
where $s=\pm$ enumerates the two solutions.
Alternatively, one can directly solve the equations 
for the corresponding $\lambda$'s.}
\[
\label{eq:cutconstraints}
\ell\indup{a}^2=\ell^2=0~, ~~~\ell\indup{b}^2=(\ell+K_1)^2=0~, ~~~\ell\indup{c}^2=(\ell+K_1+K_2)^2=0.
\]
 We now turn our attention to the Gra{\ss}mann 
integration over the delta-functions appearing in the tree-level four point amplitudes
\begin{align}
&\int \prod_{i=\mathrm{a},\mathrm{b},\mathrm{c}} d^{0|3}\eta_{i} \,
\delta^{0|6}(\Theta_{13}-\lambda_{\mathrm{b}}\eta_{\mathrm{b}}+\lambda_{\mathrm{a}}\eta_{\mathrm{a}})\,
\delta^{0|6}(\Theta_{35}-\lambda_{\mathrm{c}}\eta_{\mathrm{c}}+\lambda_{\mathrm{b}}\eta_{\mathrm{b}})\,
\delta^{0|6}(\Theta_{51}-\lambda_{\mathrm{a}}\eta_{\mathrm{a}}+\lambda_{\mathrm{c}}\eta_{\mathrm{c}})\nn\\
=&
\frac{\delta^{0|6}(Q)}{\langle \mathrm{c}\mathrm{b}\rangle^3}\,
\delta^{0|3}\bigbrk{\langle \mathrm{a}|x_{13}x_{35}|\Theta_{51}\rangle
+\langle \Theta_{13}|x_{35}x_{51}|\mathrm{a}\rangle}
.
\end{align}
In this equation we have pulled out an overall factor,  $\delta^{(6)}(Q)$, of the total supermomentum
$Q$ and 
we have repeatedly used Schouten's identity which accounts for the fact 
that $\lambda_{a}$ and $\lambda_{b,c}$ appear with different weights. Finally we have used the 
notation, $x_{jk}=\sum_{i=j}^{k-1}p_i$ and  $\langle \Theta_{jk}|=\sum_{i=j}^{k-1}\eta_i\langle i|$. Thus, including the denominator
factors from the four point amplitudes, we find the coefficient for the cut 
\begin{align}
d_1&=\frac{i}{2}\sum_{s=\pm}\frac{\delta^3(P)\,\delta^{0|6}(Q)}{\langle \mathrm{b}\mathrm{c}\rangle^3}
\frac{\delta^{0|3}(\langle \mathrm{a}|x_{13}x_{35}|\Theta_{51}\rangle
+\langle \Theta_{13}|x_{35}x_{51}|\mathrm{a}\rangle)}{
\langle 12 \rangle \langle 3 4\rangle \langle 56\rangle \langle   \mathrm{b} 2\rangle  \langle  \mathrm{c} 4\rangle  \langle \mathrm{a} 6\rangle}\,.
\end{align}
The complete contribution to the one-loop six-point amplitude from this scalar amplitude is thus
\[
d_1 {\cal I}_{3,1}=
-\frac{i \pi}{2}\,
\frac{d_1}
{\sqrt{-\langle 12\rangle^2-i\epsilon} \,
 \sqrt{-\langle 34\rangle^2-i\epsilon} \,
 \sqrt{-\langle 56\rangle^2-i\epsilon}}\,
.
\]
We can evaluate the square roots carefully to find
\[
\sqrt{-x-i\epsilon}=-i\cabs{x}=-ix\csgn(x)
\]
where the 
generalizations of the absolute value 
and sign functions were defined in \eqref{eq:csgncabs}.
We end up with 
\[
d_1 {\cal I}_{3,1}=
-\frac{\pi}{2}\,
\frac{d_1}{\langle 12\rangle\langle 34\rangle\langle 56\rangle}\,
\csgn\langle 12\rangle
 \csgn\langle 34\rangle
 \csgn\langle 56\rangle,
\]
which is proportional to the shifted tree-level scattering amplitude
\eqref{eq:shifted6amp}
up to some sign factors,%
\footnote{This comparison is done by taking specific values 
for the external momenta and evaluating various component amplitudes numerically.}
\[
d_1 {\cal I}_{3,1}=
\tfrac{i}{4}\pi A\indup{6}^{(0)}(\bar 6,1,\bar 2,3,\bar 4,5)
\csgn\langle 12\rangle
 \csgn\langle 34\rangle
 \csgn\langle 56\rangle.
\]
In fact, and as discussed in \secref{sec:anomaly6}, the tree-level
six-point amplitude of \cite{Bargheer:2010hn}
is the sum of two terms $Y_1$, $Y_2$ related by $\Lambda_k\to-\Lambda_k$ parity
transformations.
These two terms are exactly the $s=\pm1$ terms appearing in the cut. 
Here it is important to note that the shifted tree-level amplitude
$A\indup{6}^{(0)}(\bar 6,1,\bar 2,3,\bar 4,5)$
has the opposite assignment of conjugate particles
compared to the one-loop amplitude
$A\indup{6}^{(1)}(\bar 1,2,\bar 3,4,\bar 5,6)$.
Superficially this leads to the wrong sign under 
any of the transformations $\Lambda_k\to -\Lambda_k$,
which is compensated by the sign functions involving all of the external
$\lambda_k$.

The scalar triangle integral ${\cal I}_{3,2}$
is captured by the right-hand cut in \figref{fig:max_cut_triangle},
\[
d_{2}=\frac{1}{2} \sum_{\rm sol}\int 
\prod_{i=\mathrm{a},\mathrm{b},\mathrm{c}}
d^{0|3}\eta_i
~
A_{4}(\bar 1, \mathrm{b}, i\mathrm{\bar a}, 6)\,
A_{4}(\bar 3, \mathrm{c}, i\mathrm{\bar b}, 2)\,
A_{4}(\bar 5, \mathrm{a}, i\mathrm{\bar c}, 4).
\]
The calculation is identical to the previous case
and the result is 
\[
d_2=\frac{i}{2}\sum_{s=\pm}\frac{\delta^3(P)\,\delta^{0|6}(Q)}{\langle \mathrm{c}\mathrm{b}\rangle^3}\,
\frac{\delta^{0|3}\bigbrk{\langle \mathrm{a}|x_{62}x_{24}|\Theta_{46}\rangle
+\langle \Theta_{62}|x_{24}x_{46}|\mathrm{a}\rangle}}{
\langle 61 \rangle \langle 3 2\rangle \langle 54\rangle \langle 6 \mathrm{a}\rangle  \langle 2 \mathrm{b}\rangle  \langle 4 \mathrm{c}\rangle}~, 
\]
where the on-shell loop momenta, $\ell_i^2=0$, are related by $\ell\indup{b}=\ell\indup{a}-K_1$ and $\ell\indup{c}=\ell\indup{b}-K_2$. 
Expanding the square roots in the triangle integral we find 
\[
d_2 {\cal I}_{3,2} =\tfrac{i}{4}\pi A_6^{(0)}(\bar 6, 1, \bar 2, 3, \bar 4, 5)
\csgn\langle 61\rangle
 \csgn\langle 23\rangle
 \csgn\langle 45\rangle,
\]
Altogether our result for the one-loop six-point amplitudes reads
\[
A_6^{(1)}(\bar 1,2,\bar 3,4,\bar 5,6)=
\frac{i\pi}{4} c_6(\bar 1, \bar 2,\bar 3, \bar 4,\bar 5,\bar 6)
A\indup{6}^{(0)}(\bar 6,1,\bar 2,3,\bar 4,5)\,
\label{eq:oneloopfinal}
\]
with the piecewise constant combination of sign functions
\[
c_6(\bar 1, \bar 2,\bar 3,\bar  4,\bar 5, \bar  6)
=
\csgn\langle 12\rangle
 \csgn\langle 34\rangle
 \csgn\langle 56\rangle
+
\csgn\langle 61\rangle
 \csgn\langle 23\rangle
 \csgn\langle 45\rangle.
\]
Here the bars over the labels indicate that $c_6$ transforms odd under
sign flips of any $\lambda_k$. 
The relative plus sign between the two products of  ${\rm sgn}_c$
functions follows from the calculation, however we can also understand
it from the symmetries of the amplitudes. As previously described \eqref{eq:Ainversion}, 
the color ordered six-point superamplitudes are odd/even functions under 
an inversion of the color ordering,
$A(\bar 1, 2, \bar 3, 4, \bar 5, 6)=\mp A(\bar 1, 6, \bar 5, 4, \bar 3, 2)$.
While the tree-level amplitude and its cyclically shifted version are odd, the one-loop 
amplitude is even under this transformation. Thus the piecewise constant, $c_6$, must be an odd function
under the inversion map, and this requires the relative
plus sign between the two terms. 

\begin{figure}
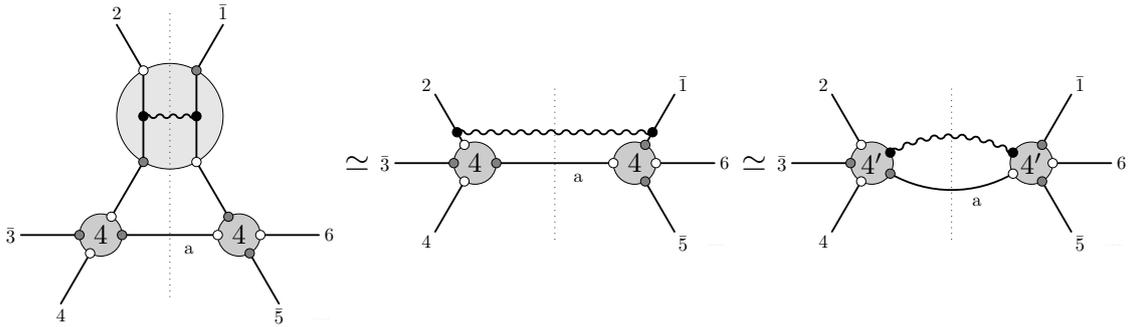
\centering
\includegraphicsbox[scale=0.7]{FigCut61CSa.mps} 
$\simeq$ \includegraphicsbox[scale=0.7]{FigCut61CSb.mps} 
$\simeq$ \includegraphicsbox[scale=0.7]{FigCut61CSc.mps} 
\caption{Cut of the one-loop six-point amplitude in the 123 channel}
\label{fig:Cut6LoopCS}
\end{figure}

To conclude, let us analyze the cuts of the above expression.
In this regard the near equality of tree and loop level amplitudes
begs for an explanation. How can the one-loop result have the same
set of discontinuities as the corresponding one at tree level when the 
cuts are obviously different?
In particular, the pole in the three-particle channel at tree level 
originates from the splitting into two four-particle trees \eqref{eq:cutsixtree}. 
The one-loop result has the same pole
but no apparent splitting into subamplitudes. 
Taking a closer look at the origin of the discontinuity one
finds that it requires the momentum transfer in one of the corners 
of the triangle to be zero. 
At this point, the four-point amplitude has a pole
which is responsible for the three-particle pole. 
This pole corresponds to the zero-mode of the Chern--Simons gauge field. 
So indeed there is a physical cut for the one-loop 
six-point amplitude in the three-particle channel, see \figref{fig:Cut6LoopCS}.%
\footnote{It is worth pointing out that unitarity in the original meaning of the word 
is different from the cutting rules of generalized unitarity:
One would not expect the non-propagating Chern--Simons field to appear in unitarity cuts.
This means that the above mentioned poles are to be evaluated 
in a principal value prescription. In this case 
the zero-mode does not appear in the imaginary part of amplitudes:
$\Im F(x) = 0$ for $F(x)=1/x$.
Conversely, in generalized unitarity the zero-mode is detectable
by the cutting rule $F(x+i\epsilon)-F(x-i\epsilon)=-2\pi i\delta(x)$, and needs to be taken into consideration.}
%
Somewhat surprisingly, this pole agrees precisely 
with the pole of the cyclically shifted tree-level amplitude
up to some sign factors.
The other relevant cuts are in two-particle channels. 
These are the natural cuts at the one-loop level but are not present 
at tree level and thus they must be associated
to the additional sign factors which we recall originate in square roots of ``masses'', 
$\sqrt{-m^2}=\sqrt{(p_i+p_{i+1})^2}$, 
occurring in the scalar triangles, 
whose branch cuts are just
next to the real axis when the two inflowing energies are aligned (positive $m^2$).

It is also quite clear that the sign factors in $c_6$ give rise to the 
correct codimension-one superconformal anomalies
discussed in \secref{sec:anomaly6}. 
The superconformal variations act as derivatives which turn the 
sign factors into delta functions supported on 
collinear configurations of any two adjacent particles as described
previously.

\section{Conclusions \& Outlook}
\label{sec:concl}
In this paper we have shown that  scattering amplitudes in three-dimensional \scs\ 
theory give rise to an anomaly of the (super)conformal symmetry in a fashion similar to \sym\ theory. 
As for the four-dimensional theory in $(2,2)$ signature \cite{Beisert:2010gn}, 
here the anomaly arises from sign functions of spinor brackets. 
These sign factors emerge when rewriting the four-point amplitude 
by making use of the different scaling behavior of bosonic 
and fermionic delta functions, schematically (cf.\ \secref{sec:anomaly}):
\begin{equation}
\frac{1}{x^2}\delta^2(x \,\mathrm{bos})\delta^3(x\, \mathrm{ferm})
=
\frac{x}{\abs{x}}\delta^2(\mathrm{bos})\delta^3(\mathrm{ferm})
=
\sgn{x}\,\delta^2(\mathrm{bos})\delta^3(\mathrm{ferm})~, 
\end{equation}
where $x$ represents the spinor brackets.
While in four dimensions the anomaly can be captured in terms of a vertex with three legs, we find a corresponding four-vertex $\gen{S}_4=\gen{S}\amp_4$ with support on collinear momentum configurations. We have employed the anomaly to predict the non-vanishing of the one-loop six-point amplitude:
Firstly, there are two Yangian invariants (up to anomalous contributions) whose linear combinations furnish the tree-level and one-loop six-point amplitude. Considering the different anomalies of the tree and one-loop term as well as discrete symmetries of physical amplitudes, shows that the proportionality factor translating between them is a non-trivial function of the external momenta. 
Consequently, the one-loop six-point amplitude is non-trivial%
\footnote{Note that a non-vanishing result for the one-loop six-point amplitude was also obtained by an independent Feynman calculation in \cite{Bianchi:2012cq}.}
and proportional to the tree-level contribution cyclically shifted: $\amp_6^{(1)}=c_6 \amp_6^{(0)}(\{ i\rightarrow i-1\})$. We have confirmed this result by a unitarity construction of the one-loop amplitude using a triple cut. It is important to note the different structure of the anomaly of the six-point amplitude at loop level when compared to \sym\ theory. While in four dimensions the anomaly is only distributional at tree-level and gets smeared in the loop integration at one loop, the three-dimensional anomaly is still distributional at one loop and only gets smeared at two loops. As a consequence, the one-loop six-point amplitude obeys the same symmetry constraints as the tree-level expression up to distributional terms. 

The non-vanishing of the one-loop six-point amplitude is particularly interesting in the light of a possible duality between scattering amplitudes and Wilson loops in ABJM theory. 
As the lightlike hexagon Wilson loop is known to vanish at one-loop order \cite{Henn:2010ps, Bianchi:2011rn},
the non-trivial one-loop six-point amplitude poses the puzzle of what a possible map could look like. 
A simple solution would be that there is no self-T-duality in ABJM theory and thus no reason for Wilson loops and scattering amplitudes to match. In order to look for further hints for either outcome one could start by stripping off the tree-level contribution and try to compare only scalar loop corrections. The most interesting question seems to be: Can one define a hexagon Wilson loop that matches our one-loop result of the six-point amplitude? Relatedly, it may be useful to find the correct analog
of the super-Wilson loop \cite{Mason:2010yk, CaronHuot:2010ek, CaronHuot:2011ky, Beisert:2012xx}
in  three dimensions to match with superamplitudes. A similar question occurs when we wish to consider Wilson loops with an odd number of 
edges, which a priori exist and are non-trivial, 
though they will be generically complex \cite{Henn:2010ps}, while 
there are only amplitudes for even numbers of external legs 
if we allow only on-shell scalars and fermions as external particles.

An important point we have not discussed here is whether the breaking 
of conformal symmetry can be cured by deformation of the free representation 
of $\alg{osp}(6|4)$ on amplitudes. This seems very plausible, in particular 
with regard to analogous considerations in \sym\ theory where 
the conformal generators of the $\alg{psu}(2,2|4)$ representation 
were deformed to compensate for the anomaly. This is particularly 
desirable since it renders the scattering matrix an exact symmetry invariant 
and thereby recursively relates amplitudes with different numbers of legs 
to each other, e.g.\ 
$ \gen{\bar S} \amp^{4d-{\rm MHV}}_n+ \gen{\bar S}_{3}^+\amp^{4d-{\rm MHV}}_{n-1}=0$. 
Here $\gen{\bar S}_{3}^+$ denotes the deformation of the generator $\gen{\bar S}$ 
corresponding to the three-point anomaly vertex in four dimensions. 
The starting point of the four-dimensional recursion is  
$\gen{\bar S}\amp^{4d}_4=0$ being consistent with a three-point 
amplitude that vanishes for physical kinematics. 
Straightforwardly translating this recursion to ABJM theory 
with only even-point amplitudes yields \begin{equation}
\label{eq:3drec}
\gen{S}\amp_n+\gen{S}_{4}^+\amp_{n-2}\stackrel{?}{=}0.
\end{equation}
 Here, however, $\gen{S}\amp_4$  is non-vanishing as discussed above 
and the inductive symmetry would require a non-trivial two-point invariant 
with $\gen{S}\amp_2=0$ as a starting point. Notably, we can construct 
a two-point $\alg{osp}(6|4)$ invariant that renders \eqref{eq:3drec} correct for $n=4$. 
It takes the form $\amp_2=\delta^{2|3}(\Lambda_1\pm i \Lambda_2)$. 
Provided the algebra of deformed generators still holds, this gives 
hope for a  construction similar to four dimensions. 
But is the two-point invariant really part of the ABJM scattering matrix?

Taking a closer look at the interaction Hamiltonian that induces the
scattering matrix, the vertex with the lowest number of points has three
legs and includes the Chern--Simons field. The Chern--Simons field, however, 
is not dynamical and should thus not appear as an external particle in scattering amplitudes. 
Could the three-vertex still give rise to a non-vanishing two-point invariant 
in the scattering matrix such that $\amp_2=\amp_3|_{p_{\mathrm{CS}} =0}$?
At first sight it seems not clear how to technically investigate this
point. The Chern--Simons field has no on-shell degrees of freedom and is
thus not captured by the ordinary on-shell superspace formulation of
scattering amplitudes. Considering the scattering matrix in terms of
oscillators corresponding to field excitations, the same problem arises.
While creation and annihilation operators live on the forward and backward
mass shell, respectively, a priori neither choice seems appropriate to
describe the Chern--Simons field. Can one still introduce a corresponding oscillator? 
What commutation relations would this imply?
Notably, the zero-mode of the Chern--Simons field already played a special
role in explaining the discontinuity of the one-loop six-point amplitude in
\secref{sec:OneLoopSixPoint} (cf.\ \figref{fig:Cut6LoopCS}). Its generic
role for the generalized unitarity construction of scattering amplitudes seems yet
unclear. A deeper understanding of these issues appears to have the
potential for new insights into the structure of Chern--Simons theories in general.

Obviously, it is very tempting to extend our results for the one-loop six-point amplitude 
to higher numbers of loops and legs. In particular this should shed light 
on the general structure of one-loop amplitudes in ABJM theory which is  
of great importance for a possible duality to Wilson loops. 
A starting point could be the investigation of constraints  
on generic one-loop amplitudes imposed by symmetry and the form of the anomaly. 
It would also be very interesting to see whether 
a similar anomaly arises in dimensions greater than four.

\pdfbookmark[2]{Acknowledgments}{acknowledgments}
\subsection*{Acknowledgments}

We thank Nima Arkani-Hamed, Marco Bianchi, Matias Leoni, Andrea Mauri,
Carlo Meneghelli, Silvia Penati and Alberto Santambrogio for interesting discussions.
The work of TB was supported by the Swedish Research Council (VR) under
grant 621-2007-4177. 
The work of NB was partially supported by grant no.\
200021-137616 from the Swiss National Science Foundation
and by grant no.\ 962 by the German-Israeli Foundation (GIF).
TB, NB and FL thank the Kavli Institute for Theoretical Physics (KITP)
for hospitality during the workshop ``Harmony of Scattering Amplitudes''. 
The research at KITP was supported in part by the U.S.\ National Science
Foundation under grant no.\ NSF PHY05-51164.
We thank NORDITA for hospitality during the workshop  
``Exact Results in Gauge-String Dualities''.

\appendix

\section{Mixed Energy Signs}
\label{app:mixedenergy}

Distributing positive/negative energies (incoming/outgoing particles) in
different ways, the analysis of \secref{sec:anomaly} gets slightly
modified.

\paragraph{Same Sign on One and Three.}

If particles $1$ and $3$ carry the same energy sign, and particles
$2$ and $4$ carry the opposite energy sign, the identities
\eqref{eq:34as12} have to be modified by appropriate factors of $i$:
\begin{align}
1&=i\abs{\sprods{12}}\int\dd\alpha_3\,\dd\beta_3\,\deltad{2}(\lambda_3-\alpha_3\lambda_1-i\beta_3\lambda_2)\,,\nn\\
1&=i\abs{\sprods{12}}\int\dd\alpha_4\,\dd\beta_4\,\deltad{2}(\lambda_4-i\alpha_4\lambda_1-\beta_4\lambda_2)\,.
\end{align}
The momentum conservation delta function becomes
\begin{align}
\deltad{3}(P)
&=
\deltad{3}\bigbrk{
	 \lambda_1\lambda_1(1+\alpha_3^2-\alpha_4^2)
	+\lambda_2\lambda_2(1-\beta_3^2+\beta_4^2)
	+i(\lambda_1\lambda_2+\lambda_2\lambda_1)(\alpha_3\beta_3+\alpha_4\beta_4)
}\nn\\
&=
\frac{1}{\abs{\sprods{12}}^3}\,\delta(1+\alpha_3^2-\alpha_4^2)\,\delta(1-\beta_3^2+\beta_4^2)\,\delta(\alpha_3\beta_3+\alpha_4\beta_4)\,,
\end{align}
such that the amplitude reads
\begin{multline}
\amp_4^{1,3\leftrightarrow2,4}(\bar 1,2,\bar 3,4)=\frac{\deltad{6}(Q)}{\abs{\sprods{12}}\sprods{12}^2}\int\dd\alpha_3\,\dd\alpha_4\,\dd\beta_3\,\dd\beta_4\,\frac{1}{\alpha_3}
	\cdot\\\cdot
	\delta(1+\alpha_3^2-\alpha_4^2)\,\delta(1-\beta_3^2+\beta_4^2)\,\delta(\alpha_3\beta_3+\alpha_4\beta_4)
	\cdot\\\cdot
	\deltad{2}(\lambda_3-\alpha_3\lambda_1-i\beta_3\lambda_2)\,\deltad{2}(\lambda_4-i\alpha_4\lambda_1-\beta_4\lambda_2)\,.
\end{multline}
The first two delta functions are each supported on a pair of hyperbolas in
$(\alpha_3,\alpha_4)$ and $(\beta_3,\beta_4)$ space. Using the
parametrization
\begin{equation}
\alpha_3=r_\alpha\sinh\alpha\,,
\qquad
\alpha_4=r_\alpha\cosh\alpha\,,
\qquad
\beta_3=r_\beta\cosh\beta\,,
\qquad
\beta_4=r_\beta\sinh\beta\,,
\end{equation}
where the radial variables $r_\alpha$, $r_\beta$ take all real values, the
radial integrals localize at $r_\alpha=\pm1$, $r_\beta=\pm1$:
\begin{multline}
\amp_4^{1,3\leftrightarrow2,4}(\bar 1,2,\bar 3,4)=\frac{\deltad{6}(Q)}{\abs{\sprods{12}}\sprods{12}^2}
	\sum_{s_\alpha,s_\beta=\pm1}\int\dd\alpha\,\dd\beta\,\frac{s_\alpha}{4\sinh\alpha}\,\delta(\sinh(\alpha+\beta))
	\cdot\\\cdot
	\deltad{2}(\lambda_3-s_\alpha\sinh\alpha\,\lambda_1-is_\beta\cosh\beta\,\lambda_2)\,
	\deltad{2}(\lambda_4-is_\alpha\cosh\alpha\,\lambda_1-s_\beta\sinh\beta\,\lambda_2)\,.
\end{multline}
The first delta function localizes the beta integral at $\beta=-\alpha$, thus
\begin{multline}
\amp_4^{1,3\leftrightarrow2,4}(\bar 1,2,\bar 3,4)=
	\frac{\deltad{6}(Q)}{\abs{\sprods{12}}\sprods{12}^2}\sum_{s_\alpha,s_\beta=\pm1}\int_{-\infty}^\infty\dd\alpha\,\frac{s_\alpha}{4\sinh\alpha}
	\cdot\\\cdot
	\deltad{2}(\lambda_3-s_\alpha\sinh\alpha\,\lambda_1-is_\beta\cosh\alpha\,\lambda_2)\,
	\deltad{2}(\lambda_4-is_\alpha\cosh\alpha\,\lambda_1+s_\beta\sinh\alpha\,\lambda_2)\,.
\end{multline}
Moving $\deltad{6}(Q)$ under the integral sign, contracting $Q$ once
with $\lambda_3$ and once with $\lambda_4$, and expanding
$\lambda_{3,4}$ in terms of $\lambda_{1,2}$ shows that
\begin{align}
\deltad{6}(Q)
&=\sprods{34}^{-3}\,\deltad{3}(\sprods{31}\eta_1+\sprods{32}\eta_2+\sprods{34}\eta_4)
                  \,\deltad{3}(\sprods{41}\eta_1+\sprods{42}\eta_2+\sprods{43}\eta_3)\nn\\
&=-s_\alpha s_\beta\sprods{12}^3\,\deltad{3}(\eta_3-s_\alpha\sinh\alpha\,\eta_1-is_\beta\cosh\alpha\,\eta_2)
	\cdot\nn\\&\mspace{200mu}\cdot
                                  \deltad{3}(\eta_4-is_\alpha\cosh\alpha\,\eta_1+s_\beta\sinh\alpha\,\eta_2)\,.
\end{align}
The four-point amplitude hence reads 
%
\begin{multline}
\amp_4^{1,3\leftrightarrow2,4}(\bar 1,2,\bar 3,4)=-i\csgn\sprods{12}\sum_{s_\alpha,s_\beta=\pm1}\int_{-\infty}^\infty\dd\alpha\,\frac{s_\beta}{4\sinh\alpha}
	\cdot\\\cdot
	\deltad{2|3}(\Lambda_3-s_\alpha\sinh\alpha\,\Lambda_1-is_\beta\cosh\alpha\,\Lambda_2)\,
	\deltad{2|3}(\Lambda_4-is_\alpha\cosh\alpha\,\Lambda_1+s_\beta\sinh\alpha\,\Lambda_2)\,,
\label{eq:A4pmpm1}
\end{multline}
where again $\Lambda=(\lambda,\eta)$.
Reverting the direction of integration in the terms with $s_\beta=-1$ gives
\eqref{eq:A4pmpm}:
\begin{multline}
\amp_4^{1,3\leftrightarrow2,4}(\bar 1,2,\bar 3,4)=-i\csgn\sprods{12}\sum_{s_\alpha,s_\beta=\pm1}\int_{-\infty}^\infty\dd\alpha\,\frac{1}{4\sinh\alpha}
	\cdot\\\cdot
	\deltad{2|3}(\Lambda_3-s_\beta(s_\alpha\sinh\alpha\,\Lambda_1+i\cosh\alpha\,\Lambda_2))\,
	\deltad{2|3}(\Lambda_4-is_\alpha\cosh\alpha\,\Lambda_1+\sinh\alpha\,\Lambda_2)\,.
\end{multline}
Alternatively, reverting the direction of integration in the terms with
$s_\alpha=-1$ results in
\begin{multline}
\amp_4^{1,3\leftrightarrow2,4}(\bar 1,2,\bar 3,4)=-i\csgn\sprods{12}\sum_{s_\alpha,s_\beta=\pm1}\int_{-\infty}^\infty\dd\alpha\,\frac{s_\alpha s_\beta}{4\sinh\alpha}
	\cdot\\\cdot
	\deltad{2|3}(\Lambda_3-\sinh\alpha\,\Lambda_1-is_\beta\cosh\alpha\,\Lambda_2)\,
	\deltad{2|3}(\Lambda_4-s_\alpha(i\cosh\alpha\,\Lambda_1-s_\beta\sinh\alpha\,\Lambda_2))\,.
\end{multline}
%

\paragraph{Same Sign on One and Four.}

If the momenta of particles $1$ and $4$ carry the same energy sign (opposed
to $1$ and $3$), then the previous derivation up to \eqref{eq:A4pmpm1}
works exactly the same, up to the following substitutions:
\begin{equation}
\alpha_3\rightarrow i\alpha_3\,,
\qquad
\alpha_4\rightarrow -i\alpha_4\,,
\qquad
\beta_3\rightarrow -i\beta_3\,,
\qquad
\beta_4\rightarrow i\beta_4\,,
\end{equation}
and accordingly
\begin{equation}
\sinh\alpha\rightarrow i\cosh\alpha\,,
\quad
\cosh\alpha\rightarrow -i\sinh\alpha\,,
\quad
\sinh\beta\rightarrow i\cosh\beta\,,
\quad
\cosh\beta\rightarrow -i\sinh\beta\,.
\end{equation}
Hence the amplitude in this case reads
\begin{multline}
\amp_4^{1,4\leftrightarrow2,3}(\bar 1,2,\bar 3,4)=-i\csgn\sprods{12}\sum_{s_\alpha,s_\beta=\pm1}\int_{-\infty}^\infty\dd\alpha\,\frac{s_\beta}{4i\cosh\alpha}
	\cdot\\\cdot
	\deltad{2|3}(\Lambda_3-is_\alpha\cosh\alpha\,\Lambda_1-s_\beta\sinh\alpha\,\Lambda_2)\,
	\deltad{2|3}(\Lambda_4-s_\alpha\sinh\alpha\,\Lambda_1+is_\beta\cosh\alpha\,\Lambda_2)\,.
\end{multline}
Substituting $s_\beta\rightarrow-s_\beta$ and subsequently reverting the
direction of integration when $s_\beta=-1$ yields \eqref{eq:A4pmmp}
\begin{multline}
\amp_4^{1,4\leftrightarrow2,3}(\bar 1,2,\bar 3,4)=i\csgn\sprods{12}\sum_{s_\alpha,s_\beta=\pm1}\int_{-\infty}^\infty\dd\alpha\,\frac{s_\beta}{4i\cosh\alpha}
	\cdot\\\cdot
	\deltad{2|3}(\Lambda_3-is_\alpha\cosh\alpha\,\Lambda_1+\sinh\alpha\,\Lambda_2)\,
	\deltad{2|3}(\Lambda_4-s_\beta(s_\alpha\sinh\alpha\,\Lambda_1+i\cosh\alpha\,\Lambda_2))\,.
\end{multline}
Alternatively, substituting $s_\beta\rightarrow-s_\beta$ and subsequently reverting the
direction of integration when $s_\alpha=-1$ gives
\begin{multline}
\amp_4^{1,4\leftrightarrow2,3}(\bar 1,2,\bar 3,4)=i\csgn\sprods{12}\sum_{s_\alpha,s_\beta=\pm1}\int_{-\infty}^\infty\dd\alpha\,\frac{s_\beta}{4i\cosh\alpha}
	\cdot\\\cdot
	\deltad{2|3}(\Lambda_3-s_\alpha(i\cosh\alpha\,\Lambda_1-s_\beta\sinh\alpha\,\Lambda_2))\,
	\deltad{2|3}(\Lambda_4-\sinh\alpha\,\Lambda_1-is_\beta\cosh\alpha\,\Lambda_2)\,.
\end{multline}


\begin{mpostcmd}
verbatimtex 
\end{document}
etex

end;
\end{mpostcmd}


\begin{bibtex}
@article{Bargheer:2011mm,
      author         = "Bargheer, Till and Beisert, Niklas and Loebbert, Florian",
      title          = "{Exact Superconformal and Yangian Symmetry of Scattering
                        Amplitudes}",
      journal        = "J.Phys.",
      volume         = "A44",
      pages          = "454012",
      doi            = "10.1088/1751-8113/44/45/454012",
      year           = "2011",
      eprint         = "1104.0700",
      archivePrefix  = "arXiv",
      primaryClass   = "hep-th",
      SLACcitation   = "
}

@article{Bianchi:2012cq,
      author         = "Bianchi, Marco S. and Leoni, Matias and Mauri, Andrea and
                        Penati, Silvia and Santambrogio, Alberto",
      title          = "{One Loop Amplitudes In ABJM}",
      journal        = "JHEP",
      volume         = "1207",
      pages          = "029",
      year           = "2012",
      eprint         = "1204.4407",
      archivePrefix  = "arXiv",
      primaryClass   = "hep-th",
      SLACcitation   = "
}

@article{Aharony:2008ug,
      author         = "Aharony, Ofer and Bergman, Oren and Jafferis, Daniel
                        Louis and Maldacena, Juan",
      title          = "{N=6 superconformal Chern-Simons-matter theories,
                        M2-branes and their gravity duals}",
      journal        = "JHEP",
      volume         = "0810",
      pages          = "091",
      doi            = "10.1088/1126-6708/2008/10/091",
      year           = "2008",
      eprint         = "0806.1218",
      archivePrefix  = "arXiv",
      primaryClass   = "hep-th",
      SLACcitation   = "
}

@article{Bargheer:2010hn,
      author         = "Bargheer, Till and Loebbert, Florian and Meneghelli,
                        Carlo",
      title          = "{Symmetries of Tree-level Scattering Amplitudes in $\mathcal{N}$ = 6
                        Superconformal Chern-Simons Theory}",
      journal        = "Phys.Rev.",
      volume         = "D82",
      pages          = "045016",
      doi            = "10.1103/PhysRevD.82.045016",
      year           = "2010",
      eprint         = "1003.6120",
      archivePrefix  = "arXiv",
      primaryClass   = "hep-th",
      SLACcitation   = "
}

@article{Agarwal:2008pu,
      author         = "Agarwal, Abhishek and Beisert, Niklas and McLoughlin,
                        Tristan",
      title          = "{Scattering in Mass-Deformed $\mathcal{N}$ $\leq$ 4 Chern-Simons
                        Models}",
      journal        = "JHEP",
      volume         = "0906",
      pages          = "045",
      doi            = "10.1088/1126-6708/2009/06/045",
      year           = "2009",
      eprint         = "0812.3367",
      archivePrefix  = "arXiv",
      primaryClass   = "hep-th",
      SLACcitation   = "
}

@article{Lee:2010du,
      author         = "Lee, Sangmin",
      title          = "{Yangian Invariant Scattering Amplitudes in
                        Supersymmetric Chern-Simons Theory}",
      journal        = "Phys.Rev.Lett.",
      volume         = "105",
      pages          = "151603",
      doi            = "10.1103/PhysRevLett.105.151603",
      year           = "2010",
      eprint         = "1007.4772",
      archivePrefix  = "arXiv",
      primaryClass   = "hep-th",
      SLACcitation   = "
}

@article{Beisert:2010gn,
      author         = "Beisert, Niklas and Henn, Johannes and McLoughlin,
                        Tristan and Plefka, Jan",
      title          = "{One-Loop Superconformal and Yangian Symmetries of
                        Scattering Amplitudes in $\mathcal{N}$ = 4 Super Yang-Mills}",
      journal        = "JHEP",
      volume         = "1004",
      pages          = "085",
      doi            = "10.1007/JHEP04(2010)085",
      year           = "2010",
      eprint         = "1002.1733",
      archivePrefix  = "arXiv",
      primaryClass   = "hep-th",
      SLACcitation   = "
}

@article{Bargheer:2009qu,
      author         = "Bargheer, Till and Beisert, Niklas and Galleas,
                        Wellington and Loebbert, Florian and McLoughlin, Tristan",
      title          = "{Exacting $\mathcal{N}$ = 4 Superconformal Symmetry}",
      journal        = "JHEP",
      volume         = "0911",
      pages          = "056",
      doi            = "10.1088/1126-6708/2009/11/056",
      year           = "2009",
      eprint         = "0905.3738",
      archivePrefix  = "arXiv",
      primaryClass   = "hep-th",
      SLACcitation   = "
}

@article{CaronHuot:2011kk,
      author         = "Caron-Huot, Simon and He, Song",
      title          = "{Jumpstarting the All-Loop S-Matrix of Planar $\mathcal{N}$ = 4 Super
                        Yang-Mills}",
      journal        = "JHEP",
      volume         = "1207",
      pages          = "174",
      doi            = "10.1007/JHEP07(2012)174",
      year           = "2012",
      eprint         = "1112.1060",
      archivePrefix  = "arXiv",
      primaryClass   = "hep-th",
      SLACcitation   = "
}

@article{Bullimore:2011kg,
      author         = "Bullimore, Mathew and Skinner, David",
      title          = "{Descent Equations for Superamplitudes}",
      year           = "2011",
      eprint         = "1112.1056",
      archivePrefix  = "arXiv",
      primaryClass   = "hep-th",
      SLACcitation   = "
}

@article{Colgain:2012ca,
      author         = "\'O Colg\'ain, Eoin",
      title          = "{Self-duality of the D1-D5 near-horizon}",
      journal        = "JHEP",
      volume         = "1204",
      pages          = "047",
      doi            = "10.1007/JHEP04(2012)047",
      year           = "2012",
      eprint         = "1202.3416",
      archivePrefix  = "arXiv",
      primaryClass   = "hep-th",
      SLACcitation   = "
}

@article{Gang:2010gy,
      author         = "Gang, Dongmin and Huang, Yu-tin and Koh, Eunkyung and
                        Lee, Sangmin and Lipstein, Arthur E.",
      title          = "{Tree-level Recursion Relation and Dual Superconformal
                        Symmetry of the ABJM Theory}",
      journal        = "JHEP",
      volume         = "1103",
      pages          = "116",
      doi            = "10.1007/JHEP03(2011)116",
      year           = "2011",
      eprint         = "1012.5032",
      archivePrefix  = "arXiv",
      primaryClass   = "hep-th",
      SLACcitation   = "
}

@article{Buchbinder:2005wp,
      author         = "Buchbinder, Evgeny I. and Cachazo, Freddy",
      title          = "{Two-loop amplitudes of gluons and octa-cuts in $\mathcal{N}$ = 4 super
                        Yang-Mills}",
      journal        = "JHEP",
      volume         = "0511",
      pages          = "036",
      doi            = "10.1088/1126-6708/2005/11/036",
      year           = "2005",
      eprint         = "hep-th/0506126",
      archivePrefix  = "arXiv",
      primaryClass   = "hep-th",
      SLACcitation   = "
}

@article{Britto:2004nc,
      author         = "Britto, Ruth and Cachazo, Freddy and Feng, Bo",
      title          = "{Generalized unitarity and one-loop amplitudes in $\mathcal{N}$ = 4
                        super-Yang-Mills}",
      journal        = "Nucl.Phys.",
      volume         = "B725",
      pages          = "275-305",
      doi            = "10.1016/j.nuclphysb.2005.07.014",
      year           = "2005",
      eprint         = "hep-th/0412103",
      archivePrefix  = "arXiv",
      primaryClass   = "hep-th",
      SLACcitation   = "
}

@Article{Bern:1994zx,
      author         = "Bern, Zvi and Dixon, Lance J. and Dunbar, David C. and Kosower, David A.",
      title          = "One loop n point gauge theory amplitudes, unitarity and collinear limits",
      journal        = "Nucl.Phys.",
      volume         = "B425",
      pages          = "217-260",
      doi            = "10.1016/0550-3213(94)90179-1",
      year           = "1994",
      eprint         = "hep-ph/9403226",
      archivePrefix  = "arXiv",
      primaryClass   = "hep-ph",
      reportNumber   = "SLAC-PUB-6415, SACLAY-SPH-T-94-20, UCLA-TEP-94-4, SWAT-94-17",
      SLACcitation   = "
}

@article{Bern:1993kr,
      author         = "Bern, Zvi and Dixon, Lance J. and Kosower, David A.",
      title          = "{Dimensionally regulated pentagon integrals}",
      journal        = "Nucl.Phys.",
      volume         = "B412",
      pages          = "751-816",
      doi            = "10.1016/0550-3213(94)90398-0",
      year           = "1994",
      eprint         = "hep-ph/9306240",
      archivePrefix  = "arXiv",
      primaryClass   = "hep-ph",
      SLACcitation   = "
}

@article{Boos:1990rg,
      author         = "Boos, E. E. and Davydychev, Andrei I.",
      title          = "{A Method of evaluating massive Feynman integrals}",
      journal        = "Theor.Math.Phys.",
      volume         = "89",
      pages          = "1052-1063",
      doi            = "10.1007/BF01016805",
      year           = "1991",
      SLACcitation   = "
}

@article{Boos:1987bg,
      author         = "Boos, E. E. and Davydychev, Andrei I.",
      title          = "{A Method Of The Evaluation Of The Vertex Type Feynman
                        Integrals}",
      journal        = "Moscow Univ.Phys.Bull.",
      volume         = "42N3",
      pages          = "6-10",
      year           = "1987",
      SLACcitation   = "
}


@article{Berkovits:2008ic,
      author         = "Berkovits, Nathan and Maldacena, Juan",
      title          = "{Fermionic T-Duality, Dual Superconformal Symmetry, and
                        the Amplitude/Wilson Loop Connection}",
      journal        = "JHEP",
      volume         = "0809",
      pages          = "062",
      doi            = "10.1088/1126-6708/2008/09/062",
      year           = "2008",
      eprint         = "0807.3196",
      archivePrefix  = "arXiv",
      primaryClass   = "hep-th",
      SLACcitation   = "
}

@article{Beisert:2008iq,
      author         = "Beisert, Niklas and Ricci, Riccardo and Tseytlin, Arkady
                        A. and Wolf, Martin",
      title          = "{Dual Superconformal Symmetry from AdS$_5$ $\times$ S$^5$
                        Superstring Integrability}",
      journal        = "Phys.Rev.",
      volume         = "D78",
      pages          = "126004",
      doi            = "10.1103/PhysRevD.78.126004",
      year           = "2008",
      eprint         = "0807.3228",
      archivePrefix  = "arXiv",
      primaryClass   = "hep-th",
      SLACcitation   = "
}

@article{Grassi:2009yj,
      author         = "Grassi, Pietro Antonio and Sorokin, Dmitri and Wulff,
                        Linus",
      title          = "{Simplifying superstring and D-brane actions in AdS$_4$ $\times$
                        CP$^3$ superbackground}",
      journal        = "JHEP",
      volume         = "0908",
      pages          = "060",
      doi            = "10.1088/1126-6708/2009/08/060",
      year           = "2009",
      eprint         = "0903.5407",
      archivePrefix  = "arXiv",
      primaryClass   = "hep-th",
      SLACcitation   = "
}

@article{Adam:2009kt,
      author         = "Adam, Ido and Dekel, Amit and Oz, Yaron",
      title          = "{On Integrable Backgrounds Self-dual under Fermionic
                        T-duality}",
      journal        = "JHEP",
      volume         = "0904",
      pages          = "120",
      doi            = "10.1088/1126-6708/2009/04/120",
      year           = "2009",
      eprint         = "0902.3805",
      archivePrefix  = "arXiv",
      primaryClass   = "hep-th",
      SLACcitation   = "
}

@article{Adam:2010hh,
      author         = "Adam, Ido and Dekel, Amit and Oz, Yaron",
      title          = "{On the fermionic T-duality of the AdS$_4$ $\times$ CP$^3$
                        sigma-model}",
      journal        = "JHEP",
      volume         = "1010",
      pages          = "110",
      doi            = "10.1007/JHEP10(2010)110",
      year           = "2010",
      eprint         = "1008.0649",
      archivePrefix  = "arXiv",
      primaryClass   = "hep-th",
      SLACcitation   = "
}

@article{Dekel:2011qw,
      author         = "Dekel, Amit and Oz, Yaron",
      title          = "{Self-Duality of Green-Schwarz Sigma-Models}",
      journal        = "JHEP",
      volume         = "1103",
      pages          = "117",
      doi            = "10.1007/JHEP03(2011)117",
      year           = "2011",
      eprint         = "1101.0400",
      archivePrefix  = "arXiv",
      primaryClass   = "hep-th",
      SLACcitation   = "
}

@article{Bakhmatov:2010fp,
      author         = "Bakhmatov, Ilya",
      title          = "{On AdS$_4$ $\times$ CP$^3$ T-duality}",
      journal        = "Nucl.Phys.",
      volume         = "B847",
      pages          = "38-53",
      doi            = "10.1016/j.nuclphysb.2011.01.020",
      year           = "2011",
      eprint         = "1011.0985",
      archivePrefix  = "arXiv",
      primaryClass   = "hep-th",
      SLACcitation   = "
}

@article{Bakhmatov:2011aa,
      author         = "Bakhmatov, Ilya and \'O Colg\'ain, Eoin and Yavartanoo,
                        Hossein",
      title          = "{Fermionic T-duality in the pp-wave limit}",
      journal        = "JHEP",
      volume         = "1110",
      pages          = "085",
      doi            = "10.1007/JHEP10(2011)085",
      year           = "2011",
      eprint         = "1109.1052",
      archivePrefix  = "arXiv",
      primaryClass   = "hep-th",
      SLACcitation   = "
}

@article{Huang:2010qy,
      author         = "Huang, Yu-tin and Lipstein, Arthur E.",
      title          = "{Dual Superconformal Symmetry of $\mathcal{N}$ = 6 Chern-Simons
                        Theory}",
      journal        = "JHEP",
      volume         = "1011",
      pages          = "076",
      doi            = "10.1007/JHEP11(2010)076",
      year           = "2010",
      eprint         = "1008.0041",
      archivePrefix  = "arXiv",
      primaryClass   = "hep-th",
      SLACcitation   = "
}

@article{Chen:2011vv,
      author         = "Chen, Wei-Ming and Huang, Yu-tin",
      title          = "{Dualities for Loop Amplitudes of $\mathcal{N}$ = 6 Chern-Simons Matter
                        Theory}",
      journal        = "JHEP",
      volume         = "1111",
      pages          = "057",
      doi            = "10.1007/JHEP11(2011)057",
      year           = "2011",
      eprint         = "1107.2710",
      archivePrefix  = "arXiv",
      primaryClass   = "hep-th",
      SLACcitation   = "
}
@article{Bianchi:2011dg,
      author         = "Bianchi, Marco S. and Leoni, Matias and Mauri, Andrea and
                        Penati, Silvia and Santambrogio, Alberto",
      title          = "{Scattering Amplitudes/Wilson Loop Duality In ABJM
                        Theory}",
      journal        = "JHEP",
      volume         = "1201",
      pages          = "056",
      doi            = "10.1007/JHEP01(2012)056",
      year           = "2012",
      eprint         = "1107.3139",
      archivePrefix  = "arXiv",
      primaryClass   = "hep-th",
      SLACcitation   = "
}

@article{Henn:2010ps,
      author         = "Henn, Johannes M. and Plefka, Jan and Wiegandt,
                        Konstantin",
      title          = "{Light-like polygonal Wilson loops in 3d Chern-Simons and
                        ABJM theory}",
      journal        = "JHEP",
      volume         = "1008",
      pages          = "032",
      doi            = "10.1007/JHEP08(2010)032",
      year           = "2010",
      eprint         = "1004.0226",
      archivePrefix  = "arXiv",
      primaryClass   = "hep-th",
      SLACcitation   = "
}

@article{Wiegandt:2011uu,
      author         = "Wiegandt, Konstantin",
      title          = "{Equivalence of Wilson Loops in $\mathcal{N}$ = 6 super
                        Chern-Simons matter theory and $\mathcal{N}$ = 4 SYM Theory}",
      journal        = "Phys.Rev.",
      volume         = "D84",
      pages          = "126015",
      doi            = "10.1103/PhysRevD.84.126015",
      year           = "2011",
      eprint         = "1110.1373",
      archivePrefix  = "arXiv",
      primaryClass   = "hep-th",
      SLACcitation   = "
}

@article{Bianchi:2011rn,
      author         = "Bianchi, Marco S. and Leoni, Matias and Mauri, Andrea and
                        Penati, Silvia and Ratti, CarloAlberto and Santambrogio, Alberto",
      title          = "{From Correlators to Wilson Loops in Chern-Simons Matter
                        Theories}",
      journal        = "JHEP",
      volume         = "1106",
      pages          = "118",
      doi            = "10.1007/JHEP06(2011)118",
      year           = "2011",
      eprint         = "1103.3675",
      archivePrefix  = "arXiv",
      primaryClass   = "hep-th",
      SLACcitation   = "
}

@article{CaronHuot:2010ek,
      author         = "Caron-Huot, Simon",
      title          = "{Notes on the scattering amplitude / Wilson loop
                        duality}",
      journal        = "JHEP",
      volume         = "1107",
      pages          = "058",
      doi            = "10.1007/JHEP07(2011)058",
      year           = "2011",
      eprint         = "1010.1167",
      archivePrefix  = "arXiv",
      primaryClass   = "hep-th",
      SLACcitation   = "
}

@article{CaronHuot:2011ky,
      author         = "Caron-Huot, Simon",
      title          = "{Superconformal symmetry and two-loop amplitudes in
                        planar $\mathcal{N}$ = 4 super Yang-Mills}",
      journal        = "JHEP",
      volume         = "1112",
      pages          = "066",
      doi            = "10.1007/JHEP12(2011)066",
      year           = "2011",
      eprint         = "1105.5606",
      archivePrefix  = "arXiv",
      primaryClass   = "hep-th",
      SLACcitation   = "
}

@article{Alday:2007hr,
      author         = "Alday, Luis F. and Maldacena, Juan Martin",
      title          = "{Gluon scattering amplitudes at strong coupling}",
      journal        = "JHEP",
      volume         = "0706",
      pages          = "064",
      doi            = "10.1088/1126-6708/2007/06/064",
      year           = "2007",
      eprint         = "0705.0303",
      archivePrefix  = "arXiv",
      primaryClass   = "hep-th",
      SLACcitation   = "
}

@article{Drummond:2006rz,
      author         = "Drummond, J. M. and Henn, J. and Smirnov, V. A. and
                        Sokatchev, E.",
      title          = "{Magic identities for conformal four-point integrals}",
      journal        = "JHEP",
      volume         = "0701",
      pages          = "064",
      doi            = "10.1088/1126-6708/2007/01/064",
      year           = "2007",
      eprint         = "hep-th/0607160",
      archivePrefix  = "arXiv",
      primaryClass   = "hep-th",
      SLACcitation   = "
}

@article{Brandhuber:2008pf,
      author         = "Brandhuber, Andreas and Heslop, Paul and Travaglini,
                        Gabriele",
      title          = "{A Note on dual superconformal symmetry of the $\mathcal{N}$ = 4 super
                        Yang-Mills S-matrix}",
      journal        = "Phys.Rev.",
      volume         = "D78",
      pages          = "125005",
      doi            = "10.1103/PhysRevD.78.125005",
      year           = "2008",
      eprint         = "0807.4097",
      archivePrefix  = "arXiv",
      primaryClass   = "hep-th",
      SLACcitation   = "
}

@article{Drummond:2008vq,
      author         = "Drummond, J. M. and Henn, J. and Korchemsky, G. P. and
                        Sokatchev, E.",
      title          = "{Dual superconformal symmetry of scattering amplitudes in
                        $\mathcal{N}$ = 4 super-Yang-Mills theory}",
      journal        = "Nucl.Phys.",
      volume         = "B828",
      pages          = "317-374",
      doi            = "10.1016/j.nuclphysb.2009.11.022",
      year           = "2010",
      eprint         = "0807.1095",
      archivePrefix  = "arXiv",
      primaryClass   = "hep-th",
      SLACcitation   = "
}

@article{Drummond:2009fd,
      author         = "Drummond, James M. and Henn, Johannes M. and Plefka, Jan",
      title          = "{Yangian symmetry of scattering amplitudes in $\mathcal{N}$ = 4 super
                        Yang-Mills theory}",
      journal        = "JHEP",
      volume         = "0905",
      pages          = "046",
      doi            = "10.1088/1126-6708/2009/05/046",
      year           = "2009",
      eprint         = "0902.2987",
      archivePrefix  = "arXiv",
      primaryClass   = "hep-th",
      SLACcitation   = "
}

@article{Drummond:2007aua,
      author         = "Korchemsky, G. P. and Drummond, J. M. and Sokatchev, E.",
      title          = "{Conformal properties of four-gluon planar amplitudes and
                        Wilson loops}",
      journal        = "Nucl.Phys.",
      volume         = "B795",
      pages          = "385-408",
      doi            = "10.1016/j.nuclphysb.2007.11.041",
      year           = "2008",
      eprint         = "0707.0243",
      archivePrefix  = "arXiv",
      primaryClass   = "hep-th",
      SLACcitation   = "
}

@article{Drummond:2007cf,
      author         = "Drummond, J. M. and Henn, J. and Korchemsky, G. P. and
                        Sokatchev, E.",
      title          = "{On planar gluon amplitudes/Wilson loops duality}",
      journal        = "Nucl.Phys.",
      volume         = "B795",
      pages          = "52-68",
      doi            = "10.1016/j.nuclphysb.2007.11.007",
      year           = "2008",
      eprint         = "0709.2368",
      archivePrefix  = "arXiv",
      primaryClass   = "hep-th",
      SLACcitation   = "
}

@article{Korchemsky:2009hm,
      author         = "Korchemsky, G. P. and Sokatchev, E.",
      title          = "{Symmetries and analytic properties of scattering
                        amplitudes in $\mathcal{N}$ = 4 SYM theory}",
      journal        = "Nucl.Phys.",
      volume         = "B832",
      pages          = "1-51",
      doi            = "10.1016/j.nuclphysb.2010.01.022",
      year           = "2010",
      eprint         = "0906.1737",
      archivePrefix  = "arXiv",
      primaryClass   = "hep-th",
      SLACcitation   = "
}

@article{Brandhuber:2007yx,
      author         = "Brandhuber, Andreas and Heslop, Paul and Travaglini,
                        Gabriele",
      title          = "{MHV amplitudes in $\mathcal{N}$ = 4 super Yang-Mills and Wilson
                        loops}",
      journal        = "Nucl.Phys.",
      volume         = "B794",
      pages          = "231-243",
      doi            = "10.1016/j.nuclphysb.2007.11.002",
      year           = "2008",
      eprint         = "0707.1153",
      archivePrefix  = "arXiv",
      primaryClass   = "hep-th",
      SLACcitation   = "
}

@article{Sever:2009aa,
      author         = "Sever, Amit and Vieira, Pedro",
      title          = "{Symmetries of the $\mathcal{N}$ = 4 SYM S-matrix}",
      year           = "2009",
      eprint         = "0908.2437",
      archivePrefix  = "arXiv",
      primaryClass   = "hep-th",
      SLACcitation   = "
}

@article{Mason:2010yk,
      author         = "Mason, L. J. and Skinner, David",
      title          = "{The Complete Planar S-matrix of $\mathcal{N}$ = 4 SYM as a Wilson Loop
                        in Twistor Space}",
      journal        = "JHEP",
      volume         = "1012",
      pages          = "018",
      doi            = "10.1007/JHEP12(2010)018",
      year           = "2010",
      eprint         = "1009.2225",
      archivePrefix  = "arXiv",
      primaryClass   = "hep-th",
      SLACcitation   = "
}

@article{Eden:2011yp,
      author         = "Eden, Burkhard and Heslop, Paul and Korchemsky, Gregory
                        P. and Sokatchev, Emery",
      title          = "{The super-correlator/super-amplitude duality: Part I}",
      year           = "2011",
      eprint         = "1103.3714",
      archivePrefix  = "arXiv",
      primaryClass   = "hep-th",
      SLACcitation   = "
}

@article{Eden:2011ku,
      author         = "Eden, Burkhard and Heslop, Paul and Korchemsky, Gregory
                        P. and Sokatchev, Emery",
      title          = "{The super-correlator/super-amplitude duality: Part II}",
      year           = "2011",
      eprint         = "1103.4353",
      archivePrefix  = "arXiv",
      primaryClass   = "hep-th",
      SLACcitation   = "
}

@article{Drummond:2007au,
      author         = "Drummond, J. M. and Henn, J. and Korchemsky, G. P. and
                        Sokatchev, E.",
      title          = "{Conformal Ward identities for Wilson loops and a test of
                        the duality with gluon amplitudes}",
      journal        = "Nucl.Phys.",
      volume         = "B826",
      pages          = "337-364",
      doi            = "10.1016/j.nuclphysb.2009.10.013",
      year           = "2010",
      eprint         = "0712.1223",
      archivePrefix  = "arXiv",
      primaryClass   = "hep-th",
      SLACcitation   = "
}

@article{Beisert:2012xx,
      author         = "Beisert, Niklas and He, Song and Schwab, Burkhard U. W.
                        and Vergu, Cristian",
      title          = "{Null Polygonal Wilson Loops in Full $\mathcal{N}$ = 4 Superspace}",
      journal        = "J.Phys.",
      volume         = "A45",
      pages          = "265402",
      doi            = "10.1088/1751-8113/45/26/265402",
      year           = "2012",
      eprint         = "1203.1443",
      archivePrefix  = "arXiv",
      primaryClass   = "hep-th",
      SLACcitation   = "
}

\end{bibtex}

\pdfbookmark[1]{\refname}{references}
\bibliographystyle{nb}
\bibliography{confanom3d}

\begin{thebibliography}{10}
\providecommand{\href}[2]{#2}
\providecommand{\arxivref}[2]{\href{http://arxiv.org/abs/#1}{#2}}
\providecommand{\doiref}[2]{\href{http://dx.doi.org/#1}{#2}}
\providecommand{\nbbstauthor}[1]{#1}
\providecommand{\nbbstjournal}[1]{\textsf{#1}}
\providecommand{\nbbsttitle}[1]{\textit{#1}}
\providecommand{\nbbsturl}[1]{\texttt{#1}}
\providecommand{\nbbsteprint}[1]{\texttt{#1}}
\providecommand{\nbbststyle}{\raggedright\small\parskip0pt}
\nbbststyle

\bibitem{Bargheer:2009qu}
\nbbstauthor{T.~Bargheer, N.~Beisert, W.~Galleas, F.~Loebbert and
  T.~McLoughlin},
\nbbsttitle{``{Exacting $\mathcal{N}$ = 4 Superconformal Symmetry}''},
\nbbstjournal{\doiref{10.1088/1126-6708/2009/11/056}{JHEP~0911,~056~(2009)}},
\nbbsteprint{\arxivref{0905.3738}{arxiv:0905.3738}}.

\bibitem{Beisert:2010gn}
\nbbstauthor{N.~Beisert, J.~Henn, T.~McLoughlin and J.~Plefka},
\nbbsttitle{``{One-Loop Superconformal and Yangian Symmetries of Scattering
  Amplitudes in $\mathcal{N}$ = 4 Super Yang-Mills}''},
\nbbstjournal{\doiref{10.1007/JHEP04(2010)085}{JHEP~1004,~085~(2010)}},
\nbbsteprint{\arxivref{1002.1733}{arxiv:1002.1733}}.

\bibitem{Bargheer:2011mm}
\nbbstauthor{T.~Bargheer, N.~Beisert and F.~Loebbert},
\nbbsttitle{``{Exact Superconformal and Yangian Symmetry of Scattering
  Amplitudes}''},
\nbbstjournal{\doiref{10.1088/1751-8113/44/45/454012}{J.Phys.~A44,~454012~(2011)}},
\nbbsteprint{\arxivref{1104.0700}{arxiv:1104.0700}}.

\bibitem{Aharony:2008ug}
\nbbstauthor{O.~Aharony, O.~Bergman, D.~L.~Jafferis and J.~Maldacena},
\nbbsttitle{``{N=6 superconformal Chern-Simons-matter theories, M2-branes and
  their gravity duals}''},
\nbbstjournal{\doiref{10.1088/1126-6708/2008/10/091}{JHEP~0810,~091~(2008)}},
\nbbsteprint{\arxivref{0806.1218}{arxiv:0806.1218}}.

\bibitem{Alday:2007hr}
\nbbstauthor{L.~F.~Alday and J.~M.~Maldacena},
\nbbsttitle{``{Gluon scattering amplitudes at strong coupling}''},
\nbbstjournal{\doiref{10.1088/1126-6708/2007/06/064}{JHEP~0706,~064~(2007)}},
\nbbsteprint{\arxivref{0705.0303}{arxiv:0705.0303}}.

\bibitem{Drummond:2006rz}
\nbbstauthor{J.~M.~Drummond, J.~Henn, V.~A.~Smirnov and E.~Sokatchev},
\nbbsttitle{``{Magic identities for conformal four-point integrals}''},
\nbbstjournal{\doiref{10.1088/1126-6708/2007/01/064}{JHEP~0701,~064~(2007)}},
\nbbsteprint{\arxivref{hep-th/0607160}{hep-th/0607160}}.

\bibitem{Drummond:2008vq}
\nbbstauthor{J.~M.~Drummond, J.~Henn, G.~P.~Korchemsky and E.~Sokatchev},
\nbbsttitle{``{Dual superconformal symmetry of scattering amplitudes in
  $\mathcal{N}$ = 4 super-Yang-Mills theory}''},
\nbbstjournal{\doiref{10.1016/j.nuclphysb.2009.11.022}{Nucl.Phys.~B828,~317~(2010)}},
\nbbsteprint{\arxivref{0807.1095}{arxiv:0807.1095}}.

\bibitem{Brandhuber:2008pf}
\nbbstauthor{A.~Brandhuber, P.~Heslop and G.~Travaglini},
\nbbsttitle{``{A Note on dual superconformal symmetry of the $\mathcal{N}$ = 4
  super Yang-Mills S-matrix}''},
\nbbstjournal{\doiref{10.1103/PhysRevD.78.125005}{Phys.Rev.~D78,~125005~(2008)}},
\nbbsteprint{\arxivref{0807.4097}{arxiv:0807.4097}}.

\bibitem{Drummond:2009fd}
\nbbstauthor{J.~M.~Drummond, J.~M.~Henn and J.~Plefka},
\nbbsttitle{``{Yangian symmetry of scattering amplitudes in $\mathcal{N}$ = 4
  super Yang-Mills theory}''},
\nbbstjournal{\doiref{10.1088/1126-6708/2009/05/046}{JHEP~0905,~046~(2009)}},
\nbbsteprint{\arxivref{0902.2987}{arxiv:0902.2987}}.

\bibitem{Berkovits:2008ic}
\nbbstauthor{N.~Berkovits and J.~Maldacena},
\nbbsttitle{``{Fermionic T-Duality, Dual Superconformal Symmetry, and the
  Amplitude/Wilson Loop Connection}''},
\nbbstjournal{\doiref{10.1088/1126-6708/2008/09/062}{JHEP~0809,~062~(2008)}},
\nbbsteprint{\arxivref{0807.3196}{arxiv:0807.3196}}.

\bibitem{Beisert:2008iq}
\nbbstauthor{N.~Beisert, R.~Ricci, A.~A.~Tseytlin and M.~Wolf},
\nbbsttitle{``{Dual Superconformal Symmetry from AdS$_5$ $\times$ S$^5$
  Superstring Integrability}''},
\nbbstjournal{\doiref{10.1103/PhysRevD.78.126004}{Phys.Rev.~D78,~126004~(2008)}},
\nbbsteprint{\arxivref{0807.3228}{arxiv:0807.3228}}.

\bibitem{Drummond:2007aua}
\nbbstauthor{G.~P.~Korchemsky, J.~M.~Drummond and E.~Sokatchev},
\nbbsttitle{``{Conformal properties of four-gluon planar amplitudes and Wilson
  loops}''},
\nbbstjournal{\doiref{10.1016/j.nuclphysb.2007.11.041}{Nucl.Phys.~B795,~385~(2008)}},
\nbbsteprint{\arxivref{0707.0243}{arxiv:0707.0243}}.

\bibitem{Brandhuber:2007yx}
\nbbstauthor{A.~Brandhuber, P.~Heslop and G.~Travaglini},
\nbbsttitle{``{MHV amplitudes in $\mathcal{N}$ = 4 super Yang-Mills and Wilson
  loops}''},
\nbbstjournal{\doiref{10.1016/j.nuclphysb.2007.11.002}{Nucl.Phys.~B794,~231~(2008)}},
\nbbsteprint{\arxivref{0707.1153}{arxiv:0707.1153}}.

\bibitem{Drummond:2007cf}
\nbbstauthor{J.~M.~Drummond, J.~Henn, G.~P.~Korchemsky and E.~Sokatchev},
\nbbsttitle{``{On planar gluon amplitudes/Wilson loops duality}''},
\nbbstjournal{\doiref{10.1016/j.nuclphysb.2007.11.007}{Nucl.Phys.~B795,~52~(2008)}},
\nbbsteprint{\arxivref{0709.2368}{arxiv:0709.2368}}.

\bibitem{Mason:2010yk}
\nbbstauthor{L.~J.~Mason and D.~Skinner},
\nbbsttitle{``{The Complete Planar S-matrix of $\mathcal{N}$ = 4 SYM as a
  Wilson Loop in Twistor Space}''},
\nbbstjournal{\doiref{10.1007/JHEP12(2010)018}{JHEP~1012,~018~(2010)}},
\nbbsteprint{\arxivref{1009.2225}{arxiv:1009.2225}}.

\bibitem{CaronHuot:2010ek}
\nbbstauthor{S.~Caron-Huot},
\nbbsttitle{``{Notes on the scattering amplitude / Wilson loop duality}''},
\nbbstjournal{\doiref{10.1007/JHEP07(2011)058}{JHEP~1107,~058~(2011)}},
\nbbsteprint{\arxivref{1010.1167}{arxiv:1010.1167}}.

\bibitem{Eden:2011yp}
\nbbstauthor{B.~Eden, P.~Heslop, G.~P.~Korchemsky and E.~Sokatchev},
\nbbsttitle{``{The super-correlator/super-amplitude duality: Part I}''},
\nbbsteprint{\arxivref{1103.3714}{arxiv:1103.3714}}.

\bibitem{Eden:2011ku}
\nbbstauthor{B.~Eden, P.~Heslop, G.~P.~Korchemsky and E.~Sokatchev},
\nbbsttitle{``{The super-correlator/super-amplitude duality: Part II}''},
\nbbsteprint{\arxivref{1103.4353}{arxiv:1103.4353}}.

\bibitem{Korchemsky:2009hm}
\nbbstauthor{G.~P.~Korchemsky and E.~Sokatchev},
\nbbsttitle{``{Symmetries and analytic properties of scattering amplitudes in
  $\mathcal{N}$ = 4 SYM theory}''},
\nbbstjournal{\doiref{10.1016/j.nuclphysb.2010.01.022}{Nucl.Phys.~B832,~1~(2010)}},
\nbbsteprint{\arxivref{0906.1737}{arxiv:0906.1737}}.

\bibitem{Sever:2009aa}
\nbbstauthor{A.~Sever and P.~Vieira},
\nbbsttitle{``{Symmetries of the $\mathcal{N}$ = 4 SYM S-matrix}''},
\nbbsteprint{\arxivref{0908.2437}{arxiv:0908.2437}}.

\bibitem{Drummond:2007au}
\nbbstauthor{J.~M.~Drummond, J.~Henn, G.~P.~Korchemsky and E.~Sokatchev},
\nbbsttitle{``{Conformal Ward identities for Wilson loops and a test of the
  duality with gluon amplitudes}''},
\nbbstjournal{\doiref{10.1016/j.nuclphysb.2009.10.013}{Nucl.Phys.~B826,~337~(2010)}},
\nbbsteprint{\arxivref{0712.1223}{arxiv:0712.1223}}.

\bibitem{CaronHuot:2011ky}
\nbbstauthor{S.~Caron-Huot},
\nbbsttitle{``{Superconformal symmetry and two-loop amplitudes in planar
  $\mathcal{N}$ = 4 super Yang-Mills}''},
\nbbstjournal{\doiref{10.1007/JHEP12(2011)066}{JHEP~1112,~066~(2011)}},
\nbbsteprint{\arxivref{1105.5606}{arxiv:1105.5606}}.

\bibitem{CaronHuot:2011kk}
\nbbstauthor{S.~Caron-Huot and S.~He},
\nbbsttitle{``{Jumpstarting the All-Loop S-Matrix of Planar $\mathcal{N}$ = 4
  Super Yang-Mills}''},
\nbbstjournal{\doiref{10.1007/JHEP07(2012)174}{JHEP~1207,~174~(2012)}},
\nbbsteprint{\arxivref{1112.1060}{arxiv:1112.1060}}.

\bibitem{Bullimore:2011kg}
\nbbstauthor{M.~Bullimore and D.~Skinner},
\nbbsttitle{``{Descent Equations for Superamplitudes}''},
\nbbsteprint{\arxivref{1112.1056}{arxiv:1112.1056}}.

\bibitem{Grassi:2009yj}
\nbbstauthor{P.~A.~Grassi, D.~Sorokin and L.~Wulff},
\nbbsttitle{``{Simplifying superstring and D-brane actions in AdS$_4$ $\times$
  CP$^3$ superbackground}''},
\nbbstjournal{\doiref{10.1088/1126-6708/2009/08/060}{JHEP~0908,~060~(2009)}},
\nbbsteprint{\arxivref{0903.5407}{arxiv:0903.5407}}.

\bibitem{Adam:2009kt}
\nbbstauthor{I.~Adam, A.~Dekel and Y.~Oz},
\nbbsttitle{``{On Integrable Backgrounds Self-dual under Fermionic
  T-duality}''},
\nbbstjournal{\doiref{10.1088/1126-6708/2009/04/120}{JHEP~0904,~120~(2009)}},
\nbbsteprint{\arxivref{0902.3805}{arxiv:0902.3805}}.

\bibitem{Adam:2010hh}
\nbbstauthor{I.~Adam, A.~Dekel and Y.~Oz},
\nbbsttitle{``{On the fermionic T-duality of the AdS$_4$ $\times$ CP$^3$
  sigma-model}''},
\nbbstjournal{\doiref{10.1007/JHEP10(2010)110}{JHEP~1010,~110~(2010)}},
\nbbsteprint{\arxivref{1008.0649}{arxiv:1008.0649}}.

\bibitem{Dekel:2011qw}
\nbbstauthor{A.~Dekel and Y.~Oz},
\nbbsttitle{``{Self-Duality of Green-Schwarz Sigma-Models}''},
\nbbstjournal{\doiref{10.1007/JHEP03(2011)117}{JHEP~1103,~117~(2011)}},
\nbbsteprint{\arxivref{1101.0400}{arxiv:1101.0400}}.

\bibitem{Bakhmatov:2010fp}
\nbbstauthor{I.~Bakhmatov},
\nbbsttitle{``{On AdS$_4$ $\times$ CP$^3$ T-duality}''},
\nbbstjournal{\doiref{10.1016/j.nuclphysb.2011.01.020}{Nucl.Phys.~B847,~38~(2011)}},
\nbbsteprint{\arxivref{1011.0985}{arxiv:1011.0985}}.

\bibitem{Bakhmatov:2011aa}
\nbbstauthor{I.~Bakhmatov, E.~\'O~Colg\'ain and H.~Yavartanoo},
\nbbsttitle{``{Fermionic T-duality in the pp-wave limit}''},
\nbbstjournal{\doiref{10.1007/JHEP10(2011)085}{JHEP~1110,~085~(2011)}},
\nbbsteprint{\arxivref{1109.1052}{arxiv:1109.1052}}.

\bibitem{Colgain:2012ca}
\nbbstauthor{E.~\'O~Colg\'ain},
\nbbsttitle{``{Self-duality of the D1-D5 near-horizon}''},
\nbbstjournal{\doiref{10.1007/JHEP04(2012)047}{JHEP~1204,~047~(2012)}},
\nbbsteprint{\arxivref{1202.3416}{arxiv:1202.3416}}.

\bibitem{Bargheer:2010hn}
\nbbstauthor{T.~Bargheer, F.~Loebbert and C.~Meneghelli},
\nbbsttitle{``{Symmetries of Tree-level Scattering Amplitudes in $\mathcal{N}$
  = 6 Superconformal Chern-Simons Theory}''},
\nbbstjournal{\doiref{10.1103/PhysRevD.82.045016}{Phys.Rev.~D82,~045016~(2010)}},
\nbbsteprint{\arxivref{1003.6120}{arxiv:1003.6120}}.

\bibitem{Huang:2010qy}
\nbbstauthor{Y.-t.~Huang and A.~E.~Lipstein},
\nbbsttitle{``{Dual Superconformal Symmetry of $\mathcal{N}$ = 6 Chern-Simons
  Theory}''},
\nbbstjournal{\doiref{10.1007/JHEP11(2010)076}{JHEP~1011,~076~(2010)}},
\nbbsteprint{\arxivref{1008.0041}{arxiv:1008.0041}}.

\bibitem{Chen:2011vv}
\nbbstauthor{W.-M.~Chen and Y.-t.~Huang},
\nbbsttitle{``{Dualities for Loop Amplitudes of $\mathcal{N}$ = 6 Chern-Simons
  Matter Theory}''},
\nbbstjournal{\doiref{10.1007/JHEP11(2011)057}{JHEP~1111,~057~(2011)}},
\nbbsteprint{\arxivref{1107.2710}{arxiv:1107.2710}}.

\bibitem{Bianchi:2011dg}
\nbbstauthor{M.~S.~Bianchi, M.~Leoni, A.~Mauri, S.~Penati and A.~Santambrogio},
\nbbsttitle{``{Scattering Amplitudes/Wilson Loop Duality In ABJM Theory}''},
\nbbstjournal{\doiref{10.1007/JHEP01(2012)056}{JHEP~1201,~056~(2012)}},
\nbbsteprint{\arxivref{1107.3139}{arxiv:1107.3139}}.

\bibitem{Wiegandt:2011uu}
\nbbstauthor{K.~Wiegandt},
\nbbsttitle{``{Equivalence of Wilson Loops in $\mathcal{N}$ = 6 super
  Chern-Simons matter theory and $\mathcal{N}$ = 4 SYM Theory}''},
\nbbstjournal{\doiref{10.1103/PhysRevD.84.126015}{Phys.Rev.~D84,~126015~(2011)}},
\nbbsteprint{\arxivref{1110.1373}{arxiv:1110.1373}}.

\bibitem{Henn:2010ps}
\nbbstauthor{J.~M.~Henn, J.~Plefka and K.~Wiegandt},
\nbbsttitle{``{Light-like polygonal Wilson loops in 3d Chern-Simons and ABJM
  theory}''},
\nbbstjournal{\doiref{10.1007/JHEP08(2010)032}{JHEP~1008,~032~(2010)}},
\nbbsteprint{\arxivref{1004.0226}{arxiv:1004.0226}}.

\bibitem{Bianchi:2011rn}
\nbbstauthor{M.~S.~Bianchi, M.~Leoni, A.~Mauri, S.~Penati, C.~Ratti and
  A.~Santambrogio},
\nbbsttitle{``{From Correlators to Wilson Loops in Chern-Simons Matter
  Theories}''},
\nbbstjournal{\doiref{10.1007/JHEP06(2011)118}{JHEP~1106,~118~(2011)}},
\nbbsteprint{\arxivref{1103.3675}{arxiv:1103.3675}}.

\bibitem{Bianchi:2012cq}
\nbbstauthor{M.~S.~Bianchi, M.~Leoni, A.~Mauri, S.~Penati and A.~Santambrogio},
\nbbsttitle{``{One Loop Amplitudes In ABJM}''},
\nbbstjournal{\doiref{10.1007/JHEP07(2012)029}{JHEP~1207,~029~(2012)}},
\nbbsteprint{\arxivref{1204.4407}{arxiv:1204.4407}}.

\bibitem{Agarwal:2008pu}
\nbbstauthor{A.~Agarwal, N.~Beisert and T.~McLoughlin},
\nbbsttitle{``{Scattering in Mass-Deformed $\mathcal{N}$ $\leq$ 4 Chern-Simons
  Models}''},
\nbbstjournal{\doiref{10.1088/1126-6708/2009/06/045}{JHEP~0906,~045~(2009)}},
\nbbsteprint{\arxivref{0812.3367}{arxiv:0812.3367}}.

\bibitem{Gang:2010gy}
\nbbstauthor{D.~Gang, Y.-t.~Huang, E.~Koh, S.~Lee and A.~E.~Lipstein},
\nbbsttitle{``{Tree-level Recursion Relation and Dual Superconformal Symmetry
  of the ABJM Theory}''},
\nbbstjournal{\doiref{10.1007/JHEP03(2011)116}{JHEP~1103,~116~(2011)}},
\nbbsteprint{\arxivref{1012.5032}{arxiv:1012.5032}}.

\bibitem{Lee:2010du}
\nbbstauthor{S.~Lee},
\nbbsttitle{``{Yangian Invariant Scattering Amplitudes in Supersymmetric
  Chern-Simons Theory}''},
\nbbstjournal{\doiref{10.1103/PhysRevLett.105.151603}{Phys.Rev.Lett.~105,~151603~(2010)}},
\nbbsteprint{\arxivref{1007.4772}{arxiv:1007.4772}}.

\bibitem{Britto:2004nc}
\nbbstauthor{R.~Britto, F.~Cachazo and B.~Feng},
\nbbsttitle{``{Generalized unitarity and one-loop amplitudes in $\mathcal{N}$ =
  4 super-Yang-Mills}''},
\nbbstjournal{\doiref{10.1016/j.nuclphysb.2005.07.014}{Nucl.Phys.~B725,~275~(2005)}},
\nbbsteprint{\arxivref{hep-th/0412103}{hep-th/0412103}}.

\bibitem{Buchbinder:2005wp}
\nbbstauthor{E.~I.~Buchbinder and F.~Cachazo},
\nbbsttitle{``{Two-loop amplitudes of gluons and octa-cuts in $\mathcal{N}$ = 4
  super Yang-Mills}''},
\nbbstjournal{\doiref{10.1088/1126-6708/2005/11/036}{JHEP~0511,~036~(2005)}},
\nbbsteprint{\arxivref{hep-th/0506126}{hep-th/0506126}}.

\bibitem{Bern:1993kr}
\nbbstauthor{Z.~Bern, L.~J.~Dixon and D.~A.~Kosower},
\nbbsttitle{``{Dimensionally regulated pentagon integrals}''},
\nbbstjournal{\doiref{10.1016/0550-3213(94)90398-0}{Nucl.Phys.~B412,~751~(1994)}},
\nbbsteprint{\arxivref{hep-ph/9306240}{hep-ph/9306240}}.

\bibitem{Boos:1987bg}
\nbbstauthor{E.~E.~Boos and A.~I.~Davydychev},
\nbbsttitle{``{A Method Of The Evaluation Of The Vertex Type Feynman
  Integrals}''},
\nbbstjournal{Moscow~Univ.Phys.Bull.~42N3,~6~(1987)}.

\bibitem{Boos:1990rg}
\nbbstauthor{E.~E.~Boos and A.~I.~Davydychev},
\nbbsttitle{``{A Method of evaluating massive Feynman integrals}''},
\nbbstjournal{\doiref{10.1007/BF01016805}{Theor.Math.Phys.~89,~1052~(1991)}}.

\bibitem{Bern:1994zx}
\nbbstauthor{Z.~Bern, L.~J.~Dixon, D.~C.~Dunbar and D.~A.~Kosower},
\nbbsttitle{``One loop n point gauge theory amplitudes, unitarity and collinear
  limits''},
\nbbstjournal{\doiref{10.1016/0550-3213(94)90179-1}{Nucl.Phys.~B425,~217~(1994)}},
\nbbsteprint{\arxivref{hep-ph/9403226}{hep-ph/9403226}}.

\bibitem{Beisert:2012xx}
\nbbstauthor{N.~Beisert, S.~He, B.~U.~W.~Schwab and C.~Vergu},
\nbbsttitle{``{Null Polygonal Wilson Loops in Full $\mathcal{N}$ = 4
  Superspace}''},
\nbbstjournal{\doiref{10.1088/1751-8113/45/26/265402}{J.Phys.~A45,~265402~(2012)}},
\nbbsteprint{\arxivref{1203.1443}{arxiv:1203.1443}}.

\end{thebibliography}

\end{document}